\documentclass[reprint,superscriptaddress,aps,prd,nofootinbib]{revtex4-2}

\usepackage{graphicx}
\usepackage{dcolumn}
\usepackage{bm}
\usepackage{amsmath}
\usepackage{graphicx}

\usepackage{amsthm,amssymb}
\usepackage{mathrsfs,subfigure}
\usepackage{color,multirow,hyperref}
\usepackage{verbatim}
\usepackage{booktabs}

\newcommand \p {\partial}

\allowdisplaybreaks[3]

\def \bal#1\eal  {\begin{align} #1 \end{align}}
\def\({\left(}
\def\){\right)}
\def\[{\left[}
\def\]{\right]}
\def\<{\langle}
\def\>{\rangle}
\def\d{\mathrm{d}}

\newcommand{\f}[2]{\frac{#1}{#2}}
\newcommand{\bim} {\begin{itemize}[noitemsep]}

\newcommand{\eim}{\end{itemize}}
\newcommand{\be} {\begin{equation}}
\newcommand{\ee} {\end{equation}}
\newcommand{\bc}{\begin{center}}
\newcommand{\ec}{\end{center}}

\newcommand{\nn} {\nonumber\\}

\newcommand{\pd} {\partial}
\newcommand{\mc} {\mathcal}

\newcommand{\oi}{\omega}
\newcommand{\epi}{\epsilon}
\newcommand{\thi}{\theta}

\begin{document}

\hfill {{\footnotesize USTC-ICTS/PCFT-23-17}}

\title{Spinning $Q$-ball Superradiance in 3+1D}

\author{Guo-Dong Zhang}
\email[]{guodongz@mail.ustc.edu.cn}
\affiliation{Interdisciplinary Center for Theoretical Study, University of Science and Technology of China, Hefei, Anhui 230026, China}
\author{Fu-Ming Chang}
\email[]{changfum@mail.ustc.edu.cn}
\affiliation{Interdisciplinary Center for Theoretical Study, University of Science and Technology of China, Hefei, Anhui 230026, China}
\author{Paul M. Saffin}
\email[]{paul.saffin@nottingham.ac.uk}
\affiliation{School of Physics and Astronomy, University of Nottingham, University Park, Nottingham NG7 2RD, United Kingdom}
\author{Qi-Xin Xie}
\email[]{xqx2018@mail.ustc.edu.cn}
\affiliation{Interdisciplinary Center for Theoretical Study, University of Science and Technology of China, Hefei, Anhui 230026, China}
\author{Shuang-Yong Zhou}
\email[]{zhoushy@ustc.edu.cn}
\affiliation{Interdisciplinary Center for Theoretical Study, University of Science and Technology of China, Hefei, Anhui 230026, China}
\affiliation{Peng Huanwu Center for Fundamental Theory, Hefei, Anhui 230026, China}
\affiliation{Theoretical Physics, Blackett Laboratory, Imperial College, London, SW7 2AZ, UK}

\date{\today}

\begin{abstract}

Recently, it has been found that a $Q$-ball can amplify waves incident upon it, due to rotation in the internal space and the interaction of the two modes in the complex scalar field. While the spherically symmetric 3D case has been investigated previously, here we explore the 3D axi-symmetric case, which is numerically much more challenging. The difficulty comes because a partial wave expansion is needed, and the different partial waves can not be separated, for either the background spinning $Q$-ball solution or  the perturbative scattering on top of it. A relaxation method and a high dimensional shooting method are applied to compute the $Q$-ball solutions and the amplification factors respectively. We also classify the behavior of the amplification factors and we discuss their bounds and the superradiance criteria. 

\end{abstract}

\maketitle

\tableofcontents

\section{Introduction}
\label{sec:level1}

$Q$-balls represent a class of solitonic solutions in field theories, characterized by configurations that are localized in space with the field rotating in the internal space \cite{Friedberg:1976me,Coleman:1985ki}. The conditions for $Q$-balls to arise are quite broad, including the case of self-interacting complex scalar field theories, in which the potential grows slower than the quadratic mass term away from its minimum, facilitating the condensation of charges. In contrast to topological solitons, whose stability arises from topological charges, $Q$-balls are non-topological solitons. Their stability is exclusively due to Noether charges. While initial investigations of $Q$-balls focused on non-rotating, spherically symmetric forms, subsequent studies have revealed the existence of spinning $Q$-balls \cite{Volkov:2002aj,Kleihaus:2005me,Radu:2008pp,Kleihaus:2011sx,Herdeiro:2014pka,Almumin:2023wwi}. Notably, it was found that the angular momentum of spinning $Q$-balls is proportional to the associated Noether charge. Multiple $Q$-balls can form composite structures, which can be either stable or long-lived \cite{Copeland:2014qra,Xie:2021glp,Hou:2022jcd}. Typically, classical theories are employed to investigate the properties of $Q$-balls and their composite counterparts. Nevertheless, recent research has also made progress in exploring these phenomena within the framework of quantum theories \cite{Tranberg:2013cka,Xie:2023psz,Kovtun:2020udn}. Aside from their theoretical interest, $Q$-balls have found uses in a cosmological setting, such as the Affleck-Dine baryogenesis scenario in supersymmetric extensions of the Standard Model. After formation these $Q$-balls decay into baryonic matter or remain stable and act as self-interacting dark matter, influencing the evolution of the universe \cite{Kusenko:1997si,Enqvist:1997si,Fujii:2002kr,Enqvist:2003gh,Roszkowski:2006kw,Shoemaker:2009kg,Zhou:2015yfa,Kawasaki:2019ywz,Gouttenoire:2021jhk,Kasai:2022vhq,ElBourakadi:2023pue}. When the gravitational effect is significant, analogous soliton solutions are dubbed boson stars \cite{Kaup:1968zz,Colpi:1986ye}. If the constituent scalar field is sufficiently light, boson stars can achieve astronomical masses comparable to generic fermionic stars and become extremely compact. Boson stars may serve as dark matter sources and black hole 
candidates \cite{Schunck:2003kk,Chavanis:2011zm,Liebling:2012fv,Herdeiro:2014goa,Cardoso:2019rvt,Visinelli:2021uve}. In addition to cosmology and relativistic field theory, $Q$-balls are also produced and investigated experimentally in cold atom systems \cite{Enqvist:2003zb,Bunkov:2007fe}. 


The concept of superradiance, originally introduced by Dicke to describe radiation enhancement in a coherent medium \cite{Dicke:1954zz}, has been extended to encompass a wide range of phenomena associated with enhanced radiation. A famous example is the rotational superradiance discovered by Zel'dovich, demonstrating that the energy and angular momentum of a rotating cylinder with absorbing surfaces can be extracted via the scattering of incident waves \cite{Zeld1,Zeld2}. In the realms of relativity and astrophysics, Reissner-Nordstr\"{o}m black holes and Kerr black holes have been found to be capable of inducing superradiance and in some cases superradiant instabilities can arise. This property is employed in the search for dark matter and many related questions are investigated in recent years \cite{Cardoso:2004hs,Dolan:2007mj,Arvanitaki:2009fg,Bredberg:2009pv,Arvanitaki:2010sy,Pani:2012vp,Berti:2015itd,Marsh:2015xka,Cardoso:2015zqa,East:2017ovw,Baryakhtar:2017ngi,Baumann:2018vus,Berti:2019wnn,Zhu:2020tht,Zhang:2020sjh,Stott:2020gjj,Baryakhtar:2020gao,Mehta:2021pwf,Roy:2021uye,Chen:2022nbb,Siemonsen:2022yyf}. Reviews of this rapidly developing field can be found in \cite{Bekenstein:1998nt,Brito:2015oca}. 


Recently, superradiance in $Q$-ball systems has been identified, enabling energy, charge and angular momentum to be extracted from a $Q$-ball through incident waves \cite{Saffin:2022tub}. The underlying reason is that the coherent internal rotation of the scalar field in the complex plane allows the transfer of energy and other quantities among coupled wave modes occur. For certain ranges of wave frequency, the system experiences a net extraction of energy that exceeds the energy input from the incident waves, leading to the manifestation of superradiance. The rotation of the background $Q$-ball further influences the superradiance process. It was soon realized that this mechanism can apply to generic $Q$-ball-like, time-periodic solitons and specific cases of boson stars and their Newtonian limits have been investigated \cite{Cardoso:2023dtm, Gao:2023gof}. This emerging phenomenon may offer novel approaches to the search for new particles and the identification of exotic compact objects. 


Although some key ingredients of $Q$-ball superradiance have been identified in previous studies, they primarily delved into relatively straightforward cases, such as $Q$-balls in 2+1D and non-rotating, spherical $Q$-balls in 3+1D. In this paper we study the more realistic case of solitions in three dimensions, with the inclusion of angular momentum. As we delve into our analysis, it becomes clear that the introduction of angular momentum, which specifies a preferred direction in space, makes the phenomena significantly more complex and diverse. Moreover, this complexity also introduces numerous technical challenges. When doing a mode analysis, both the equations governing the background $Q$-balls and the perturbations form infinite sets of coupled equations, categorized by spherical harmonics $Y_l^m(\cos\theta)$. Achieving solutions with high accuracy necessitates truncating these equations at a relatively high degree $l$, which means that a large number of equations have to be solved simultaneously. Furthermore, the dimension of the parameter space is also greatly increased, posing an additional challenge. Addressing these difficulties is a focus of our work, and we shall investigate these intricate phenomena using representative cases. 


The paper is organized as follows. In Section \ref{sec:level2}, we introduce the fiducial field model to be studied, review basic properties of spinning $Q$-balls in 3+1D and explain the methods used to solve for spinning $Q$-ball solutions. In Section \ref{sec:level3}, we derive the equations expanded with spherical harmonics for perturbation waves and define various amplification factors, some of which depend on the polar angle. The details of the numerical methods used to solve spinning $Q$-balls and perturbative waves are explained in Appendix \ref{sec:levela}, while the accuracy of the results and the convergence test are presented in Appendix \ref{sec:levelb}. In Section \ref{sec:level4}, we show the numerical results of spinning $Q$-ball superradiance. Addtionally, we perform an analytic examination of the asymptotic behaviour of the amplification factors as the wave frequency approaches infinity or the mass gap. We identify the criteria for superradiance and establish bounds on the amplification factors that we use in this paper. We summarize in Section \ref{sec:level5}.

\section{Spinning $Q$-balls}
\label{sec:level2}

In this section, we construct the spinning $Q$-ball solution in 3+1D \cite{Volkov_2002}, which provides a background for the waves to scatter on, and from which energies and angular momentum can be extracted. While a spinning $Q$-ball in 2+1D can be easily obtained by solving an ordinary differential equation (ODE) with a 1D shooting method, a 3+1D spinning $Q$-ball necessitates solving a system of coupled ODEs involving several partial waves. This can be handled with a more intricate higher dimensional shooting method or relaxation method.

\begin{figure*}
	\centering
	\subfigure[$\ell_{\max}=11$]{
		\includegraphics[height=3.6cm]{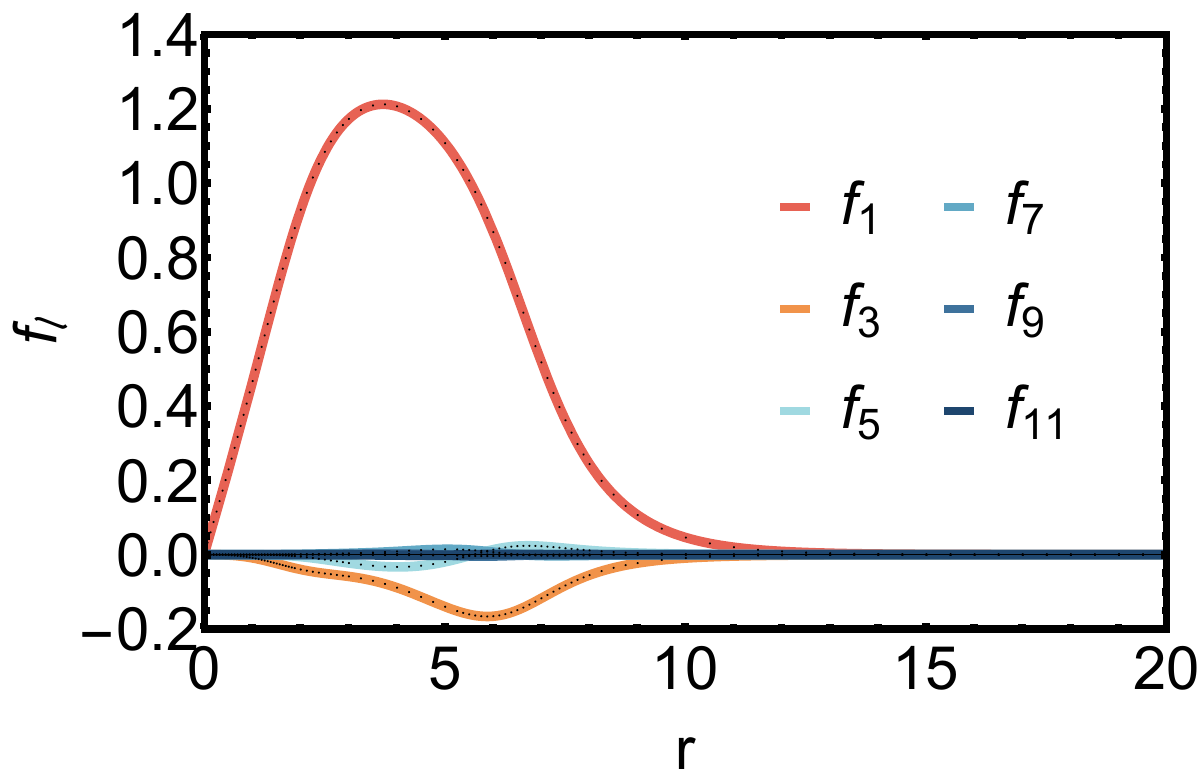}
	}
	\subfigure[$$]{
		\includegraphics[height=4.3cm]{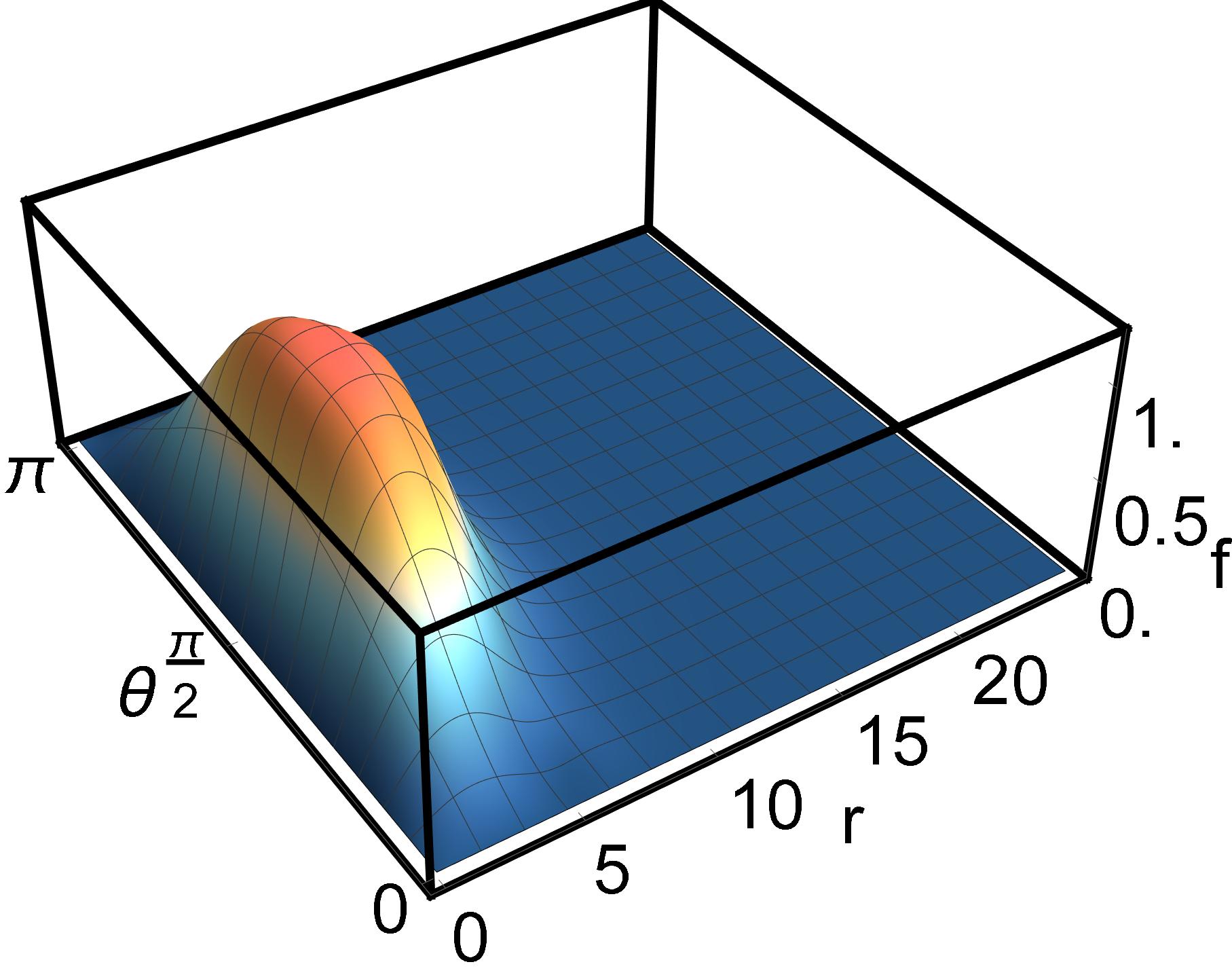}
	}
	\subfigure[$$]{
		\includegraphics[height=3.6cm]{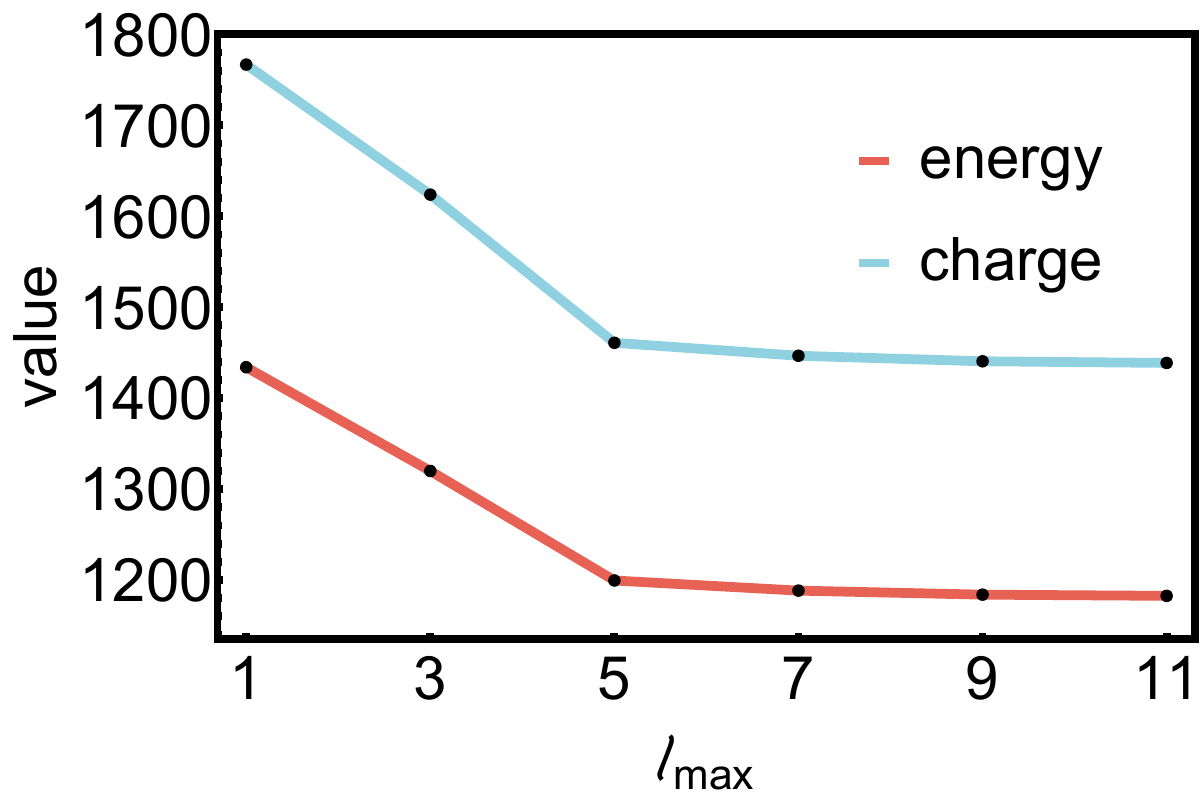}
	}
	\caption{Radial amplitudes $f_\ell$ for the even parity  spinning $Q$-ball ($\Phi$ being a scalar) in subfigure (a). The mode cut-off is $\ell_{\max} = 11$ or, equivalently,  $k_{\max}^Q+1=6$ modes in the partial wave expansion. The other parameters are $m_Q = 1, \omega_Q = 0.7, g = 1/3$, and all calculations in this paper use this set of parameters unless otherwise stated. The (b) subfigure plots the $r$ and $\thi$ dependence for $\ell_{\rm max}$. The (c) subfigure plots how the total energy and charge converge with $\ell_{\rm max}$. }
 \label{Fig:Qballs_even}
\end{figure*}

We will consider a U(1) symmetric complex field in 3+1 dimensional spacetime, whose effective Lagrangian density\footnote{We use a mostly positive signature for the spacetime metric throughout.} is given by
\be
        \widetilde{\mathcal{ L }} = - \widetilde\p^{{\mu}} \widetilde{\Phi}^* \widetilde\p_{{\mu}} \widetilde{\Phi} - V  ,~~~
        V = \widetilde{m}^2 \big| \widetilde{\Phi} \big|^2 - \widetilde\lambda \big| \widetilde{\Phi} \big|^4 + \widetilde{g} \big| \widetilde{\Phi} \big|^6 ,
\ee
where the parameters are chosen such that $\widetilde{\Phi} = 0$ is the true vacuum. We can reduce the number of relevant parameters to one by introducing the dimensionless variables
\be
  x_\mu=  \widetilde{m} \widetilde{x}_{\mu},~~  \Phi =\sqrt{\widetilde\lambda} \f{\widetilde{\Phi}}{\widetilde{m}}  , ~~ g =  \widetilde{g} \f{\widetilde{m}^2}{\widetilde\lambda^2} ,  
\ee
such that we may work with the rescaled Lagrangian density
\be
\label{Lagstarting}
\mathcal{ L } =-\p^{\mu} {\Phi}^* \p_{{\mu}}{\Phi} - V,~~~
V = \left| \Phi \right|^2 - \left| \Phi \right|^4 + g \left| \Phi \right|^6 .
\ee
The conserved charge associated with the global U(1) symmetry is 
\begin{align}
    Q = i \int {\rm d}^3x \left( \Phi^* \Dot{\Phi} - \Phi \Dot{\Phi}^* \right) , 
\end{align}
where a dot means a time derivative $\Dot{\Phi} = {\partial\Phi}/{\partial t}$, and the energy-momentum tensor for the scalar field has components
\begin{align}
	T_{\mu\nu} = \p_{\mu} \Phi^* \p_\nu \Phi + \p_{\mu} \Phi \p_\nu \Phi^* + g_{\mu\nu} \mathcal{ L } ,  
\end{align}
where $g_{\mu\nu}$ is the Minkowski metric.

In spherical coordinates $(t,r,\thi,\varphi)$, the ansatz for a spinning $Q$-ball has the form
\begin{align}
    \Phi_Q =  f(r,\theta) e^{-i(\omega_Q t -m_Q \varphi)} , \label{ans_qball}
\end{align}
where $m_Q$ is an integer, $m_Q = 0$ for non-spinning configurations and $m_Q \ne 0$ for spinning configurations, and $f(r,\theta)$ is the profile function or amplitude that depends on the polar angle $\thi$. 
For a stable, spherical $Q$-ball to exist, $\omega_Q$ must be real and satisfy the following conditions \cite{Coleman:1985ki}: 
\begin{align}
    \omega_Q^2 & > \omega_{min}^2  \equiv \min_f \left( \frac{V}{f^2} \right) = 1 - \frac{1}{4g} , \\
    \omega_Q^2 & < \omega_{max}^2  \equiv \frac{1}{2} V''(0) = 1 .
\end{align}
 For the spinning case, the stability range seems to be roughly the same, as can be checked numerically, and 
we can choose $\omega_Q>0$ without loss of generality. With a nonzero $\omega_Q$, the $Q$-ball rotates in the internal field space with angular velocity $\omega_Q$; with additionally a nonzero $m_Q$, the $Q$-ball also rotates in real space with angular phase velocity $\Omega_Q\equiv \omega_Q/m_Q$. For a given profile $f(r,\thi)$,
the U(1) charge, energy and angular momentum of the spinning $Q$-ball are respectively given by
\begin{align}
	Q &= 4 \pi \omega_Q \int {\rm d}r {\rm d}\theta r^2 \sin \theta f^2 , 
	\\
    E  &= 2 \pi \int {\rm d}r {\rm d}\theta r^2 \sin \theta {T_{tt}} , 
    \\
    L &= 2 \pi \int {\rm d}r {\rm d}\theta r^2 \sin \theta {T_{t\varphi}} = m_Q Q .
\end{align}
where ${T_{tt}} =  (\p_r f)^2 + \frac{1}{r^2} (\p_\theta f)^2  + \frac{m_Q^2 f^2}{r^2 \sin^2 \theta} + (\omega_Q f)^2 + V $.

With the $Q$-ball ansatz Eq.~\eqref{ans_qball}, the field equation reduces to 
\begin{align}
    \bigg( \p_r^2 + \frac{2}{r} \p_r +   \f{\p_\theta^2}{r^2} + \frac{\cos \theta}{\sin \theta } \f{\p_\theta}{r^2} &- \frac{m_Q^2}{{r^2}\sin^2 \theta } + \omega_Q^2 \bigg) f 
    \nonumber\\
     &~~~= f - 2f^3 + 3g f^5 .
    \label{equqball}
\end{align}
To obtain a $Q$-ball solution, which has a finite energy, we also need to supply the field equation with appropriate boundary conditions. The profile function of a spinning $Q$-ball $f(r,\thi)$ must decay to zero as $r$ goes to 0 or infinity, the exact form depending on the parity of the scalar $\Phi$.

\begin{figure*}
	\centering
	\subfigure[$\ell_{\max}=12$]{
		\includegraphics[height=3.6cm]{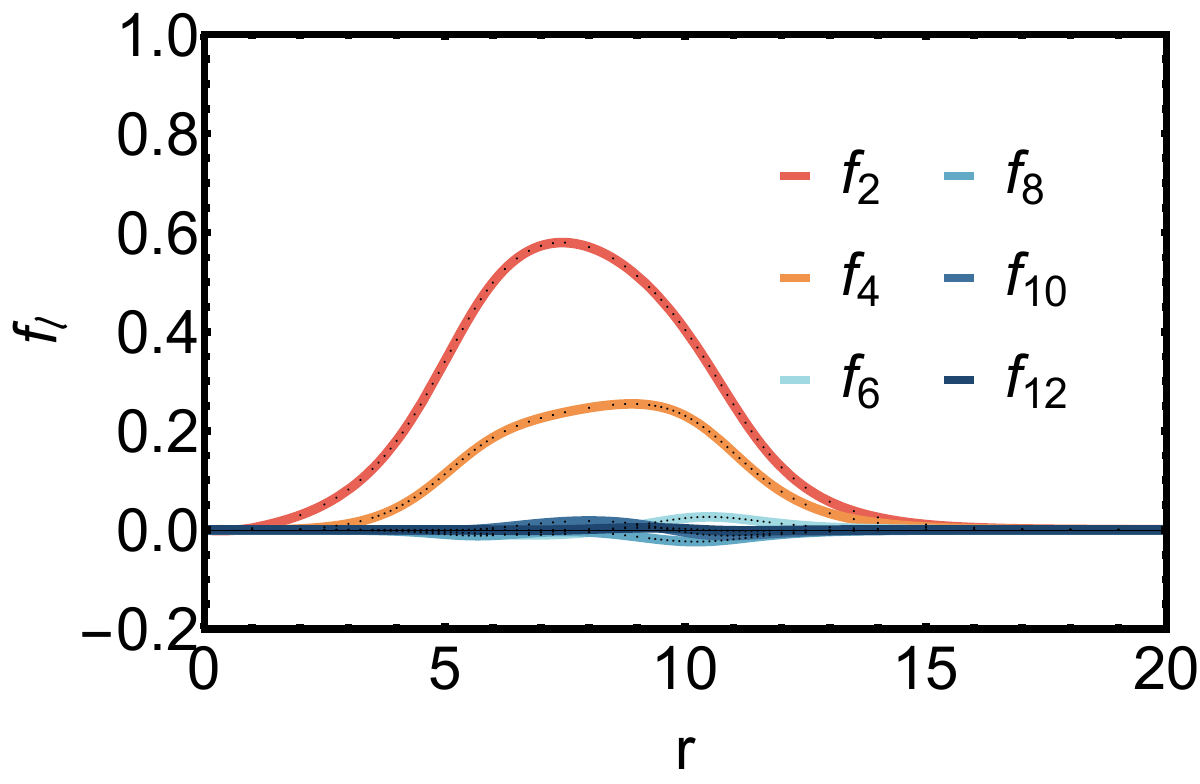}
	}
	\subfigure[$$]{
		\includegraphics[height=4.3cm]{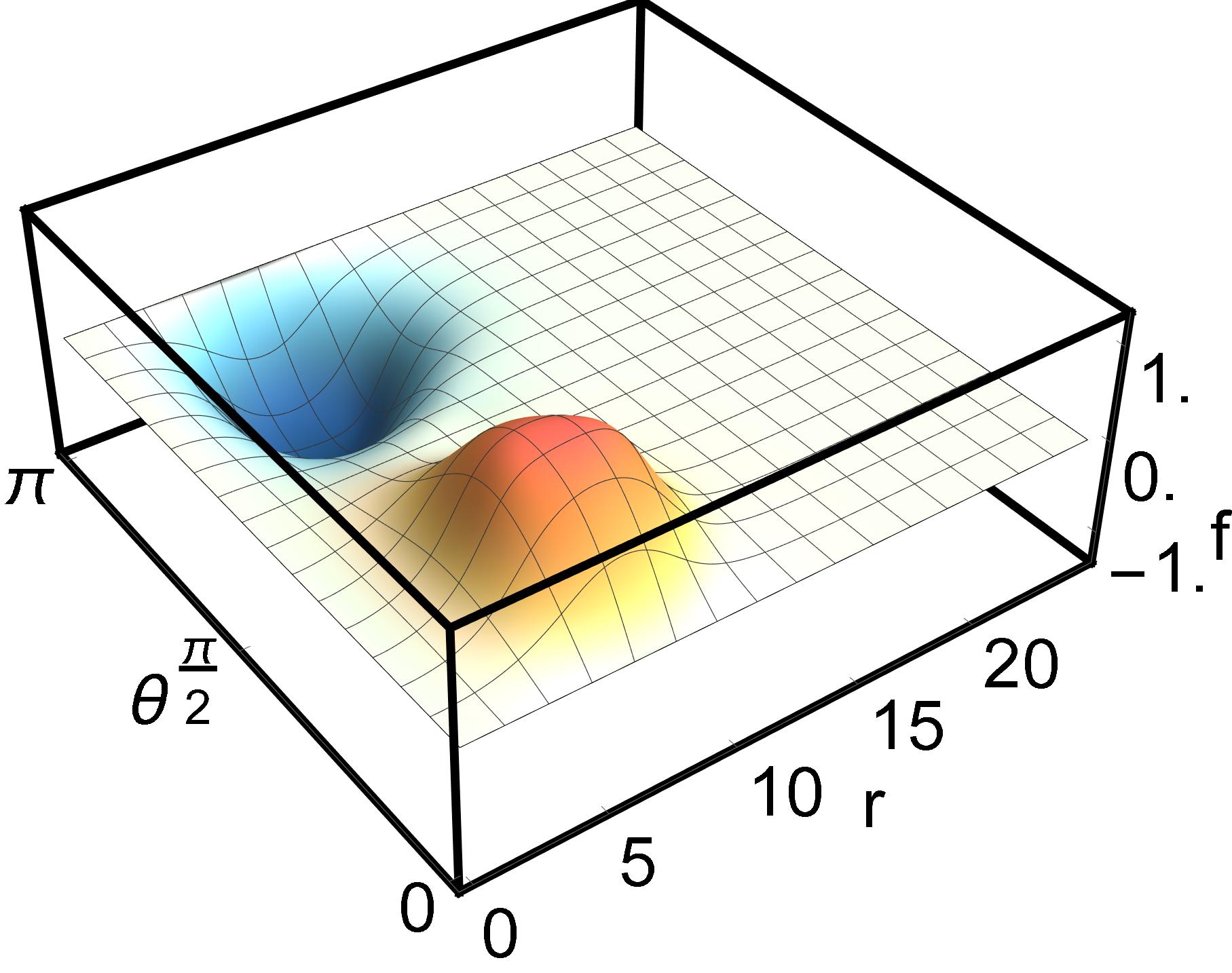}
	}
	\subfigure[$$]{
		\includegraphics[height=3.6cm]{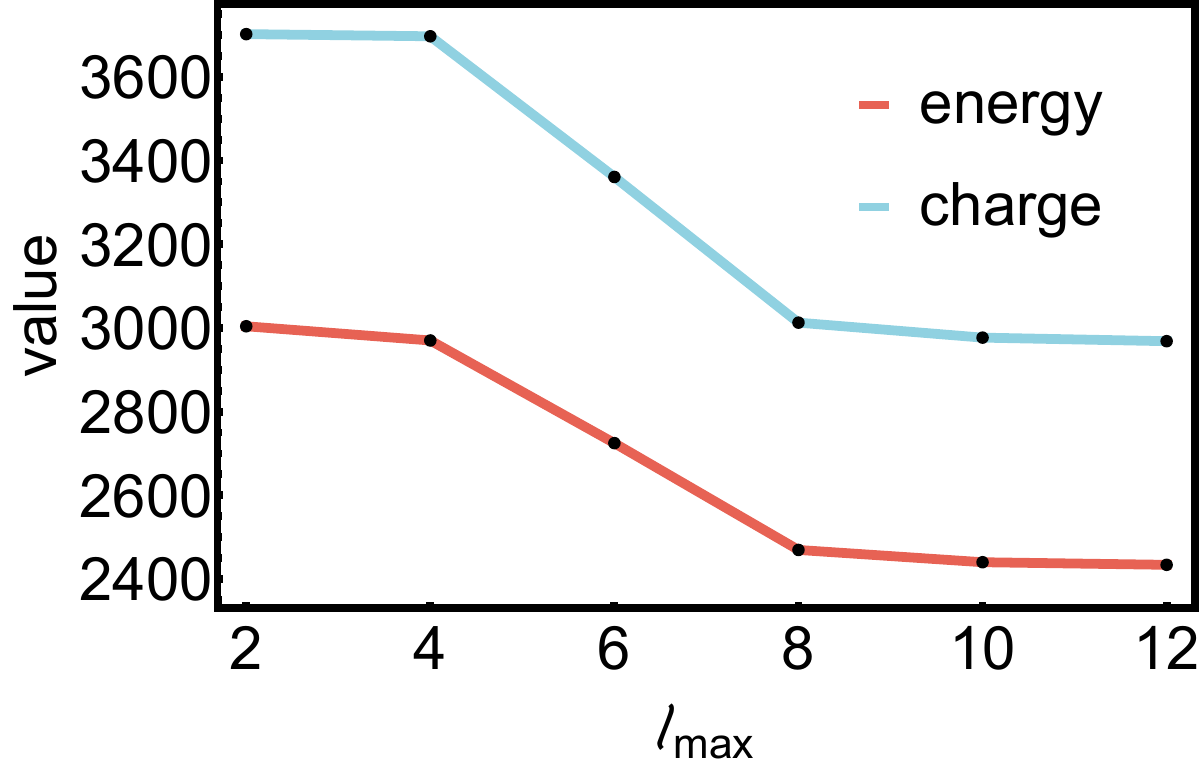}
	}
	\caption{Radial amplitudes $f_\ell$ for the odd parity  spinning $Q$-ball ($\Phi$ being a pseudo-scalar) in subfigure (a). The mode cut-off is $\ell_{\max}=12$ or,  equivalently, $k_{\rm max}^{Q}+1=6$ modes in the partial wave expansion. The other parameters are $m_Q = 1, \omega_Q = 0.7, g = 1/3$. From subfigure (b), we see that the profile function $f(r,\thi)$ has two peaks, compared to one peak for the even parity case. The (c) subfigure plots how the total energy and charge converge with $\ell_{\rm max}$.}
 \label{Fig:Qballs_odd}
\end{figure*}

To solve the PDE \eqref{equqball} along with the boundary conditions, we can perform an angular mode expansion on $f(r,\thi)$:
\begin{align}
    f(r,\theta) = \sum_{\ell={\Bar{m}}_Q}^{\infty} {f_{\ell} (r) } P^{m_Q}_{\ell} (\cos \theta)  , 
    \label{qballsansatz}
\end{align}
where $P^{m_Q}_{\ell} (\cos \theta)$ are the associated Legendre functions and the over bar denotes the absolute value 
\be
{\Bar{m}}_Q \equiv \left| m_Q \right| .
\ee
Note that the field equation \eqref{equqball} is invariant under parity $(r,\thi)\to (r,\pi-\thi)$ and the associated Legendre function satisfies $P^{m_Q}_{\ell} (-\cos \theta)=(-1)^{\ell+m_Q}P^{m_Q}_{\ell} (\cos \theta)$. So if $\Phi$ is a scalar, meaning $f(r,\pi-\thi)=f(r,\thi)$, we only have even modes 
\be\label{eq:f_scalar}
f(r,\theta) = \sum_{k=0}^{\infty} {f_{{\Bar{m}}_Q+2k} (r) } P^{m_Q}_{{\Bar{m}}_Q+2k} (\cos \theta) ,
\ee
and if $\Phi$ is a pseudo-scalar, meaning $f(r,\pi-\thi)=-f(r,\thi)$, we only have odd modes
\be\label{eq:f_pseudoscalar}
f(r,\theta) = \sum_{k=0}^{\infty} {f_{{\Bar{m}}_Q+1+2k} (r) } P^{m_Q}_{{\Bar{m}}_Q+1+2k} (\cos \theta)  . 
\ee
Thus, a clear distinction is that the profile function $f(r,\thi)$ vanishes at the equatorial plan $\theta = \pi/2$ for the odd parity/pseudo-scalar case, but is nonzero for the even parity/scalar case. The two cases should be explored separately.

Plugging the angular mode expansion into the field equation for $f(r,\thi)$ and expanding the nonlinear terms $f^n=(\sum_\ell f_\ell P^{m_Q}_\ell)^n$ in the basis of $P^{m_Q}_{\ell}$, the PDE field equation can be split into a system of coupled ODEs. In  our numerical evaluations, we cut off $\ell$ at some finite $\ell_{\rm max}$ where the energy of the spinning $Q$-ball has reached convergence, and the coupled ODEs are of the following form: 
\begin{align}
\label{fellequation}
    &\! \left(\p^2_r + \frac{2}{r} \p_r +  \omega_Q^2 - 1 - \frac{\ell (\ell+1)}{r^2} \right)  f_{\ell} + \mathcal{V}_{\ell} (f_{\ell'}) = 0 , \!
\end{align} 
where $\ell = \bar m_Q + s + 2 k$, $k=0,1,2,\cdots$, $s=0$ for even parity and $s=1$ for odd parity, and $\mathcal{V}_{\ell} (f_{\ell'})$ are nonlinear polynomial functions of $f_{\ell'}$ with multiple $\ell'$ that are generally different from $\ell$. As $r$ goes to zero or infinity, the nonlinear terms are negligible for a spinning $Q$-ball solution,  
so we can infer from \eqref{fellequation} that the mode functions $f_{\ell}$ should have the following asymptotic form
\begin{align}
	 f_{\ell} (r \to \infty)  \to 0 ,~~~
 f_{\ell} (r \to 0) \to (\kappa_{\ell} r)^{\ell} , 
\end{align}
where $\kappa_{\ell}$ are constants.

The above set of coupled ODEs can be solved with either a high dimensional shooting method or a relaxation method \cite{NRC}. Later, for the perturbative waves scattering on the $Q$-ball, we will use a high dimensional shooting method. For the background $Q$-ball solution, we will present the solutions with the relaxation method, which produces more accurate results with high efficiency. The Relaxation method transforms, via finite-difference, the coupled ODEs into a set of matrix equations that include every discretized point in the solving range, and solves it with an initial guess and subsequent iterations. By comparison, the high dimensional shooting method solves the coupled ODEs from a point near $r=0$ to a large $r_*$ to establish a numerical map between the ``initial'' conditions near $r=0$ and the ``final'' conditions $r_*$, and then, by matching to the correct boundary conditions, this map can be turned into a set of coupled equations that can be solved to find the desired $Q$-ball. Both of the two methods will be explained in Appendix \ref{sec:levela}.

We note from (\ref{eq:f_scalar}) and (\ref{eq:f_pseudoscalar}) that the $k$ index of the sum is related to the angular momentum number, $\ell$, by $\ell_{\rm max} = \Bar{m}_Q +2k+ s$, and while the exact solution would require $k$ to range from zero to infinity, in numerical solutions we must impose a cut-off. Convergence is ensured by varying this cut-off. The number of the cut-off modes is $N_{max}=k_{\rm max}^{Q}+1 = (\ell_{\rm max} - \Bar{m}_Q - s +2 )/2$. In Figure \ref{Fig:Qballs_even} and Figure \ref{Fig:Qballs_odd}, we present two spinning $Q$-ball solutions for the same number of modes, $k_{\rm max}^{Q}+1=6$, one for the even parity and one for the odd parity.
The parameters used are $m_Q = 1, \omega_Q = 0.7 , g = 1/3$, which will be our fiducial case unless otherwise stated. We see that the fiducial spinning $Q$-ball solutions converges very quickly with $\ell_{\rm max}$. The first mode profile $f_1$ for the even parity case changes only slightly as $\ell_{\rm max}$ increases, while the first mode for the odd parity case $f_2$ reduces significantly when $\ell_{\rm max}\geq 4$, which seems to be due to the fact that the second mode $f_4$ is sizable for the odd case. For both the even and odd case, the first couple of $\ell$ modes contain most of the energy. The odd parity $Q$-ball is more energetic compared to the event parity one --- both the energy and charge of the odd parity one is about twice the even parity one. This is mainly because the even parity case has one peak in the profile $f(r,\thi)$, while the odd parity case has two peaks.

\section{Perturbative scattering}
\label{sec:level3}

Having constructed the spinning $Q$-ball solutions, in this section we study the waves scattering on top of the $Q$-ball background. We will work with small scattering waves and so will only keep the linear perturbations in the equations of motion. Note that the linear approximation still takes into account the backreaction of the scattering waves on the $Q$-ball. That is, the perturbative fields contain the information about both the scattering waves and the corrections to the $Q$-ball background. The various amplification factors will be defined, which can be used to describe the energy and angular momentum enhancements in the scattering from various angles.

\subsection{Perturbative waves}
\label{sec:pertwaves}

Now, we consider small perturbations $\phi$ on top of a $Q$-ball background solution $\Phi_Q$
\begin{align}
    \Phi = \Phi_Q + \phi , 
\end{align}
The linear perturbations satisfy the following equation of motion
\bal
\label{phimaineq}
\Box \phi & = \left. \frac{\p^2 V}{\p \Phi^* \p \Phi} \right|_{\Phi_Q} \phi + \left. \frac{\p^2 V}{\p (\Phi^{*})^2} \right|_{\Phi_Q} \phi^* , 
\nonumber\\
	& = \left( 1 + U \right) \phi + W e^{-2i(\omega_Q t - m_Q \varphi)} \phi^* , 
\eal
where $U$ and $W$ are determined by the background $Q$-ball solution
\begin{align}
	U(r,\theta) & = \frac{\p }{\p (f^2)} \left( f^2 \frac{\p V}{\p (f^2)} \right) - 1 = - 4 f^2 + 9 g f^4  , \\
	W(r,\theta) & = f^2 \frac{\p^2 V}{(\p (f^2))^2} = -2 f^2 + 6 g f^4 .
\end{align}
Here, $\Box$ is the Minkowski d'Alembertian, $U$ and $W$ depend only on the background $Q$-ball, and they both approach zero as $r \to \infty$, thanks to the asymptotic behavior of the spinning $Q$-ball amplitude $f$. As have been shown in the last section, for a spinning $Q$-ball where $m_Q\neq 0$, $f$ also approach zero as $r \to 0$. Given that the scalar is complex, we may compliment  Eq.~\eqref{phimaineq} with its complex conjugate.

The perturbative equations of motion become easier to solve by Fourier transforming to the frequency domain, ``factoring out'' the time dependence. Equivalently, since $\Phi$ or $\phi$ inherently contains two coupled modes, we can minimally consider a scattering involving the following two modes for $\phi$:
\begin{align}
\label{phietaeta}
    \phi &= \eta^+\!(r,\theta) e^{-i(\omega_+ t - m_+ \varphi)} + \eta^-\!(r,\theta) e^{-i(\omega_- t - m_- \varphi)} , 
\end{align}
where 
\be
 \omega_{\pm} = \omega_Q \pm \omega ,~~~ m_{\pm} = m_Q \pm m .
\ee 
(Although it would be more explicit to express $\eta^\pm$ as $\eta^{\omega_\pm,m_\pm}$, we opt for the former to minimize clutter, and the same rationale applies to other quantities defined in the following.)
Using this ansatz (or by Fourier transform), the equations of motion become  
\begin{align}
	\left( \Box + k_{\pm}^2 \right) \eta^{\pm} = U(r,\theta) \eta^{\pm} + W(r,\theta) \eta^{\mp*}\label{emo} ,
\end{align}
where $ k_{\pm}^2 = \omega_{\pm}^2 - 1 $. As we seek a propagating solution, we can impose a physical condition on the wave numbers
\begin{align}
	\left| \omega_Q \pm \omega \right| > 1 . 
\end{align}

Again, similar to solving the background $Q$-ball solution, since $\eta^\pm$ are functions of $r,\theta$, we can perform an angular mode expansion for them to turn the PDE into a set of ODEs:
\begin{align}
\label{etapmExp}
	\eta^{\pm} ( r,\theta ) & = \sum\limits_{\ell=\Bar{ m}_{\pm} }^{\infty} \eta^{\pm}_{\ell}(r) P^{m_{\pm}}_{\ell} ( \cos \theta ) , 
\end{align}
where we have defined $\Bar{{m}}_{\pm} = \left| m_Q \pm m \right|$. Since the $\eta^\pm$ equation of motion is linear and invariant under parity, the even and odd parity modes are decoupled, so we can discuss them separately. 
Substituting Eq.~\eqref{etapmExp} into the equation of motion and integrating both sides against $P^{m_{\mp}}_{\ell}$, the PDE for $\eta^\pm$ splits into a set of ODEs for $\eta^{\pm}_\ell(r)$: 
\begin{align}
     & \left( \p_r^2 + \frac{2}{r} \p_r - \frac{\ell (\ell + 1 )}{r^2} + k_{\pm}^2 \right) \eta^{\pm}_{\ell} \nonumber \\ 
     & \qquad \qquad = \sum\limits_{\ell'=\Bar{m}_{\pm} }^{\ell_{max}^{\pm}} \left( \mathcal{U}^{\pm}_{\ell' \ell}(r)  \eta^{\pm}_{\ell'}  + \mathcal{W}^{\mp}_{\ell' \ell}(r) \eta^{\mp*}_{\ell'} \right) ,
     \label{modes-equ}
\end{align}
where  $\ell_{max}^{\pm} = \Bar{m}_{\pm} + s + 2 N_{\rm max}^{\pm} - 2 $ , $N_{\rm max}^{\pm}$ is the number of the cut-off modes and  
\begin{align}
		\mathcal{U}^{\pm}_{\ell' \ell}  &= C^{\pm}_{\ell}  \int_{-1}^{1} {\rm d} c_{\thi} U(r,\theta) P^{m_{\pm}}_{\ell'}(c_{\thi}) P^{m_{\pm}}_{\ell}(c_{\thi}) , \\
		\mathcal{W}^{\mp}_{\ell' \ell}  &= C^{\pm}_{\ell}  \int_{-1}^{1} {\rm d} c_{\thi} W(r,\theta) P^{m_{\mp}}_{\ell'}(c_{\thi}) P^{m_{\pm}}_{\ell}(c_{\thi}) ,
		\\
		c_{\thi} &\equiv \cos \theta, ~~~~  ( C^{\pm}_\ell )^{-1} = \int_{-1}^{1} {\rm d}c_{\thi} (P^{m_{\pm}}_{\ell}(c_{\thi}))^2 .
\end{align}
We see that the $W$ (or $\mathcal{W}$) function couples the $\eta^+$ and $\eta^-$ modes so that they can not be separated. The coupling of the $+$ and $-$ modes is essential for the energy enhancement to happen.   

The perturbative equations of motion also need to be supplied with appropriate boundary conditions. The background spinning $Q$-ball solution asymptotes to the Minkowski vacuum both as $r\to 0$ and $r\to \infty$, implying that $U,W$ and thus $\mathcal{U},\mathcal{W}$ asymptotes to zero as $r\to 0$ and  $r \to \infty$. Thus,  the $+$ and $-$ modes are decoupled asymptotically. In the $r \to \infty$ region, they can be described by spherical waves. In the $r\to 0$ region, we still have regularity conditions, similar to the case of the background $Q$-ball solution. Therefore, we have the following boundary conditions at the origin and the infinity
\begin{align}
		\eta^{\pm}_\ell &\to  F^{\pm}_{\ell} (k_{\pm} r)^\ell ,\quad r \to 0 , \\
		\eta^{\pm}_\ell &\to \frac{A^{\pm}_\ell}{k_{\pm}r} e^{ik_{\pm}r} + \frac{B^{\pm}_\ell}{k_{\pm}r} e^{-ik_{\pm}r} ,\quad r \to \infty , \label{Eq:modesinfity}
\end{align}
where $F^{\pm}_{\ell}, A^{\pm}_\ell, B^{\pm}_\ell$ are complex constants.

\begin{figure}
	\centering
		\includegraphics[height=4.1cm]{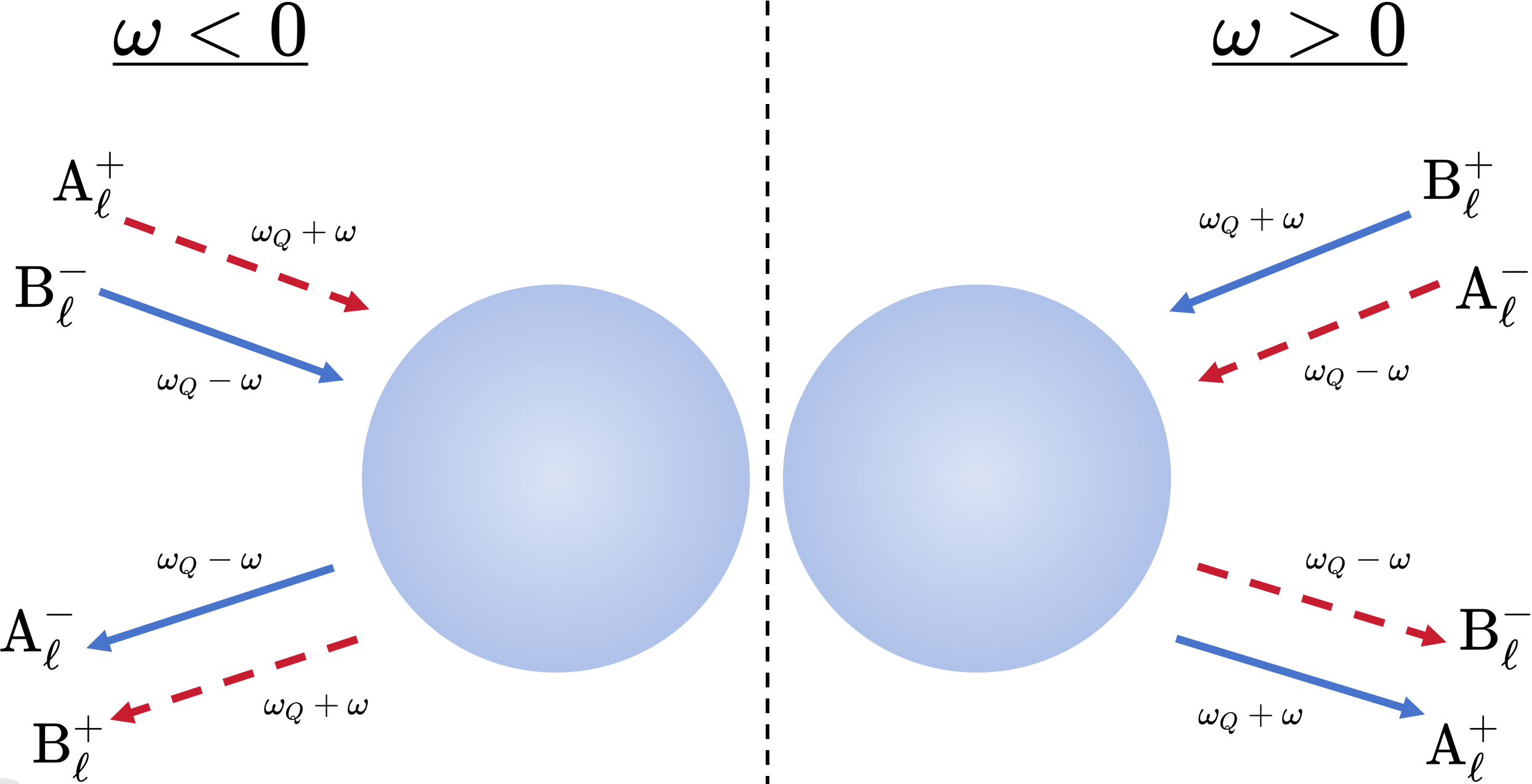}
	\caption{ Ingoing and outgoing waves scattering on and off a $Q$-ball (cf. Eqs. \eqref{phimaineq}, \eqref{phietaeta} and \eqref{Eq:modesinfity}).  
Solid (dashed) lines represent positive (negative) charge. }
 \label{fig:inout}
\end{figure}

Let us now describe the interpretation of the asymptotic solutions. We are considering a scattering problem where {\it in general} there can be two ingoing wave modes and two outgoing wave modes, with frequencies $\omega_+$ and $\omega_-$ respectively. By the asymptotic solutions at $r\to \infty$ and Eq.~\eqref{phietaeta}, we can recognize the physical meaning of $A_\ell^{\pm}$ and $B_\ell^\pm$ as giving ingoing or outgoing waves, depending on the sign of $\oi$, as shown in Figure ~\ref{fig:inout}.  

With the above setup, we can now treat Eq.~\eqref{modes-equ} as an initial value problem with $r$ playing the role of time. In other words, for a given set of $F^{\pm}_{\ell}$ near $r=0$, we can evolve Eq.~\eqref{modes-equ} to a large $r$, which will produce a set of $A^{\pm}_\ell$ and $B^{\pm}_\ell$ at the large $r$. From this point of view, if we truncate the angular modes to $\ell_{\rm max}^{\pm}$ and we choose the same cut-off mode number $N_{\rm max} = N_{\rm max}^+ = N_{\rm max}^-$, the naive number-count of required initial data is $4 N_{\rm max}$, since $F^{+}_{\ell}$ and $F^{-}_{\ell}$ are complex. However, since Eq.~\eqref{modes-equ} is linear, there is an overall complex scaling of the solution that can not be fixed, which reduces the number-count by 2 to $4 N_{\rm max}-2$. This complex scaling can be used to, say, fix 
\be
F^{+}_{{\Bar{m}}_{+}} = 1 ,
\ee
which will be the choice we make in this paper.

It is instructive to do a counting of the dimensions of the solution space from another point of view. After the $\ell_{\rm max}^{\pm}$ truncation, Eq.~\eqref{modes-equ} is a linear system of $2 N_{\rm max}$ (the 2 coming from the $+$ and $-$ modes) complex second order ODEs. So if one starts to evolve the system from a large $r$ inwards, naively, we require $8 N_{\rm max}-2$ parameters (again the $-2$ coming from the unfixed overall complex scaling and the 8 being due to 4 types of complex constants $A^\pm_\ell,B^\pm_\ell$), and this seems to match the freedom in choosing $A^{\pm}_{\ell}$ and $B^{\pm}_{\ell}$. However, solving from large $r$ inwards is not purely an initial value problem, we have $4 N_{\rm max}$ regularity conditions at $r=0$, which counts as boundary conditions and reduces the degrees of freedom to $4 N_{\rm max}-2$. This agrees with the counting from $r=0$ outwards.

\subsection{Amplification factors}
\label{sec:3b}

Let us now define the amplification factors between the outgoing and ingoing scattering waves. Because the system inherently contains two coupled modes with different phase velocities, we can define a number of alternative amplification factors for the relevant physical quantities.

First, since the amplitude of the background $Q$-ball solution decays to zero exponentially at large $r$, the $\mc{O}(\phi^0)$ and $\mc{O}(\phi^1)$ terms are negligible at large $r$. So the energy density of the scattering waves as $r\to \infty$ can be approximated by\footnote{We will use the = symbol even though this and other similar equations only hold to leading order asympototically.}
\be
    {T_{tt}} = \left| \p_t \phi \right|^2  + \left| \p_r \phi \right|^2 + \frac{\left| \p_\theta \phi \right|^2}{r^2}   + \frac{\left| \p_\varphi \phi \right|^2}{r^2 \sin^2\thi} + V(|\phi|^2) .
\ee
The first two terms and the $\left| \phi \right|^2$ term of the potential are $\mathcal{O}\left({1}/{r^2}\right)$ and the remaining terms are $\mathcal{O}\left({1}/{r^4}\right)$. Therefore, as $r \to \infty$, the leading order of the perturbative wave energy density can be obtained by evaluating 
\be
{T_{tt}} = \left| \p_t \phi \right|^2 + \left| \p_r \phi \right|^2  + \left| \phi \right|^2 . 
\ee
Similarly, for the scattering waves at large $r$, the energy flux $T_{rt}$, the $z$ component of the angular momentum density $T_{t \varphi}$ and the $z$ component of the angular momentum flux $T_{r \varphi}$ are respectively
\bal
    {T_{rt}} &= \p_r \phi^* \p_t \phi + \p_r \phi \p_t \phi^* , \\
    {T_{t\varphi}} & = \p_t \phi^* \p_\varphi \phi + \p_t \phi \p_\varphi \phi^* ,\\  
    {T_{r\varphi}} & = \p_r \phi^* \p_\varphi \phi + \p_r \phi \p_\varphi \phi^* .
\eal
To see why $T_{t \varphi}$ is the $z$ component of the angular momentum density,  note that the angular momentum densities in Cartesian coordinates are $J^L_{ij} = x_i T_{tj} - x_{j} T_{ti}$. Transforming to the spherical coordinates, we get $J^L_{xy}  = x T_{ty} - y T_{tx} = T_{t\varphi }$, $J^L_{xz}  = - \cos \varphi T_{t\theta} + \cot \theta \sin \varphi T_{t\varphi }$ and $J^L_{yz}  = - \sin \varphi T_{t\theta} - \cot \theta \cos \varphi T_{t\varphi}$. For our ansatz \eqref{phietaeta}, since $T_{t\theta}$ and $T_{t\varphi}$ are independent of $\varphi$, $\int \d\varphi J^L_{xz}=\int \d\varphi J^L_{yz}=0$. So we will only evaluate the amplification for $J^L_{xy}$.

To determine the enhancement of physical observables, let us first define the integration of the averaged energy density, energy flux, angular momentum density, and angular momentum flux respectively over a spherical shell region from $r_1$ to $r_2$ as $r_1, r_2\to \infty$: 
    \begin{align}
    E_\circledcirc & = \frac{1}{r_2-r_1} \int_{r_1}^{r_2} {\rm d}r r^2 \left \langle {T_{tt}}  \right \rangle_{T\Omega} ,  \\
     &= 2 \frac{\omega_+^2}{k_+^2} \left(  A^2_+ + B^2_+ \right) +  2 \frac{\omega_-^2}{k_-^2} \left( A^2_- +  B^2_- \right) , \nn
    P^{tr}_\circledcirc & = \frac{-1}{r_2-r_1} \int_{r_1}^{r_2} {\rm d}r r^2 \left \langle {T_{rt}}  \right \rangle_{T\Omega} ,  \\
    & = 2 \frac{\omega_+}{k_+} \left( A^2_+ - B^2_+ \right) +  2 \frac{\omega_-}{k_-} \left( A^2_- -  B^2_- \right) , \nn
    L_\circledcirc^{xy} &  = \frac{1}{r_2-r_1} \int_{r_1}^{r_2} {\rm d}r r^2 \left \langle {T_{t\varphi}}  \right \rangle_{T\Omega} , \\
    & = 2 \frac{\omega_+ m_+}{k_+^2} \left( A^2_+ + B^2_+ \right) +  2 \frac{\omega_- m_-}{k_-^2} \left( A^2_- +  B^2_- \right) , \nn 
    P_\circledcirc^{r\varphi} & = \frac{1}{r_2-r_1} \int_{r_1}^{r_2} {\rm d}r r^2 \left \langle {T_{r\varphi}}  \right \rangle_{T\Omega} , \\
    & = 2 \frac{m_+}{k_+} \left( A^2_+ - B^2_+ \right) +  2 \frac{m_-}{k_-} \left( A^2_- -  B^2_- \right) , \nonumber
    \end{align}
where the shell region from $r_1$ to $r_2$ must include at least a full spatial oscillation of the longest wave, $\langle \ \rangle_{T\Omega}$ denotes the average over several temporal oscillations and over the whole 2-sphere, and we have defined
\begin{align}
\begin{split}
    A^2_\pm & = \frac{1}{2} \int_{-1}^{1} {\rm d}c_{\thi}   \left|\sum\limits_{\ell} A^{\pm}_{\ell} P^{m_\pm}_{\ell} (c_{\thi}) \right| ^2, \\
	B^2_\pm & = \frac{1}{2} \int_{-1}^{1} {\rm d}c_{\thi}  \left| \sum\limits_{\ell} B^{\pm}_{\ell} P^{m_\pm}_{\ell} (c_{\thi}) \right|^2 .
\end{split}
\end{align}
$E_\circledcirc$, $P^{tr}_\circledcirc$, $L_\circledcirc^{xy}$ and $P_\circledcirc^{r\varphi}$ contain both the ingoing and the outgoing waves, but it is easy to identify the ingoing and the outgoing waves with help of Figure ~\ref{fig:inout} . Therefore, for a generic scattering, we find that the amplification factors for the energy density, energy flux, angular momentum density, and angular momentum flux in a large $r$ spherical shell are respectively
\begin{align}
    \mathcal{A}_E & = \left( \frac{ \frac{\omega_+^2}{k_+^2}  A_+^2 + \frac{\omega_-^2}{k_-^2} B_-^2}{ \frac{\omega_-^2}{k_-^2}  A_-^2 +  \frac{\omega_+^2}{k_+^2} B_+^2 }  \right) ^{ {\rm sign}(\omega) } ,    \label{AEdef} \\
    \mathcal{A}_{tr} & = \left( \left| \frac{ \frac{\omega_+}{k_+}  A_+^2 - \frac{\omega_-}{k_-} B_-^2}{ \frac{\omega_-}{k_-}  A_-^2 -  \frac{\omega_+}{k_+} B_+^2 } \right|  \right) ^{ {\rm sign}(\omega) } ,    \label{Atrdef} \\
    \mathcal{A}_L & = \left( \frac{ \frac{\omega_+ m_+}{k_+^2}  A_+^2 + \frac{\omega_- m_-}{k_-^2}  B_-^2}{ \frac{\omega_- m_-}{k_-^2}   A_-^2 +  \frac{\omega_+ m_+}{k_+^2}  B_+^2 }  \right) ^{ {\rm sign}(\omega) } ,    \label{ALdef} \\
    \mathcal{A}_{r\varphi} & = \left( \left| \frac{ \frac{m_+}{k_+}  A_+^2 - \frac{m_-}{k_-} B_-^2}{ \frac{m_-}{k_-}  A_-^2 - \frac{m_+}{k_+} B_+^2 } \right| \right) ^{ {\rm sign}(\omega) } .
    \label{Arphidef}
\end{align}

Apart from the quantities constructed from the energy momentum tensor, there are also observables associated with the U(1) symmetry. In particularly, we can look at the radial current $J^{Q}_r = \phi^* \p_r \phi - \phi \p_r \phi^*$. 
Again, averaging over several temporal oscillations and over the whole 2-sphere in from $r_1$ to $r_2$, we get the charge density in a far away region:
\begin{align}
    P_\circledcirc^{Q} & = \frac{i}{r_2-r_1} \int_{r_1}^{r_2} {\rm d}r r^2 \left \langle J^{Q}_r \right \rangle_{T\Omega} ,  \\ 
	& =  \frac{2}{k_+} \left( -  A_+^2 + B_+^2 \right) +   \frac{2}{k_-} \left( - A_-^2 + B_-^2 \right) . \nonumber
\end{align}
Since both one positive charge and one negative charge give rise to one particle number, we can also define the particle number density in a far away region as
\begin{align}
	N_\circledcirc  \!=\! \frac{2}{k_+} \big( A_+^2 + B_+^2 \big) +   \frac{2}{k_-} \big( A_-^2 + B_-^2 \big) . 
\end{align}
Then, we can define the amplification factors for the particle number respectively in the scattering as follows
\begin{align}
    \mathcal{A}_N & = \left( \frac{ \frac{1}{k_+}  A_+^2 + \frac{1}{k_-} B_-^2}{ \frac{1}{k_-}  A_-^2 +  \frac{1}{k_+} B_+^2 }  \right) ^{ {\rm sign}(\omega) } .
\end{align}

An important feature in the scattering for a U(1) field is that the particle number is conserved \cite{Saffin:2022tub,Gao:2023gof} , which means  $\mathcal{A}_N = 1$, or 
\begin{align}
    \frac{1}{k_+} A_+^2 + \frac{1}{k_-} B_-^2 = \frac{1}{k_-} A_-^2 + \frac{1}{k_+} B_+^2 .
    \label{ampn1}
\end{align}
This relation is useful to derive the superradiance criteria for the various observables as well as bounds on the amplification factors. It also serves as a consistency check for our numerical results. 

Given that the $Q$-ball is not spherically symmetric, it is also interesting to look at how the amplification factors vary according to the $\theta$ angle. One way to examine the $\thi$ distribution is to define the $\theta$ dependent amplification factors $\mc{A}^\thi_*$ by computing the counterparts of $E_\circledcirc, P^{tr}_\circledcirc, L_\circledcirc^{xy}, P_\circledcirc^{r\varphi}$ with the average operation $\langle \ \rangle_{T\varphi}$, which does not include the average over $\theta$, {\it both} for the ingoing and outgoing modes:
\bal
\label{thetadef1}
\langle \ \rangle_{T\Omega}  &\to  \langle \ \rangle_{T\varphi},
\\
A^2_\pm &\to (A^\thi_\pm)^2 = \left|\sum\limits_{\ell} A^{\pm}_{\ell} P^{m_\pm}_{\ell} (c_{\thi}) \right| ^2 \sin\thi\d \thi , 
\\
\label{thetadef3}
B^2_\pm &\to (B^\thi_\pm)^2 = \left|\sum\limits_{\ell} B^{\pm}_{\ell} P^{m_\pm}_{\ell} (c_{\thi}) \right| ^2 \sin \thi \d \thi .
\eal
That is,  $\mc{A}^\thi_*$ measures the amplification between waves incident on and emitted from the $Q$-ball in the region $\thi\to\thi+\d\thi$. Note that in this case the $\mc{A}^\thi_*$  factors will be independent of the choice of $\d c_\thi$, as $\d c_\thi$ cancels out between the ingoing and the outgoing modes. 

Another set of complementary $\theta$ dependent amplification factors $\tilde{\mc{A}}^\thi_*$ can be defined with the average operation $\langle \ \rangle_{T\varphi}$ for the outgoing modes and with the average operation $\langle \ \rangle_{T\Omega}$ for the ingoing modes. That is,  $\tilde{\mc{A}}^\thi_*$ measures the amplification between the waves incident on the $Q$-ball from all angles and the waves emitted from the $Q$-ball in the angle $\d \thi$ from the direction $\thi$. In this case, the results do depend on the choice of $\d\thi$, but this simply re-scales all the amplification factors by an overall factor.

\section{$Q$-ball Superradiance}
\label{sec:level4}

In this section, we will present the numerical results of waves scattering around a 3+1D spinning $Q$-ball, the energy and angular momentum enhancements in the process and some salient features of the superradiance. As mentioned, to solve the perturbative equations numerically, we will use the high dimensional shooting method, as the convergence of the relaxation method is difficult to achieve for the perturbative equations.

\subsection{Numerical results}

As discussed towards the end of Section \ref{sec:pertwaves}, since the different partial wave modes ({\it i.e., modes with different $\ell$}) are inherently coupled in this 3+1D system, there are many free parameters to choose in a generic scattering, even with a relatively small $\ell_{\rm max}^{\pm}$ truncation. Therefore, in presenting the numerical, we will be selective in probing the parameter space, deferring a comprehensive survey for future work. The directions of the parameter space that we are going to probe in this paper include:

\begin{itemize}

\item {\it Background}: First of all, we will divide the numerical results into two sub-categories: (1) the background $Q$-ball has even parity and (2) the background $Q$-ball has odd parity. We will mainly focus on the even sector, since the main superradiance features of the odd sector seem to be similar. Note that when the background $Q$-ball has even (or odd) parity, the scattering waves on top of it can have the same or the opposite parity. We will focus on the $Q$-ball solutions with $m_Q=1$.

\item {\it One ingoing mode}:  We will be interested in scattering where there is only one ingoing mode with the lowest $\ell=\bar m_\pm+s=|m_Q\pm m|+s$, where $s=0$ ($s=1$) for an even (odd) parity $Q$-ball. To be more concrete, if $\omega>0$, the only ingoing mode is set to be $B^+_{\ell=\bar m_+ +s}$; if  $\omega<0$, the only ingoing mode is set to be $A^+_{\ell=\bar m_+ +s}$. Although there is just one ingoing modes,  the energy will be scattered into all available $\ell$ modes in the outgoing waves. Note that the parity of the perturbative modes can be different from that of the background $Q$-ball.

\item {\it Two ingoing modes}: We also plot figures where there are both $+$ and $-$ ingoing modes. In this case, we also only consider the scenario where only the lowest $\ell$ modes, {\it i.e.,} $\ell=\bar m_\pm+s$, are present in the ingoing waves.

\end{itemize}

\subsubsection{Even parity}

In this subsubsection, we present the numerical results for the case where the background $Q$-ball is of even parity, but both the even and odd parity perturbative modes will be explored. We will discuss the one ingoing mode case and the two ingoing modes case, as itemized above.

\begin{figure}
	\centering
		\includegraphics[height=9.8cm]{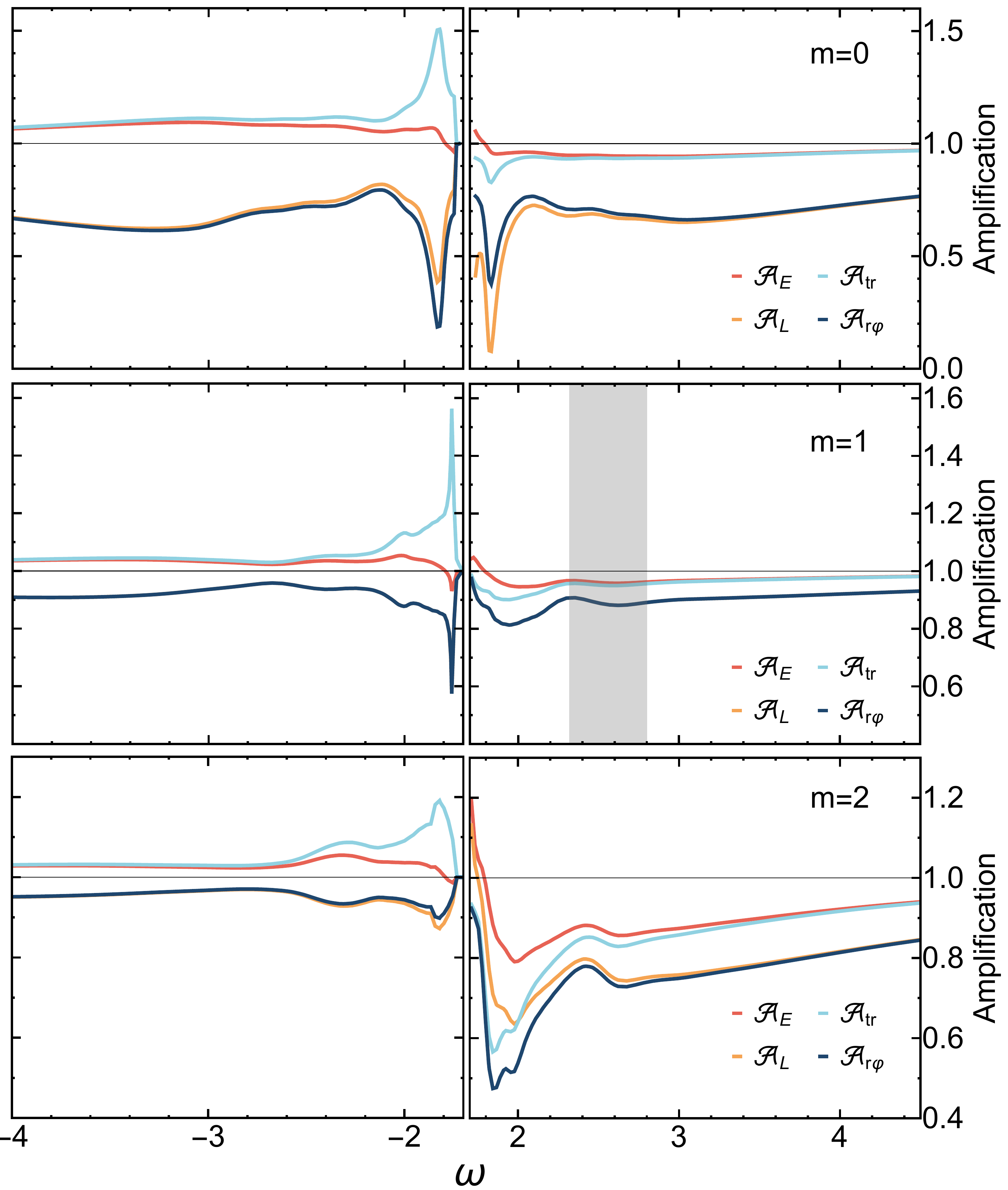}
	\caption{Spectra of the energy amplification factor $\mathcal{A}_E$, the angular momentum amplification factor $\mathcal{A}_L$, the energy current amplification factor $\mathcal{A}_{tr}$ and the angular stress current amplification factor $\mathcal{A}_{r\varphi}$. Both the $Q$-ball background (truncated up to 6th order in the $\ell$ expansion, suth that $N_{\rm max}^Q=6$; {\it i.e.,} the solution of Figure \ref{Fig:Qballs_even}) and the perturbative waves are of even parity (truncated up to 4th order in the $\ell$ expansion, suth that $N_{\rm max}=N_{\rm max}^+=N_{\rm max}^-=4$). Convergence is generally very good, but in the grey shaded region of the $m=1$ plot, the numerical errors in the perturbative scattering calculations with the 4th order truncation can exceed 2\%. In the $m=1$ plot, the line of $\mc{A}_L$ overlaps with that of $\mc{A}_{r \varphi}$. }
 \label{fig:1mode}
\end{figure}

\begin{figure*}
	\centering
		\includegraphics[height=8.5cm]{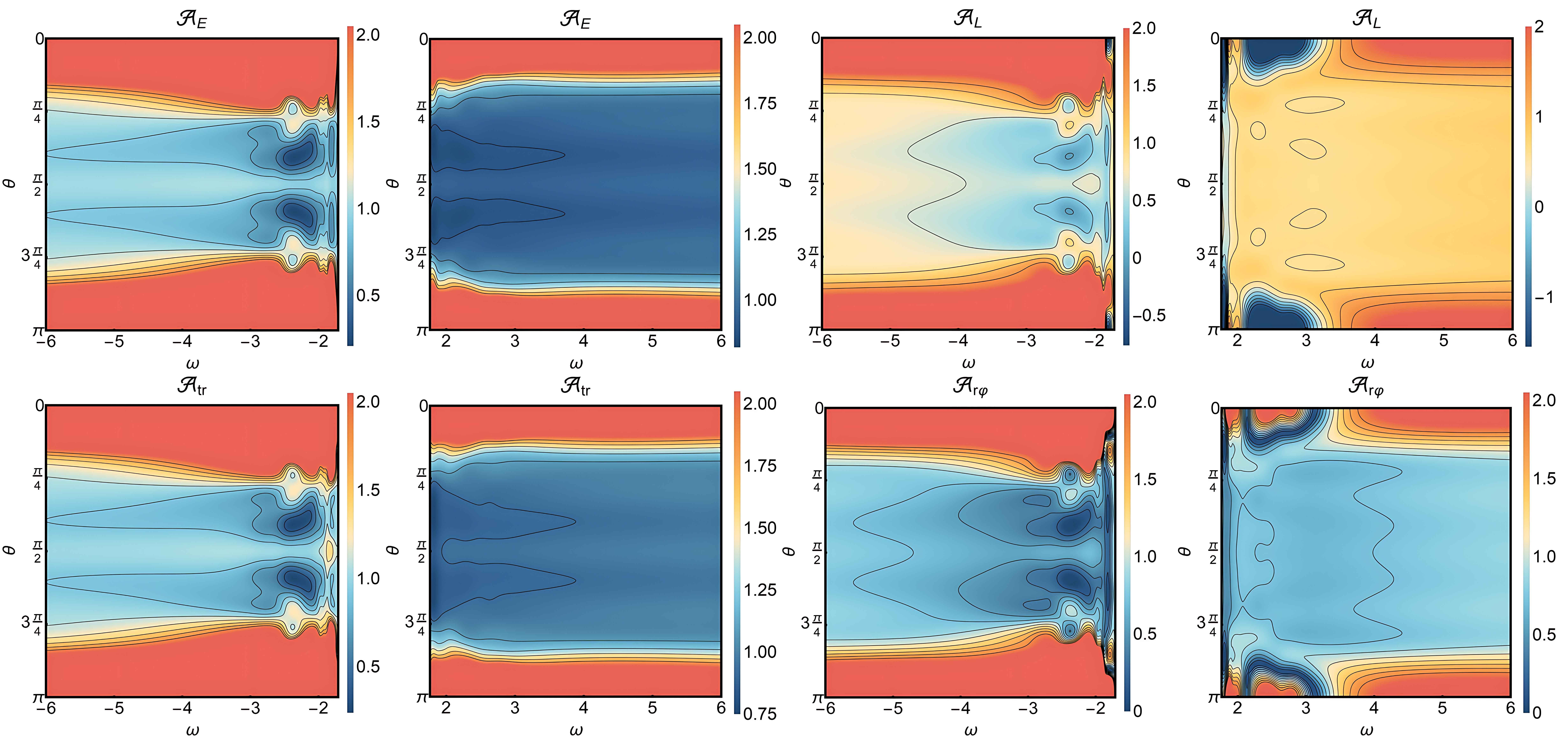}
	\caption{Angular distribution of the amplification spectra for the energy $\mathcal{A}^\thi_E$, angular momentum $\mathcal{A}^\thi_L$, energy current $\mathcal{A}^\thi_{tr}$ and angular stress current $\mathcal{A}^\thi_{r\varphi}$. $\thi$ is the polar angle, and these angular amplification factors are defined via the replacements \eqref{thetadef1}-\eqref{thetadef3}.  The numerical setup is the same as the $m=0$ case of Figure \ref{fig:1mode}. }
 \label{fig:thi1mode}
\end{figure*}

\begin{figure*}
	\centering
		\includegraphics[height=8.5cm]{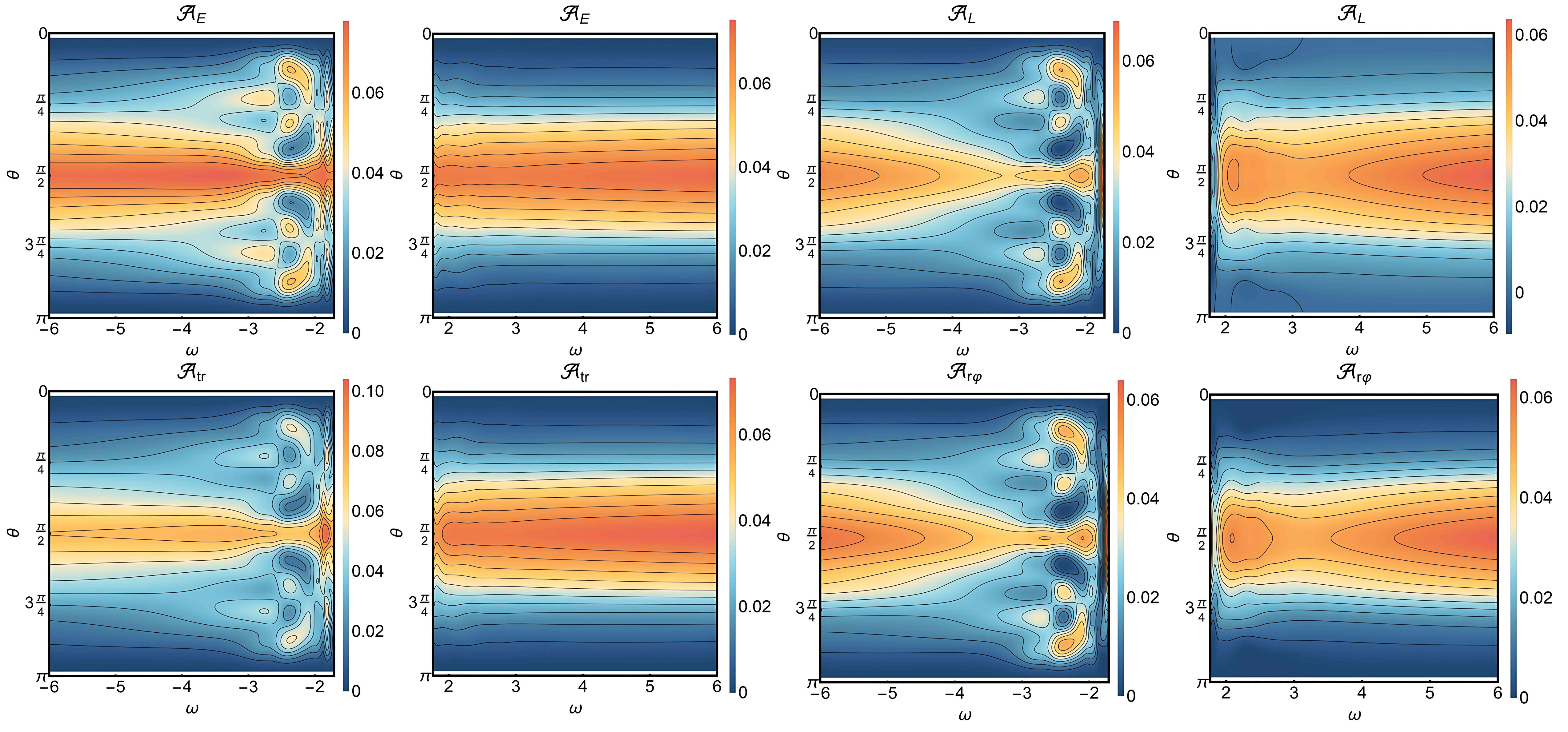}
	\caption{Angular distribution of the amplification spectra for the energy $\tilde{\mathcal{A}}^\thi_E$, angular momentum $\tilde{\mathcal{A}}^\thi_L$, energy current $\tilde{\mathcal{A}}^\thi_{tr}$ and angular stress current $\tilde{\mathcal{A}}^\thi_{r\varphi}$. The parameter used for the numerical setup is ${\rm d} \theta = 0.1$. $\thi$ is the polar angle, and these angular amplification factors are defined via the replacements \eqref{thetadef1}-\eqref{thetadef3}.  The numerical setup is the same as the $m=0$ case of Figure \ref{fig:1mode}. The definition of the $\tilde{\mathcal{A}^{\theta}}$ can be found at the end of the section \ref{sec:3b}. }
 \label{fig:thi1mode2}
\end{figure*}

In Figure \ref{fig:1mode}, we plot the spectra for the amplification factors  $\mathcal{A}_E$, $\mathcal{A}_L$, $\mathcal{A}_{tr}$ and $\mathcal{A}_{r\varphi}$, defined in the last section, for the case of one ingoing mode with $\ell=\bar m_+=1,2,3$, corresponding to $m=0,1,2$. Note that there are empty gaps, which is referred to as the mass gap, in the $\oi$ spectra simply because we are interested in propagating waves where $\oi_\pm^2 = k_\pm^2 + 1$. For this figure, we truncate $\ell_{\rm max}^{\pm}$ up to the 4th order, such that $N_{\rm max}=N_{\rm max}^+=N_{\rm max}^-=4$, which for one point of $\oi$ can be computed approximately within $10^5 \sim 10^6$ seconds by the high dimensional shooting method with \texttt{ParametricNDSolve} and \texttt{FindRoot} in Mathematica. This allows us to achieve good convergence for these amplification factors, except for the small grey region in the $m=1$ case where the errors in the amplification factor $\mathcal{A}_E$, inferred from the 3rd and 4th order, can exceed about 2\% in terms of $\mc{A}_E$, but still below 4\% for the largest error. Generally, the convergence for all of these amplification factors are very similar. The convergence of the perturbative scattering computations can also be confirmed by monitoring the particle number amplification factor $\mc{A}_N=1$. Figures \ref{fig:error} \ref{fig:1modeconvergence} depicting the the particle number conservation and the differences between the 3rd and 4th order for $\mathcal{A}_E$ can be found in Appendix \ref{sec:levelb}.

In Figure \ref{fig:1mode}, we see that the angular momentum can be amplified in the $m=2$ case near the mass gap but can not be enhanced in the $m=0,1$ cases over the whole $\oi$ spectrum. This is related to the fact that the background $Q$-ball has $m_Q=1$. That is, when the ingoing modes are of one kind ($\eta^+$ or $\eta^-$), angular momentum can only be enhanced when $m>m_Q$. As will be explained in the Section \ref{sec:asymp}, this is different from the Zel'dovich rotational superradiance condition. Also, in the $m=1$ plot, the line of $\mc{A}_L$ overlaps with that of $\mc{A}_{r \varphi}$. Technically, this is because when $m_-=m_Q-m=0$ we can see from Eqs.~\eqref{ALdef} and \eqref{Arphidef} that $\mc{A}_L=\mc{A}_{r \varphi}$. Physically, it simply means that when $m_-=0$, only the $\eta^+$ modes have angular momentum, in which case it is the same to define the amplification factor with the angular momentum in the far-away region and with the angular stress current.

Moreover, we observe that there can not be superradiance for the energy current and the angular stress current when $\omega>0$, which is consistent with the semi-analysis in Section \ref{sec:limits}. However, the energy current does get amplified from the left hand side of the mass gap, that is, when it satisfies a Zel'dovich-like rotational superradiance condition $\oi_+ < \oi_Q$ \cite{Cardoso:2023dtm}. We want to emphasize that, different from the traditional real-space rotation, here it is a rotation in the internal field space. We can also establish the superradiance criteria for other amplification factors such as $\mc{A}_E$ and $\mc{A}_L$ \cite{Saffin:2022tub,Gao:2023gof}, which will be discussed in Section \ref{sec:limits}. Our numerical results confirm all of these criteria. 

A probably unsurprising feature is that the peaks and dips of the different amplification factors often align with each other. However, near the mass gap, there are obvious differences between $\mc{A}_E$ and $\mc{A}_{tr}$ or between $\mc{A}_L$ and $\mc{A}_{r\varphi}$. This is because the scattering problem involves two kinds of modes with frequency $\oi_+$ and $\oi_-$ respectively and they have different group velocities due to the massive nature of the complex scalar field \cite{Cardoso:2023dtm}. Far away from the mass gap, the differences between the group velocities become negligible, so we expect $\mc{A}_E\simeq \mc{A}_{tr}$ and $\mc{A}_L\simeq \mc{A}_{r\varphi}$, as we can seen in Figure \ref{fig:1mode}. We also see that there are generally multiple peaks in the spectrum, and it would be interesting to understand the underlying mechanism to generate them.

A 3D spinning $Q$-ball is axisymmetric, so it is of interest to look at the polar-angle dependence of the amplification factors. We have defined two sets of such amplification factors $\mc{A}^\thi_*$ and $\tilde{\mc{A}}^\thi_*$ in the last section. Here we will see more clearly that they are complementary when the ingoing modes are not spherically symmetric. More concretely, in the setup of Figure \ref{fig:thi1mode}, the ingoing mode is given by the $(|m_+|,m_+)$ partial wave with $m_+=1+m$, $m$ being  0,1 or 2. The intensity of each of these partial waves peaks at the equator, monotonically decreases away from the equator and vanishes at the north/south pole. Since the outgoing waves have nonzero intensity at the two poles, we will find that the $\mc{A}^\thi_*$ factors tend to infinity at the two poles, as we see in Figure \ref{fig:thi1mode} for the case of $m=0$. However, there are some regions where the amplification factors do not tend to infinity at the two poles. 
 The reason is that near the two poles, the outgoing wave has a negative angular momentum, opposite to that of the ingoing wave. Therefore, $\mathcal{A}_L^{\theta}$ becomes negative at the two poles.  
The cases with $m=1$ and $m=2$, which are not displayed, exhibit a considerable degree of similarity. On the other hand, in Figure \ref{fig:thi1mode2} for the numerical setup ${\rm d}\theta = 0.1$, the $\tilde{\mc{A}}^\thi_*$ factors faithfully depict the $\thi$ dependence of the outgoing waves, but neglect the fact that the ingoing waves are not spherically symmetric. 
As we can see in Figure \ref{fig:thi1mode} and \ref{fig:thi1mode2}, near the poles the $\mathcal{A}^\theta_*$ factors tend to infinity roughly. While these figures show the amplification factors, we note that the actual amount of flux in the various physical quantities is larger in the region of the equator.

\begin{figure*}
	\centering
		\includegraphics[height=10.5cm]{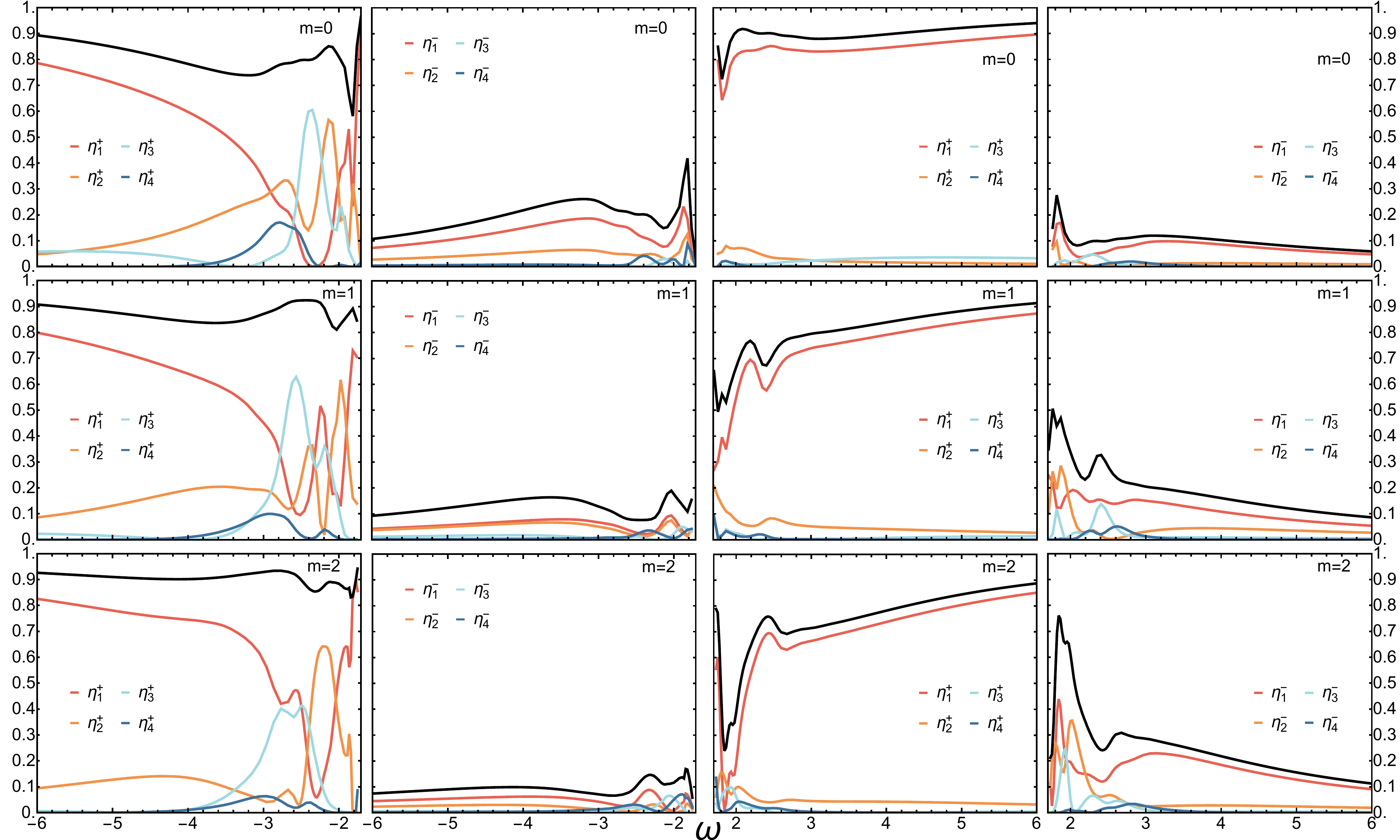}
	\caption{Fractional distribution of energy in the outgoing $\ell$ modes for various $\oi$ and $m$, when there is only one ingoing wave ($\eta^+_{1+m}$) in the scattering. The plots should be taken in pairs, for example the first two plots on the top row show the fractional distribution of energy for $m=0$ and $\oi<0$, with the first of the two giving the energy in $\eta^+$ and the second giving the energy in $\eta^-$. The black lines are the sum of all the four corresponding $\ell$ modes and we note that the energy in these outgoing $\eta^+$ and $\eta^-$ modes add up to one, when the plots are considered in pairs as described. The numerical setup is the same as Figure \ref{fig:1mode}.
    }
 \label{fig:diffellmodes}
\end{figure*}

The scattering of Figure \ref{fig:1mode} is a process where one ingoing mode ($\eta^+_{\ell=1+m}$) is scattering into various different $\ell$ modes in the outgoing waves. 
Figure \ref{fig:diffellmodes} is a break-down of the distribution of the energy in the different $\eta^\pm_\ell$ modes for the outgoing modes for different $\oi$ and $m$. Since the ingoing mode are of the $+$ type, we see that most energy in the outgoing waves remain in the $+$ modes. Also, it is clear that the lower $\ell$ modes contain more energy in the scattered outgoing waves, except for a small range of negative $\oi$ near the mass gap where the $\ell=2$ or $3$ mode can dominate the energy budget. For the $m=2$ case, there is also a small range of positive $\oi$ where the $\ell=2$ and $-$ mode can dominate. In general, it is easier to convert the $+$ modes into the $-$ modes when the ingoing mode has a larger $m$ and when $\oi$ is closer to the positive side of the mass gap. Again, since the system is symmetric with respect to the swap of the $+$ and $-$ modes, similar results also apply to the case with one $-$ ingoing mode.

\begin{figure*}
	\centering
		\includegraphics[height=8.2cm]{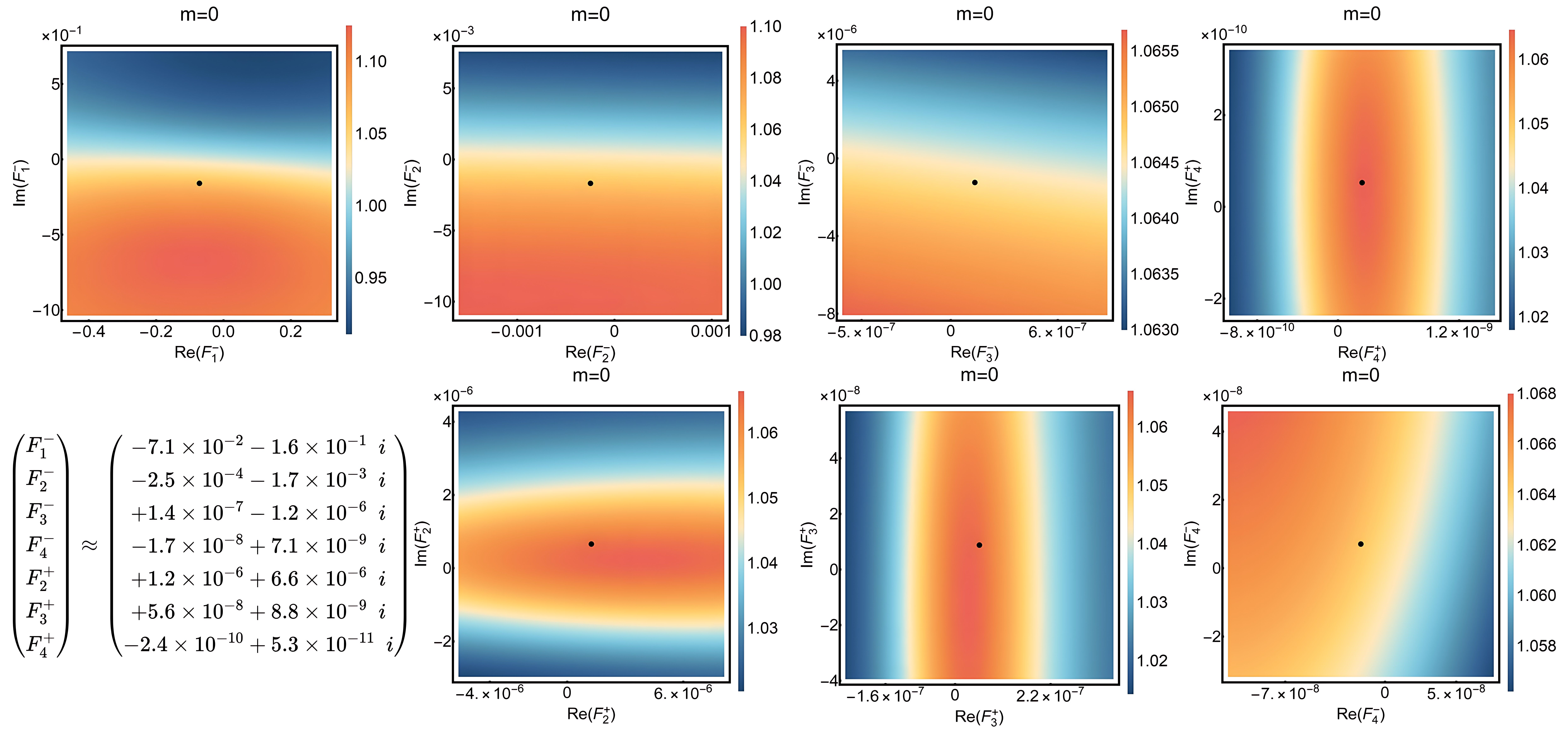}
	\caption{Amplification factor $\mathcal{A}_E$ for two ingoing modes parameterized by $F_{\ell}^{\pm}$. The other numerical setup is the same as the $m=0$ case of Figure \ref{fig:1mode}. The one ingoing mode case of Figure \ref{fig:1mode} is marked with black dots, and the corresponding  $F_{\ell}^{\pm}$ parameters are listed in the bottom left corner of the figure. The parameters used are $m=0, \omega=-4.00$.}
 \label{fig:mF012}
\end{figure*}

In general, there can be both $+$ and $-$ modes in the ingoing waves, in which case the  $+$ and $-$ modes interact to enhance the amplification of waves in the scattering. In Figure \ref{fig:mF012}, we probe the scenario of both $+$ and $-$ ingoing modes around the one ingoing mode case of Figure \ref{fig:1mode}. It is simplest to parametrize the deviation from the one ingoing mode case by the $F^\pm_i$ parameters, which parameterize all possible regularity conditions at the center of the $Q$-ball. They contain the information about the scattering waves and also the correction to the background $Q$-ball by the scattering waves. The $F^\pm_i$ parameters corresponding to the one ingoing mode case of Figure \ref{fig:1mode} are listed in the bottom left corner of Figure \ref{fig:mF012}. Note that all the $\ell$ modes are activated even for the case of one ingoing $\ell$ mode.

\subsubsection{Odd parity}

\begin{figure}
	\centering
		\includegraphics[height=13.0cm]{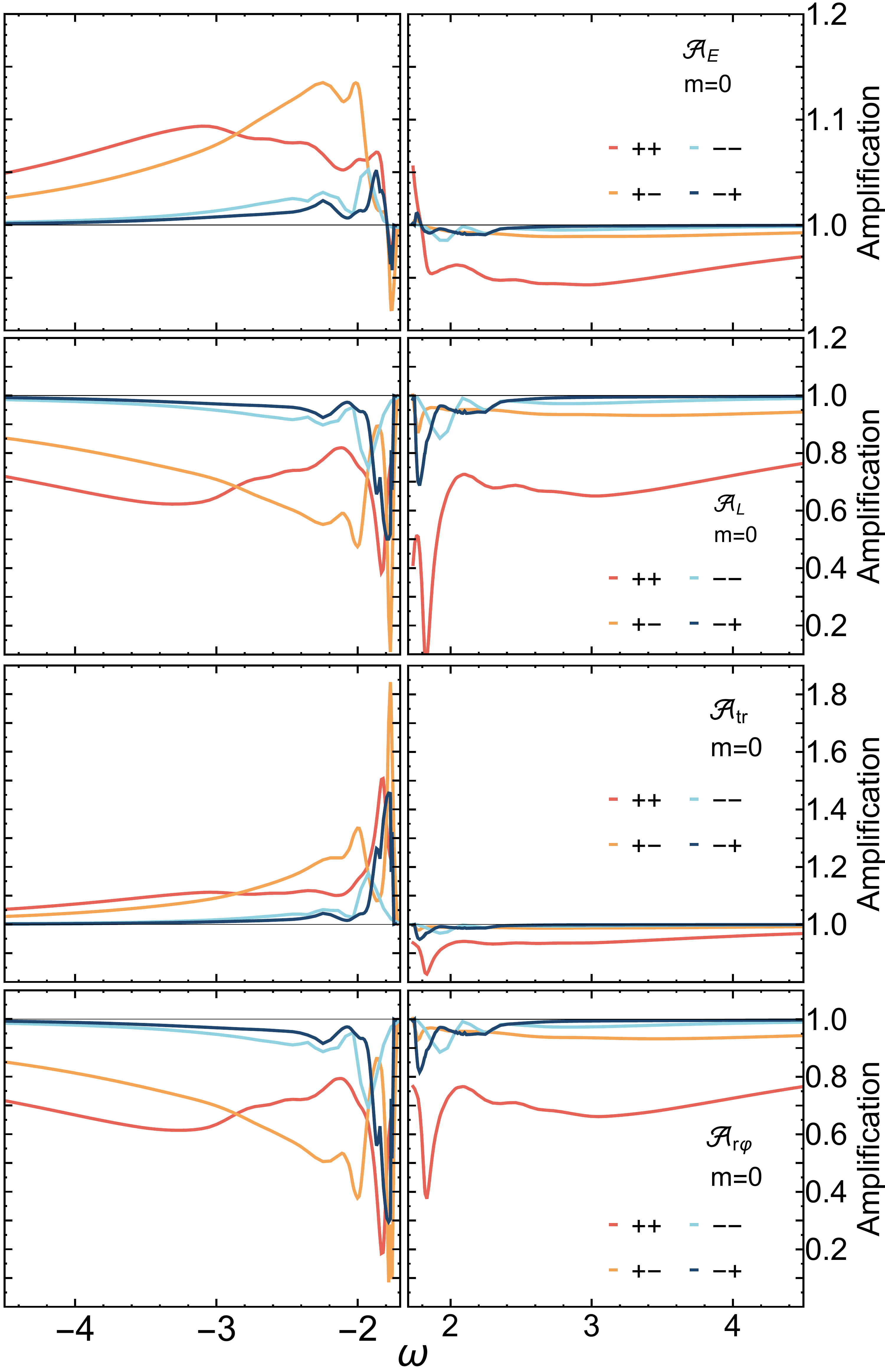}
	\caption{Amplification factors $\mathcal{A}_E$, $\mathcal{A}_{tr}$, $\mathcal{A}_L$ and $\mathcal{A}_{r\varphi}$ for mixed parities. For example, $+-$ means that the background $Q$-ball is of even parity and the scattering waves are of odd parity. The even parity $Q$-ball is that of Figure \ref{Fig:Qballs_even} and the odd parity $Q$-ball is that of Figure \ref{Fig:Qballs_odd}.}
 \label{fig:pmaeP}
\end{figure}

In the previous subsubsection, we focussed on the case where both the background $Q$-ball and the scattering waves are of even parity. Here we shall briefly explore the cases where either the background or the scattering waves are of odd parity; see Figure  \ref{fig:pmaeP}. For the same set of parameters, the odd parity $Q$-ball is more energetic, due to the double-peak structure -- the energy of the $Q$-ball in Figure \ref{Fig:Qballs_odd} is about twice that of Figure \ref{Fig:Qballs_even}. Despite that, we see that generally the even parity $Q$-ball tends to create more energy amplification than the odd parity $Q$-ball. In the event of energy reduction, the even parity $Q$-ball also absorbs more energy than the odd parity $Q$-ball. However, as for the parity of the scattering waves, there is no clear trend whether one parity can enhance or reduce the energy more than the other. In particular, a mixed parity case sometimes can produce the largest amplification effects.

\subsection{Amplification in asymptotic regions}
\label{sec:asymp}

In the previous subsection, the numerical results of the various amplification factors are presented. In this subsection, to further confirm these results, we (semi-)analytically examine the asymptotic behavior of the amplification factors as $\omega$ approaches the mass gap or $\omega \to \infty$, for the case where the ingoing modes are of the same frequency. 

If we consider a single ingoing mode we can, without lost of generality, take there to be only $\eta^+$ modes but no $\eta^-$ ingoing modes. There are two cases for this:  (i) $\omega>0$ and $A_-=0$; (ii) $\omega<0$ and $B_-=0$. For simplicity, we also assume $m \ge 0$. Let us consider the case of $\omega>0$ and $A_-=0$ first. Near the mass gap, we can write $\omega = 1 + \omega_Q + \epsilon $, where $0 < \epsilon \ll 1$ (note that we have also assumed $\omega_Q>0$). In this case, by the particle number conservation Eq.~\eqref{ampn1}, the amplification factors can be written as
\begin{align}
    \mathcal{A}_E & =  \frac{  \frac{1}{2 \epsilon} B_-^2 + \frac{\omega_+^2}{k_+^2}  A_+^2 }{   \frac{\omega_+^2}{k_+} \frac{1}{\sqrt{2 \epsilon}} B_-^2 + \frac{\omega_+^2}{k_+^2}  A_+^2 }  , \\ 
    \mathcal{A}_L &  =  \frac{ - \frac{ m_-}{2 \epsilon}  B_-^2 + \frac{\omega_+ m_+}{k_+^2}  A_+^2 }{ \frac{\omega_+ m_+}{k_+} \frac{1}{\sqrt{2 \epsilon}}  B_-^2 + \frac{\omega_+ m_+}{k_+^2}  A_+^2 }   ,  \\   
    \mathcal{A}_{tr} & = \left| \frac{  \frac{1}{\sqrt{2 \epsilon}} B_-^2 + \frac{\omega_+}{k_+}  A_+^2 }{  \frac{\omega_+}{\sqrt{2 \epsilon}} B_-^2 + \frac{\omega_+}{k_+}  A_+^2  } \right|  ,  \\
    \mathcal{A}_{r\varphi}  &= \left| \frac{ -\frac{m_-}{\sqrt{2 \epsilon}} B_-^2 + \frac{m_+}{k_+}  A_+^2 }{ \frac{m_+}{\sqrt{2 \epsilon}} B_-^2 +  \frac{m_+}{k_+}  A_+^2  } \right|  . 
    \label{ampL}
\end{align}
Note that these amplification factors now only depend on the far-away amplitudes $B_-^2$ and $A_+^2$. As we are considering linear scattering, we can always scale the amplitude of the ingoing wave $A_+$ to be order one. Generically, the outgoing $\eta^+$ wave $B_+$ should also be order one. Now, let us assume the leading behavior of $B_-^2$ goes like 
\be
B_-^2 \propto \epsilon^n, ~~~n\geq \f12,~~~0 < \epsilon \ll 1 , 
\ee
as $\epsilon$ goes to zero. The fact that $n$ must be no less than $1/2$ can be seen from the particle number conservation condition $|B_-|^2/\sqrt{2\epsilon}=(|B_+|^2-|A_+|^2)/k_+$ and the fact that the right hand side of this equation is finite. We then see that the behavior of the amplification factors near the mass gap depends on the size of $n$:
\begin{itemize}

\item $n>1$: These four amplification factors all have the same asymptotic behavior $\mathcal{A} \to 1$ as $\oi$ approaches the mass gap: $\mathcal{A}_E$ should tend to 1 from above, and $\mathcal{A}_{tr}$ $\mathcal{A}_{r\varphi}$ should tend to 1 from below.
For $\mathcal{A}_L$, it depends on the value of $m$: it should tend to 1 from below when $0 \le m \le m_Q$ and from above when $m>m_Q$.
 However, we do not observe this case in Figure \ref{fig:1mode}. 

\item $n=1$: The amplification factors for the energy and the angular momentum will asymptotically approach non-vanishing constants as $\epsilon \to 0$. This seems to happen when $m\neq m_Q$ in Figure \ref{fig:1mode}. This non-vanishing constant is greater than 1 for $\mc{A}_E$. For $\mc{A}_L$, it depends on the value of $m$: the constant is greater than 1 when $m>m_Q$ and is less than 1 when $0 \le m\leq m_Q$. Additionally,  it is also easy to obtain the following asymptotic ratio
\begin{align}
\label{AEALrel}
    \lim_{\epsilon \to 0} \frac{\mathcal{A}_L-1}{\mathcal{A}_E-1} = - \frac{\omega_+ m_-}{m_+} , 
\end{align}
This consistency relation agrees quite well with our numerical results in Figure \ref{fig:1mode}, as can be seen in Figure \ref{fig:s1}.

\begin{figure}
	\centering
		\includegraphics[height=5.9cm]{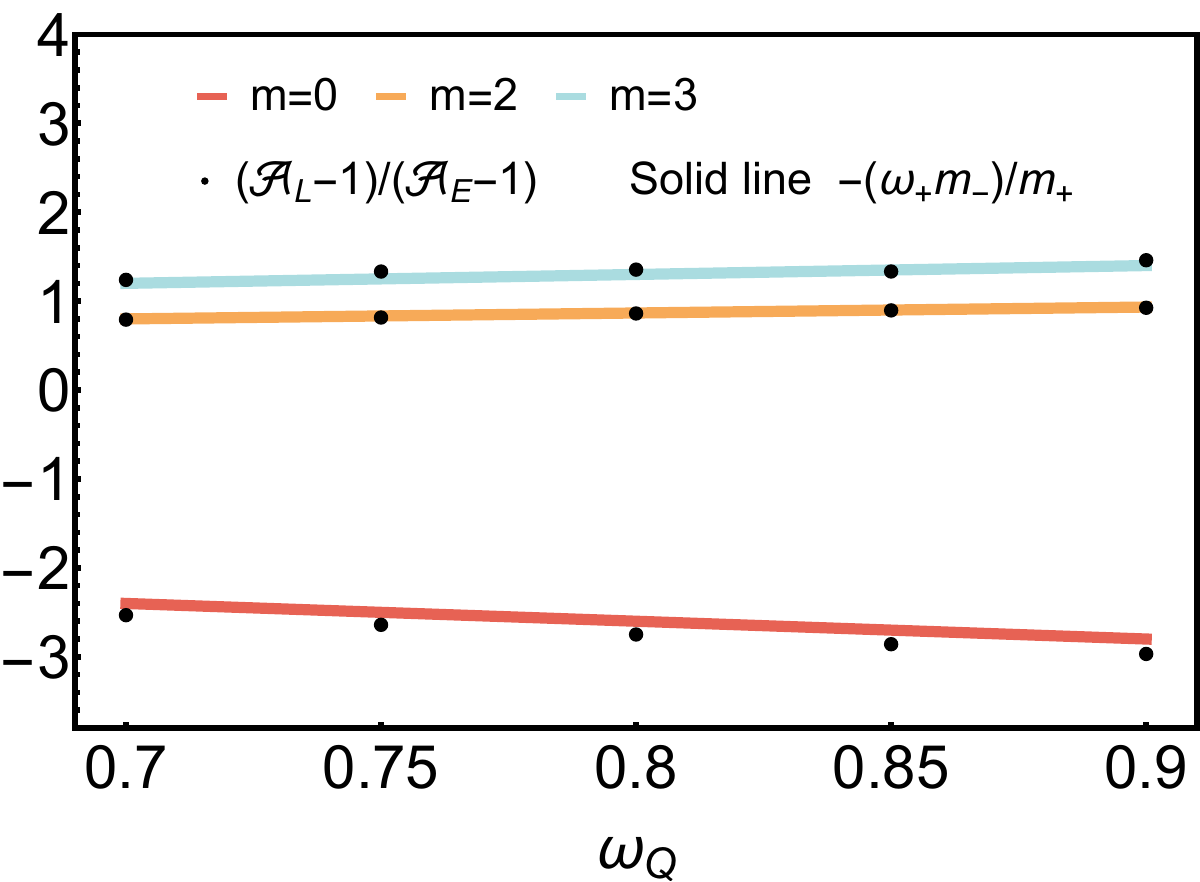}
	\caption{Numerical confirmation of the consistent relation \eqref{AEALrel} between $\mc{A}_E$ and $\mc{A}_L$ near mass gap.}
 \label{fig:s1}
\end{figure}

\item $1 > n \ge 1/2$: The amplification factor for the energy $\mc{A}_E$ will asymptotically tend to infinity as $\epsilon \to 0$. While this case is allowed in principle, we do not observe it in all of numerical evaluations. That is, when we approach the mass bound, the difference between the group velocities of $\eta_+$ and $\eta_-$ tends to infinity, the energy amplification factor may not tend to infinity, because $B_-$ tends to zero faster. On the other hand, the amplification factor for the angular momentum $\mathcal{A}_L$ depends on the value of $m$: it should tend to negative infinity when $m_Q > m \ge 0$, tend to a non-vanishing constant that is less or equal than 1 when $m = m_Q$ and tend to infinty when $m>m_Q$.

\end{itemize}
Generally, neither the energy current nor the angular stress current can be amplified near the right hand side of the mass gap, since neither $\mathcal{A}_{tr} > 1$ nor $\mathcal{A}_{r\varphi} > 1$ has a real solution for $\epsilon$. In fact, this holds for generic $\omega>0$, as we will see in the next subsection.

For the case of $\omega<0$ and $B_-=0$ (and $m \ge 0$), the analysis is similar. Assuming that 
\be
|A_-|^2\propto \epsilon^n,~~~n\geq -\f12,~~~0<\epsilon\ll 1 ,
\ee
as $\oi$ approaches the mass gap from the left $\oi\to -1-\oi_Q-\epsilon$,  we find 
\begin{itemize}
\item $n>-\f12$: $\mc{A}_{tr}$ always tends to 1 from above and $\mc{A}_E$, $\mc{A}_L$ and $\mc{A}_{r\phi}$ tend to 1 from below, as $\epsilon \to 0$. 

\item $n=-\f12$: As $\epsilon \to 0$, $\mc{A}_E$ and $\mc{A}_L$ approaches a constant less or equal than 1, and $\mc{A}_{tr}$ and $\mc{A}_{r\varphi}$ approaches 1, $\mc{A}_{tr}$ approaching 1 from above and $\mc{A}_{r\varphi}$ approaching 1 from below.
\end{itemize}

When $\omega$ is large, $\omega_+$, $-\omega_-$, $k_+ $ and $ k_- $ all tend to $\omega$, for which case all the amplification factors approach $1$. Specifically, when $\omega\to \infty$, $\mc{A}_E$, $\mc{A}_{tr}$, $\mc{A}_L$ and $\mc{A}_{t\varphi}$ approaches to 1 from below; when $\omega\to -\infty$, $\mc{A}_E$ and $\mc{A}_{tr}$ approaches to 1 from above and $\mc{A}_L$ and $\mc{A}_{t\varphi}$ approaches to 1 from below. This is expected as the couplings between the high-frequency modes and the modes of the $Q$-ball should be suppressed by the frequency hierarchy between them.

Near the mass gap, it is also easy to extract the superradiance criteria for the amplification factors.  Let us again take the case of $\omega>0$, $m \ge 0$ and $A_-=0$ for an example. First, by requiring $\mathcal{A}_E > 1$,  we can get the energy superradiance criterion: 
\begin{align}
\label{epsE}
    \epsilon < \frac{2\omega_Q (1+\omega_Q)}{(1+2\omega_Q)^4} \equiv \epsilon_E . 
\end{align}
From this, we see that the energy is always enhanced $\mathcal{A}_E > 1$ as long as $B_-$ is nonzero and $0<\epsilon<\epsilon_E$. Similarly, we can obtain the angular momentum superradiance criterion by requiring $\mathcal{A}_L > 1$, which leads to
\begin{align}
    m_Q < m \text{ and } \epsilon < \frac{2 \omega_Q (1+\omega_Q) (m_Q-m)^2 }{(1+2\omega_Q)^2 (m_Q+m)^2 } \equiv \epsilon_L . 
    \label{ALS}
\end{align}
In fact, some generic superradiance criteria can be established away from the mass gap, as we shall see in the next subsection.

As we are considering one ingoing mode here, it is instructive to compare the angular momentum superradiance condition $m>m_Q$ with the Zel'dovich rotational superradiance condition $\oi_+/m_+<\Omega_Q=\oi_Q/m_Q$. Re-writing the Zel'dovich condition near the mass gap, we get $m>(1+(1+\epi)/\oi_Q)m_Q$. Since $\epi>0$ and $\oi_Q<1$, we see that the Zel'dovich condition
is stronger than $m>m_Q$. So in the $Q$-ball case, it is not necessary for the Zel'dovich condition to be satisfied to have angular momentum superradiance.  

\subsection{Superradiance criteria and amplification limits}
\label{sec:limits}

In this subsection, we shall review the frequency criteria for the amplification factors to go above 1, {\it i.e.,} for the wave amplification to take place \cite{Saffin:2022tub,Gao:2023gof,Cardoso:2023dtm}. We will also obtain some generic upper bounds on the amplification factors, which are independent of the amplitudes of the perturbative scattering waves. This is again made possible by using the particle number conservation in the scattering. 

\begin{figure}
	\centering
		\includegraphics[height=12.0cm]{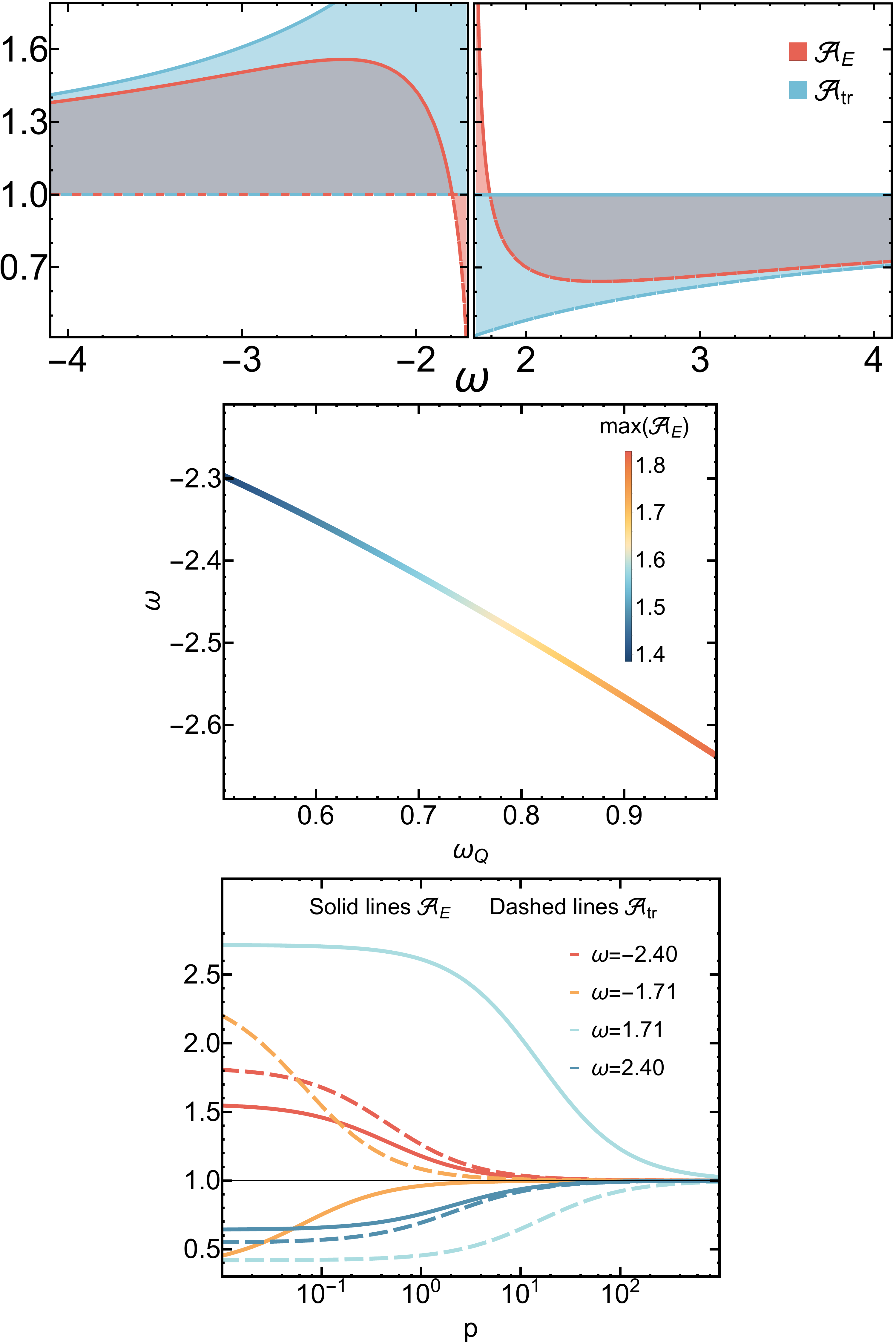}
	\caption{A guide to achieve large amplification factors, taking the case of only ingoing $\eta^+$ modes for an example. {\it Top subfigure}: Bounds on amplification factors $\mathcal{A}_E$ and $\mathcal{A}_{tr}$, solid lines being upper bounds and dashed lines being lower bounds. {\it Middle subfigure}: Possible maxima of $\mathcal{A}_E$ when $\oi<0$ and their corresponding $\oi$ and $\oi_Q$ values. {\it Bottom subfigure}: Dependence of $\mathcal{A}_E$ (solid) and $\mathcal{A}_{tr}$ (dashed) on $\oi$ and the ratio $p$; $p={A_+^2}/{B_-^2}$ when $\oi>0$ and $p=A_-^2/B_+^2$ when $\oi<0$. 
 }
 \label{rangeEt}
\end{figure}

Let us first consider the case where there are only ingoing $\eta^+$ modes. In this case, when $\omega>0 \,(\omega<0)$, we have $A_-=0\,(B_-=0)$. Making use of the particle number conservation,
the amplification factors can be rewritten as
\begin{align}
    \mathcal{A}_E & =  \frac{ \frac{\omega_+^2}{k_+^2} p + \frac{\omega_-^2}{k_-^2} }{  \frac{\omega_+^2}{k_+^2} p + \frac{\omega_+^2}{k_+ k_-}  }  , \quad 
    \mathcal{A}_{tr}  = \left| \frac{ \frac{\omega_+}{k_+} p - \frac{\omega_-}{k_-}  }{ \frac{\omega_+}{k_+} p + \frac{\omega_+}{k_-} } \right|  , \nonumber \\  
    \mathcal{A}_L & =  \frac{ \frac{\omega_+ m_+}{k_+^2} p + \frac{ m_- \omega_- }{k_-^2} }{ \frac{\omega_+ m_+}{k_+^2} p + \frac{\omega_+ m_+}{k_+ k_-}  }  , \quad 
    \mathcal{A}_{r\varphi}  = \left| \frac{ \frac{m_+}{k_+} p - \frac{m_-}{k_-} }{ \frac{m_+}{k_+} p + \frac{m_+}{k_-} } \right|  ,
\end{align}
where we have defined $p={A_+^2}/{B_-^2}$ when $\oi>0$ and $p^{-1}=A_-^2/B_+^2$ when $\oi<0$. Obviously, $p \ge 0$, so running through all possible $p$ gives us the allowed range of these amplification factors, and these ranges only depend on with $\omega_Q$, $\omega$ and $m$. Assuming that $\oi_E$ solves $\mathcal{A}_E=1$ and $\oi_L$ solves $\mathcal{A}_L=1$, they must satisfy
\bal
  \f{\omega_+^2}{k_+} &=\f{\omega_-^2 }{k_-} \bigg|_{\omega \to \omega_E } , \\
  \f{\omega_+ m_+ }{k_+} &= \f{\omega_- m_- }{k_-} \bigg|_{\omega \to \omega_L}.
\eal
It is easy to see that when $\omega<-\omega_E$ or $ 1 + \omega_Q < \omega < \omega_E $, the energy amplification factor falls within the range $1 <\mc{A}_E <{\omega_-^2 k_+}/{\omega_+^2 k_-} $. That is, we can always achieve superradiance in this range. When $\omega_E < \omega $, we have ${\omega_-^2 k_+}/{\omega_+^2 k_-} <\mc{A}_E< 1$. (Numerically, we can verify that the criteria $\epsilon_E$ established in Eqs.~\eqref{epsE} and \eqref{ALS} are conservative: $1+\omega_Q+\epsilon_E < \omega_E$ and $1+\omega_Q+\epsilon_L < \omega_L$ for $m>m_Q$.) Similarly, with the same analyses, we can find the exact superradiance ranges for all these amplification factors. The results are tabulated in Table \ref{Tab:arange} for the case where there are only ingoing $\eta^+$ modes, and the table shows the range of parameters and amplification factors that can produce superradiance in the top four rows, and the range of parameters and amplification factors that correspond to the non-superradiant regime in the bottom four rows. Note that the bounds on $\mc{A}_{E}$ and $\mc{A}_{tr}$ do not depend on $m$, and we see that the upper limits on these amplification factors are all unbounded from above. In Table \ref{Tab:arange}, when $m=-m_Q$, we have $m_+ = 0$, and the amplification factors $\mathcal{A}_L$ and $\mathcal{A}_{r\varphi}$ diverge. This is because the ingoing waves have zero angular momentum and angular momentum flux, so the outgoing waves carry away any amount of angular momentum and angular momentum flux, making $\mathcal{A}_L$ and $\mathcal{A}_{r\varphi}$ unbounded. A few visualizations of the bounds on these amplification factors are shown in Figure \ref{rangeEt}. Our numerical results all fall within these bounds. These bounds can be useful as a guide to find parameters such as $\oi_Q$, $\oi$, $m$ and $p$ to achieve larger amplifications for the scattering waves. 

Since the $\eta^+$ and $\eta^-$ modes are symmetric with respect to the reflection $\oi\to-\oi$, the bounds for the case where there are only ingoing $\eta^-$ modes can be obtained from Table \ref{Tab:arange} by replacing $\oi$ with  $-\oi$.

\begin{table*}[ht]
    \centering
    \caption{Analytical bounds on the amplification factors when there are only ingoing $\eta^+$ modes. The bounds for ingoing $\eta^-$ modes only can be obtained by replacing $\oi$ with $-\oi$. 
The table shows the range of parameters and amplification factors that can produce superradiance in the top four rows, and the range of parameters and amplification factors that correspond to the non-superradiant regime in the bottom four rows. When $m=-m_Q$, the amplification factors $\mathcal{A}_L$ and $\mathcal{A}_{r\varphi}$ tend to infinity because $m_+ = 0$. }
    \begin{tabular}{c|c|c}
    \hline \hline
          & $\omega$ ( $m_Q \ne 0$ ) & $\mathcal{A}$  \\ \hline
           $\mathcal{A}_E$   & $\omega<-\omega_E$ or $1+\omega_Q<\omega<\omega_E$  &  $\left( 1 , \frac{\omega_-^2 k_+}{\omega_+^2 k_-} \right)$   \\ 
\hline
          $\mathcal{A}_{tr}$   & $\omega<-(1+\omega_Q)$ & $\left( 1 , -\frac{\omega_-}{\omega_+} \right)$  \\ 
\hline
          \multirow{2}{*}{ $\mathcal{A}_L$ }  & ($m>m_Q$ and $1+\omega_Q<\omega<\omega_L$)  & \multirow{2}{*}{ $ \left( 1 , \frac{\omega_- m_- k_+}{\omega_+ m_+ k_-} \right)  $  }  \\ 
         & or ($m<-m_Q$ and ($\omega<\omega_L$ or $1+\omega_Q<\omega$)) &    \\
         \hline
         \multirow{2}{*}{ $\mathcal{A}_{r\varphi}$ }  & $-m_Q<m<0$ and $| \omega |>(1+\omega_Q)$  &  $\left( 0 ,\left| \frac{m_-}{m_+} \right|  \right)$   \\ 
         & $m<-m_Q$ and $| \omega |>(1+\omega_Q)$  &  $\left( 1 ,\left| \frac{m_-}{m_+} \right|  \right)$   \\ 
 \bottomrule[1.1pt]  
          $\mathcal{A}_E$  & $-\omega_E<\omega<-(1+\omega_Q)$ or $ \omega_E < \omega $  &  $\left( \frac{\omega_-^2 k_+}{\omega_+^2 k_-} , 1 \right)$  \\ 
\hline
          $\mathcal{A}_{tr}$    & $1+\omega_Q<\omega$ & $\left( -\frac{\omega_-}{\omega_+} , 1 \right)$ \\ 
\hline
         \multirow{3}{*}{ $\mathcal{A}_L$ }  & ($ m>m_Q$ and $( \omega<-(1+\omega_Q)$ or $\omega>\omega_L ) $)  &  \multirow{3}{*}{ $\left( \frac{\omega_- m_- k_+}{\omega_+ m_+ k_-} , 1 \right)$ }  \\ 
          &  or ($ m_Q \ge m > -m_Q$ and $| \omega |>(1+\omega_Q)$) &   \\
          &  or ($m<-m_Q$ and $\omega_L<\omega<-(1+\omega_Q)$) &   \\ 
\hline
          $\mathcal{A}_{r\varphi}$   & $ m \ge 0 $ and $| \omega |>(1+\omega_Q)$  &  $\left( {\rm max}\left(0,-\frac{m_-}{m_+}\right) , 1  \right)$   \\  
         \hline \hline
    \end{tabular}
    \label{Tab:arange}
\end{table*}

\begin{figure}
	\centering
		\includegraphics[height=10.0cm]{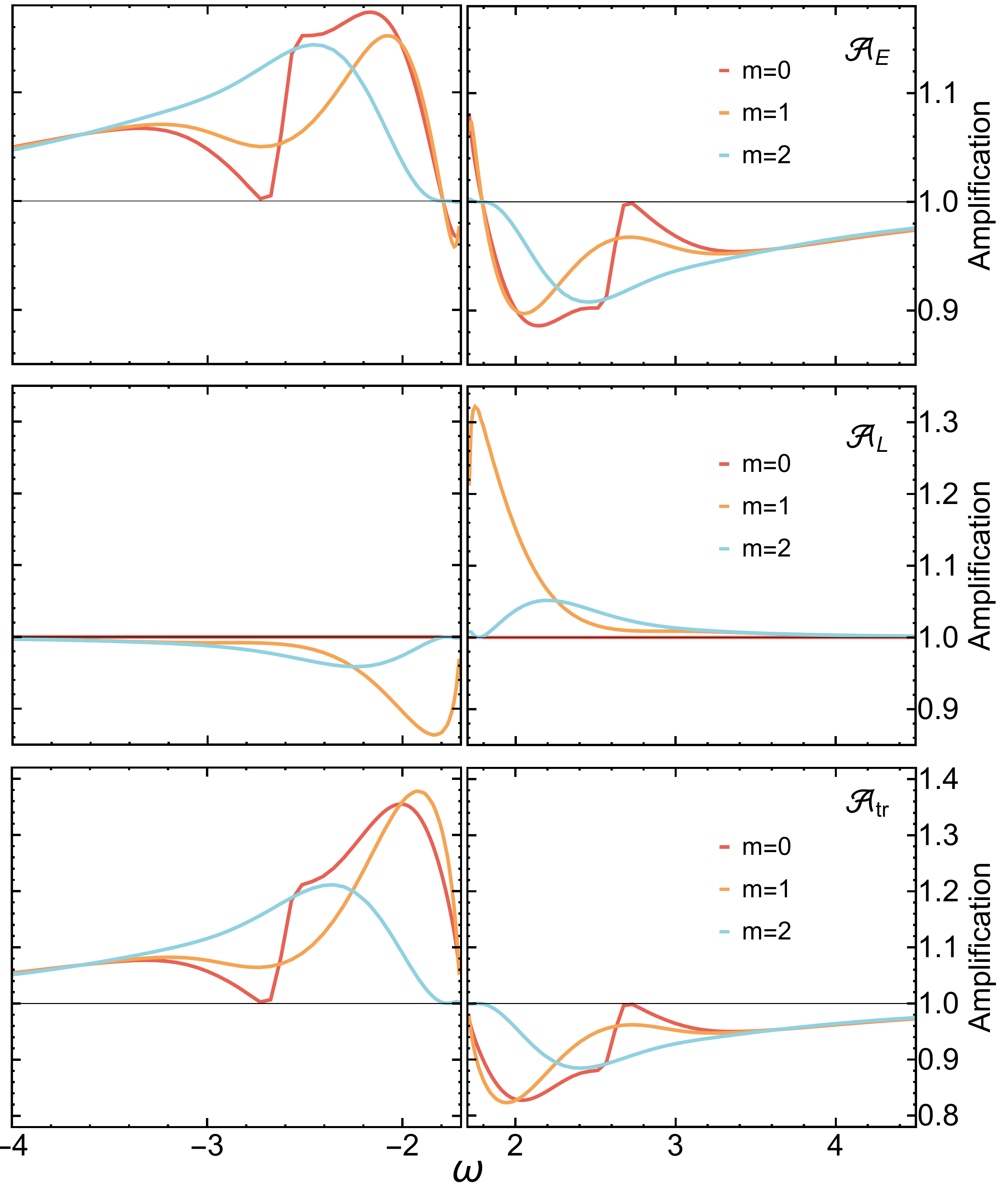}
	\caption{ Spectra of the energy amplification factor $\mathcal{A}_E$, the angular momentum amplification factor $\mathcal{A}_L$, and the energy current amplification factor $\mathcal{A}_{tr}$. The background $Q$-ball is the non-spinning case and spherical symmetric solution, and the scattering mode only keep the lowest cut-off $N^{\pm}_{\rm max}=1$ with the even parity. Since the angular momentum current is proportional to the particle number according to the numerical analysis, so $\mathcal{A}_{r\varphi}=1$ for the non-spinning background $Q$-ball. }
 \label{fig:mq0}
\end{figure}

For $\mathcal{A}_E$ and $\mathcal{A}_L$, the general amplification criteria can also be established when there are both $+$ and $-$ modes ingoing. Assuming that $\oi_E$ solves $\mathcal{A}_E=1$ and $\oi_L$ solves $\mathcal{A}_L=1$, they must satisfy
\bal
 \omega_E  & = \sqrt{ 1+\omega_Q^2 + \sqrt{ 1+ 4 \omega_Q^2 } } , \\ 
 \omega_L  & \approx \left\{\begin{matrix}
 \omega_Q + \left( 1 -  \frac{m_-^2}{m_+^2}  \left( 1 - \left( 1+2 \omega_Q \right)^{-2} \right)  \right)^{-1/2} , \\ \text{if }  m>m_Q ; \\
-\omega_Q - \left( 1 - \frac{m_+^2}{m_-^2} \left( 1 - (1+2\omega_Q)^{-2} \right) \right)^{-1/2} , \\ \text{if }  m<-m_Q .
\end{matrix}\right.
\eal

Finally, we briefly discuss the non-spinning $Q$-ball and its superradiance effect, where $m_Q = 0$ and $m_+=- m_-=m$. The background $Q$-ball only gives us a spherical symmetric solution for the non-spinning case of the complex scalar field, which has only one cut-off mode $N_{\rm max}=k^{Q}_{\rm max} + 1 = 1$. Due to the spherical symmetric background $Q$-ball, the linear perturbation gives rise to uncoupled ODEs between the different modes. In Figure \ref{fig:mq0}, we only consider the uncoupled lowest cut-off $N_{\rm max}^{\pm}=1$ with the even parity. When $m=0$, the ingoing and outgoing mode has not any angular momentum, so we define $\mathcal{A}_L=1$ for this case. When $m \neq 0$, the frequency criteria for the angular momentum tends to infinity, therefore we can achieve the superradiance of the $\mathcal{A}_L$ in the range $\omega>1+\omega_Q$. We also find the angular momentum current is proportional to the particle number for both  the ingoing and outgoing cases from the numerical analysis, and hence $\mathcal{A}_{r\varphi}=1$ for the non-spinning background $Q$-ball.

\section{Conclusion}
\label{sec:level5}

In summary, we have first reviewed the setup of the spinning $Q$-ball solutions in $3+1 {\rm D}$, which uses the expansion of the associated Legendre functions to convert the intractable partial differential equation into the coupled ODEs. The numerical solutions can be obtained by truncating the partial wave expansion. We found that the problem can be efficiently solved with a relaxation method (see Appendix \ref{sec:levela1}) with high accuracy.

Second, we have investigated the perturbative scattering on a $Q$-ball background. By again applying the expansion of the associated Legendre functions, the scattering wave equations can be obtained, which are also in a form of coupled ODEs.
We imposed appropriate boundary conditions, choosing only one ingoing mode with the lowest cut-off, and use a high dimensional shooting method (see Appendix \ref{sec:levela2}) to obtain the solution to the scattering wave equations. This enables us to obtain the amplification factors for the energy density, energy flux, angular momentum density and angular momentum flux between the outgoing and ingoing scattering waves in a large $r$ spherical shell. The particle number in the scattering is accurately conserved. The numerical accuracy and convergence are discussed in Appendix \ref{sec:levelb}.
Since the spinning $Q$-ball is axi-symmetric, angular distribution of the amplification factors is explored in details (see Figure \ref{fig:thi1mode} and \ref{fig:thi1mode2}.) We also investigated the influence of the parity in the scattering, for both the background $Q$-ball and scattering waves, and found that, in general, the even parity $Q$-ball lends to more energy amplification than the odd one. 

Next, we analyzed the amplification in asymptotic regions: near the mass gap, the amplification factors exhibit some distinct behaviors (see Section \ref{sec:asymp}); but when $\omega$ is large, all the amplification factors approach 1. 
We also derived the superradiance criteria and amplification limits (see Table \ref{Tab:arange}) and found that, for only ingoing $\eta^+$ modes, the factor $\mathcal{A}_E$ can be superradiant only when $\omega<-\omega_E$ or $ 1 + \omega_Q < \omega < \omega_E $, $\mathcal{A}_L$ can be superradiant only when $m>m_Q$ or $m<-m_Q $, $\mathcal{A}_{tr}$ can be superradiant only when $\omega<-(1+\omega_Q)$, and  $\mathcal{A}_{r \varphi}$ can be superradiant only when $m<0$.  We found that the standard Zel'dovich rotational superradiance criterion is violated for the angular momentum at least for the case where there is only one ingoing mode. In fact, for the $Q$-ball system, some Zel'dovich-violating parameter regions still allow for superradiance to happen. This is of course not surprising, as the perturbative scattering around a $Q$-ball contains two modes: when there is only one ingoing mode, the outgoing waves contain both modes.  

In this paper, we focused on scatterings where the perturbative ingoing modes contain only the lowest mode of the partial waves. One obvious generalization is to consider more generic ingoing setups and explore how the ingoing partial waves affect the amplification factors. Another intriguing aspect that is worth exploring is the time domain evolution of the scattering around the 3D $Q$-ball, which for a spinning $Q$-ball needs to be done without the spherical symmetry. Furthermore, one may also study the time domain evolution for boson stars with a $Q$-ball-like potential, \cite{Dvali:2023qlk} may give us some insight.

\acknowledgments

We would like to thank Xin Meng and Victor Jaramillo for helpful discussions. SYZ acknowledges support from the National Natural Science Foundation of China under grant No.~12075233 and 12247103, and from the National Key R\&D Program of China under grant No. 2022YFC220010. The work of PMS was funded by STFC Consolidated Grant Number ST/T000732/1.


\section*{Data Access Statement}
All data created during this research are openly available from the University of Nottingham data repository at https://rdmc.nottingham.ac.uk/

\appendix

\section{Relaxation and high dimensional shooting}
\label{sec:levela}

In this appendix, we will review the basics of the relaxation method that we use to solve for the background spinning $Q$-balls and the high dimensional shooting method that we use to solve for the perturbative scattering on top of the $Q$-ball background.

\subsection{Relaxation method}
\label{sec:levela1}

Relaxation methods solve boundary value problems by iteratively updating trial solutions across the entire solving range (or grid) until a desired accuracy is reached. 

Consider a boundary value problem for a system of first-order ODEs
\be
\label{ode001}
\p_r F(r) = Y (r;F(r)) , 
\ee 
in the range $(r_a,r_b)$ with boundary conditions
\be
B^{(a)}(r_a;F(r_a))=0,~~~B^{(b)}(r_b;F(r_b))=0 , 
\ee
where $F$, $Y$ and $B=(B^{(a)},B^{(b)})^T$ are $N$-component vectors, such as $F=(F^{(1)},F^{(2)},...,F^{(N)})^T$. Now, we want to solve this boundary value problem with a finite difference method, discretizing the solving range with a grid $r_k= r_a+k(r_b-r_a)/M,~k=0,1,2,...,M$. Denoting the value of a quantity $X$ at $r_k$ as $X_k$, a finite-difference version of \eqref{ode001} and its boundary conditions can be written as
\be
E_k(r_k,r_{k-1}; F_k,F_{k-1})=0 , 
\ee
where $k=1,2,...,M$ we have
\begin{align}
E_{k} & \equiv F_k - F_{k-1} \nonumber \\
&~~~~ + (r_{k-1} - r_{k}) Y \( \f{r_{k-1} + r_{k}}2,\f{F_k+F_{k-1}}2 \) , 
\end{align}
and we have also defined 
\be
E_0 \equiv (B^{(a)}_0,0),~E_M \equiv (0,B^{(b)}_M) , 
\ee
Note that $E_k$ only depends on $r_k$, $r_{k-1}$, $F_k$ and $F_{k-1}$. We can conveniently combine $E_k=0$ with different $k$ into an $(M+1)\times N$ matrix equation when solving them numerically. 

The relaxation method starts with an initial profile $P_k=P(r_k)$ for the whole grid and assumes the approximate solution is given by
\be
F_k =  P_k  + \Delta F_k.
\ee 
$\Delta F_k=(\Delta F^{(1)}_k,\Delta F^{(2)}_k,...,\Delta F^{(N)}_k)^T$ is to be determined by the following equation
\begin{align}
\label{E1stOrder} 
 E_{k}(P_k,P_{k-1}) +  E_{k,k} \cdot \Delta F_k +  E_{k,k-1} \cdot \Delta F_{k-1} \approx 0 , 
\end{align}
where we have Taylor-expanded the equations of motion $E_{k}(P_k+\Delta F_k,P_{k-1}+\Delta F_{k-1})=0$ to leading order and defined $E_{k,k} = {\p E_{k}}/{\p P_k}$ and $E_{k,k-1} = {\p E_{k}}/{\p P_{k-1}}$ and the dot product of vectors $A \cdot B = \sum_{i=1}^{N}A^{(i)} B^{(i)}$. Viewing $(\Delta F^{(i)}_k)$ as a vector $\Delta \mc{F}$ by enumerating both the $i$ and $k$ indices, it is easy to transform \eqref{E1stOrder} into a linear algebraic equation of the form $\mc{A}\Delta \mc{F} = \mc{B}$, which can be solved by modern efficient matrix equation solvers. After obtaining $\Delta F_k$, we use $F_k =  P_k  + \Delta F_k$ as the second profile and then iterate until a desired accuracy is reached.

To determine whether the profile has relaxed to a solution with a target accuracy, we can define the following relative error  
\begin{align}
E_{\rm rel} = \frac{1}{(M+1)N} \sum\limits_{i=1}^{N} \sum\limits_{k=0}^{M} \left| \frac{\Delta F^{(i)}_{k} }{ {\rm max}_k ({F^{(i)}_{k}})}\right|,
\end{align}
to control the iteration process. For example, in this paper, we choose $E_{\rm rel}$ to be around $10^{-15}$. Alternatively, we can also monitor the relaxation process with an absolute error.

For the background $Q$-ball solution, we need to solve a system of second-order ODEs \eqref{fellequation}.
But they can be easily reduced to first-order ODEs by introducing more variables $h_\ell(r)=\p_r f_\ell(r)$, which leads to
\begin{align}
\p_r f_\ell & = h_\ell , \\
\p_r h_\ell & =  -\frac{2}{r} h_\ell - \!\[ \omega_Q^2 - 1 - \frac{\ell (\ell+1)}{r^2} \] \! f_{\ell} - \mathcal{V}_{\ell} (f_{\ell'}) .
\end{align}
The solving range of the exact problem is from $r=0$ to $r=\infty$, but in practice, we numerically solve it from $r_a=\varepsilon$ to $r_b=50$, where $0<\epsilon\ll 1$. It is also crucial to employ a sensible initial profile; otherwise, achieving a convergent solution may not be obtained, or the process may take an extended period to reach a solution. However, when it comes to our background $Q$-ball solution, obtaining reasonable initial profiles is easily achievable through a few trial-and-error attempts, especially for a larger $\omega_Q$.

\subsection{Shooting method}
\label{sec:levela2}

The shooting method solves a boundary value problem by treating it as an initial value problem and iteratively tuning the initial conditions to achieve the desired solution. While a 1D shooting method ({\it i.e.,} with one parameter to tune) can often be done manually, a high dimensional shooting method is computationally more challenging.

Consider a boundary value problem of the following form
\be
\label{shootingeom0}
\mc{D}_r H^{(i)}(r)  + \mc{V}(r;H^{(i')}(r)) =0, ~~~i=1,2,...,N , 
\ee
in the range $(r_a,r_b)$ with boundary conditions
\bal
\label{shootingB1}
B^{(k)}_a(r_a;H(r_a), \pd_r H(r_a))&=0,~~~k=1,2,...,K,\\
\label{shootingB2}
B^{(l)}_b(r_b;H(r_b), \pd_r H(r_b))&=0,~~~l=1,2,...,L , 
\eal
where $\mc{D}_r$ is a second-order different operator, $\mc{V}$ is a function of $H^{(i')}$\footnote{In general, $\mc{V}$ can be a function of $H^{(i')}$, but it is linear for the perturbative wave scattering on top of the $Q$-ball background we are dealing in this paper.}, and $K+L=2N$. In the simplest shooting method, we effectively treat $r$ as time, $r_a$ as the ``initial time'' (for our perturbative wave scattering case, $r_a=\epsilon$ with $0<\epsilon\ll 1$) and $r_b$ as the ``final time''. The goal is find an appropriate set of initial conditions for $H^{(i)}(r_a)$ and $\pd_r H^{(i)}(r_a)$ that satisfy \eqref{shootingB1} and \eqref{shootingB2}. Often, it is the case, such as in the $Q$-ball perturbative scattering we have in this paper, that $H^{(i)}(r_a)$ and $\pd_r H^{(i)}(r_a)$ are not freely adjustable; instead, they must adhere to additional conditions relevant to the problem at hand. While $B_a^{(k)}(r_a;H(r_a), \pd_r H(r_a))=0$ can be easily factored into the choice of the initial conditions, to satisfy $B^{(k)}_b(r_a;H(r_a), \pd_r H(r_a))=0$, we essentially choose various different initial conditions at $r_a$, evolve \eqref{shootingeom0} from $r_a$ to $r_b$ and then pick out the correct set of initial conditions. Alternatively, one can also evolve the ODEs both from $r_a$ and $r_b$ to a meeting point $r_m$ and matching the solutions at $r_m$, which is sometimes more accurate.

In Mathematica, this process can be efficiently implemented with the \texttt{ParametricNDSolve} and \texttt{FindRoot} command. 
Essentially, by numerically integration from $r_a$ to $r_b$, \texttt{ParametricNDSolve} can output a set of numerical equations among the free parameters of boundary conditions. We can then feed these equations into the \texttt{FindRoot} function to find the solution of this set of equations, which gives rise to the desired solution for $H^{(i)}(r)$ between $r_a$ and $r_b$. 

For the $Q$-ball perturbative scattering in this paper, by solving \eqref{modes-equ} from $r_a=\varepsilon$ to $r_b=50$ with \texttt{ParametricNDSolve} and requiring that the matching conditions on $\eta^\pm_\ell(r_b)$ and $\pd_r\eta^\pm_\ell(r_b)$, we can establish $8 N_{\rm max}$ real relations between $F^\pm_\ell$ and $A^\pm_\ell$ and $B^\pm_\ell$:
\be
\mc{E}_I(F^\pm_\ell, A^\pm_\ell, B^\pm_\ell) =0, ~~~I=1,2,...,8 N_{\rm max} , 
\ee
To get a unique solution, we also need to, by user's choice, supply $4 N_{\rm max}$ relations among $F^\pm_\ell$ and $A^\pm_\ell$ and $B^\pm_\ell$, as there are $12 N_{\rm max}$ degrees of freedom among them. For example, if we are interested in an even-parity scattering where there is only one ingoing mode with the lowest $\ell$ for $B^+_\ell$ and $\omega>0$, we can impose $A^-_\ell=B^+_{\ell>1}=0$, which gives $4 N_{\rm max}-2$ extra conditions. Additionally, we can also choose $F^+_1=1$ due to the linearity of the scattering equation, which gives another 2 conditions. Thus, we have $12 N_{\rm max}$ equations, which can be uniquely solved for $12 N_{\rm max}$ variables $F^\pm_\ell$ and $A^\pm_\ell$ and $B^\pm_\ell$. We can then use \texttt{FindRoot} solve these equations, which gives the solution for the scattering problem.

\begin{figure}
	\centering
		\includegraphics[height=2.8cm]{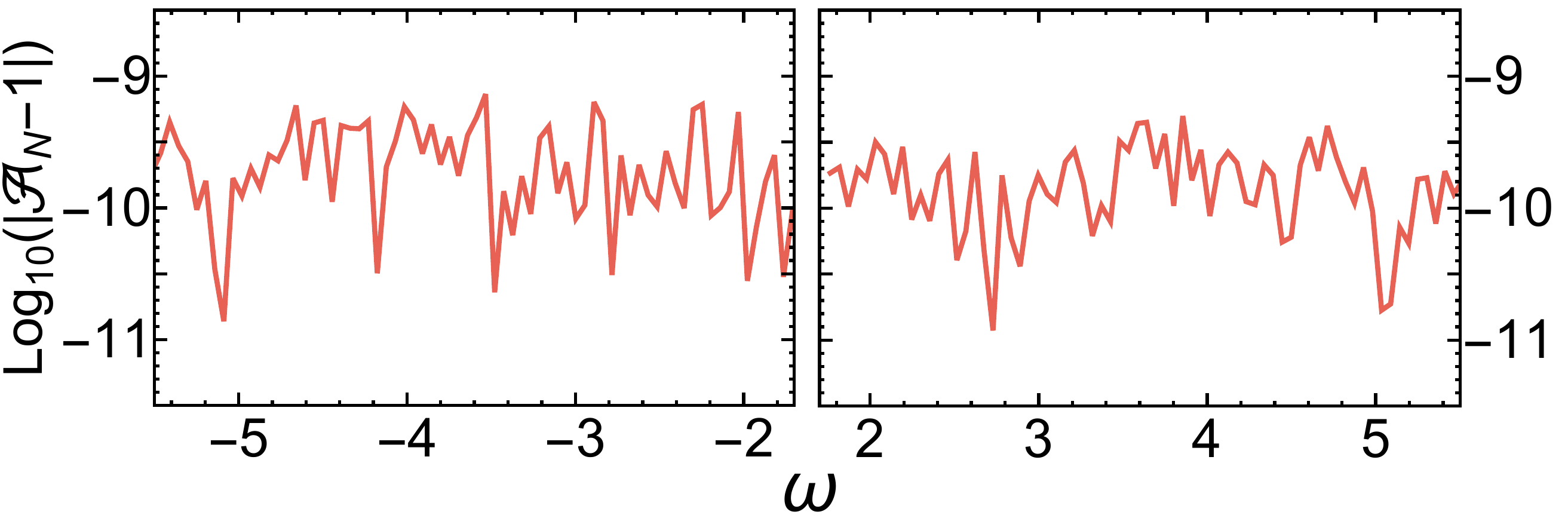}
	\caption{ Deviation of the amplification factors for the particle number ${\rm Log}( | \mathcal{A}_N - 1 | )$.
The red line is the numerical result, which has relative high accuracy, as shown in the figure, with an error of at least below $10 ^ {-9}$. 
The numerical setup is the same as the $m=0$ case of Figure \ref{fig:1mode}. }
 \label{fig:error}
\end{figure}

\begin{figure}
	\centering
		\includegraphics[height=3.6cm]{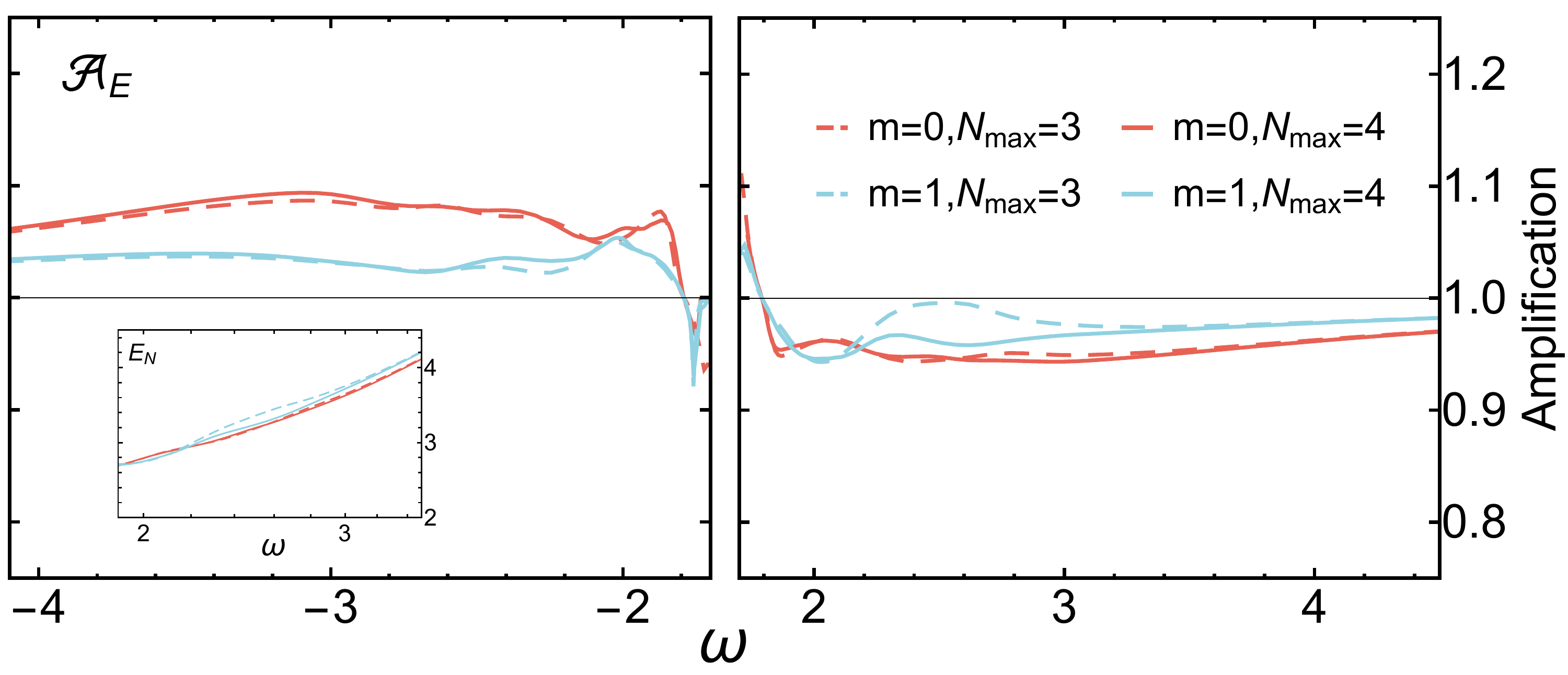}
	\caption{Differences between truncations with $N_{\rm max}=3$ and $N_{\rm max}=4$. The inset shows the averagy energy for the ingoing waves for the 3rd and 4th order truncation in $\ell$, which means that the differences between $N_{\rm max}=3$ and $N_{\rm max}=4$ are due to numerical accuracy, rather than the truncation itself. The setup is the same as Figure \ref{fig:1mode}. }
 \label{fig:1modeconvergence}
\end{figure}

\section{Numerical accuracy and convergence}
\label{sec:levelb}

In this appendix, we provide some plots to demonstrate the numerical accuracy and convergence in the obtained results.

As mentioned in the main text, a good quantity to monitor the accuracy of the numerical results is the amplification factor $\mc{A}_N$, which is guaranteed to be $1$ by the particle number conservation in the scattering on the $Q$-ball background. In Figure \ref{fig:error}, we see that in our numerical computations $\mc{A}_N=1$ is held in relative high accuracy.

In Figure \ref{fig:1modeconvergence}, we plot the spectra of the $\mc{A}_E$ amplification factor with a 3rd and 4th order truncation in partial waves respectively, for the case of one ingoing mode. We see that generally the convergence is very good with a 3rd truncation.  All the perturbative scattering results in the main text are presented with the 4th order truncation.   However, the $m=1$ lines have a small region where the differences between the 3rd and 4th order truncation can be greater than 0.02 but still smaller than 0.04. The inset inside the left of the figure shows the averagy energy for the ingoing waves, which should be independent of the $\ell_{\rm max}$ cutoff, given by
\begin{align}
E_N = \frac{\frac{\omega_+^2}{k_+^2} B^{2}_+ + \frac{\omega_-^2}{k_-^2} A^{2}_-}{\frac{2}{k_+} B^2_+ + \frac{2}{k_-} A^2_-} \text{ for } \omega>0. 
\end{align}
Therefore, the relative sizable errors near $\oi=2.6$ in the $m=1$ case seem to be largely due to the accuracy of the high dimensional shooting method, rather than the truncation in $\ell$. Recall that in our high dimensional shooting method, we need to solve a system of high dimensional algebraic equations to obtain both the amplitudes of the ingoing and outgoing waves, which in turn are used to compute the energies of the waves and the amplification factors. 


\bibliography{xref}

\begin{thebibliography}{67}%
\makeatletter
\providecommand \@ifxundefined [1]{%
 \@ifx{#1\undefined}
}%
\providecommand \@ifnum [1]{%
 \ifnum #1\expandafter \@firstoftwo
 \else \expandafter \@secondoftwo
 \fi
}%
\providecommand \@ifx [1]{%
 \ifx #1\expandafter \@firstoftwo
 \else \expandafter \@secondoftwo
 \fi
}%
\providecommand \natexlab [1]{#1}%
\providecommand \enquote  [1]{``#1''}%
\providecommand \bibnamefont  [1]{#1}%
\providecommand \bibfnamefont [1]{#1}%
\providecommand \citenamefont [1]{#1}%
\providecommand \href@noop [0]{\@secondoftwo}%
\providecommand \href [0]{\begingroup \@sanitize@url \@href}%
\providecommand \@href[1]{\@@startlink{#1}\@@href}%
\providecommand \@@href[1]{\endgroup#1\@@endlink}%
\providecommand \@sanitize@url [0]{\catcode `\\12\catcode `\$12\catcode
  `\&12\catcode `\#12\catcode `\^12\catcode `\_12\catcode `\%12\relax}%
\providecommand \@@startlink[1]{}%
\providecommand \@@endlink[0]{}%
\providecommand \url  [0]{\begingroup\@sanitize@url \@url }%
\providecommand \@url [1]{\endgroup\@href {#1}{\urlprefix }}%
\providecommand \urlprefix  [0]{URL }%
\providecommand \Eprint [0]{\href }%
\providecommand \doibase [0]{https://doi.org/}%
\providecommand \selectlanguage [0]{\@gobble}%
\providecommand \bibinfo  [0]{\@secondoftwo}%
\providecommand \bibfield  [0]{\@secondoftwo}%
\providecommand \translation [1]{[#1]}%
\providecommand \BibitemOpen [0]{}%
\providecommand \bibitemStop [0]{}%
\providecommand \bibitemNoStop [0]{.\EOS\space}%
\providecommand \EOS [0]{\spacefactor3000\relax}%
\providecommand \BibitemShut  [1]{\csname bibitem#1\endcsname}%
\let\auto@bib@innerbib\@empty
\bibitem [{\citenamefont {Friedberg}\ \emph {et~al.}(1976)\citenamefont
  {Friedberg}, \citenamefont {Lee},\ and\ \citenamefont
  {Sirlin}}]{Friedberg:1976me}%
  \BibitemOpen
  \bibfield  {author} {\bibinfo {author} {\bibfnamefont {R.}~\bibnamefont
  {Friedberg}}, \bibinfo {author} {\bibfnamefont {T.}~\bibnamefont {Lee}},\
  and\ \bibinfo {author} {\bibfnamefont {A.}~\bibnamefont {Sirlin}},\
  }\bibfield  {title} {\bibinfo {title} {{A Class of Scalar-Field Soliton
  Solutions in Three Space Dimensions}},\ }\href
  {https://doi.org/10.1103/PhysRevD.13.2739} {\bibfield  {journal} {\bibinfo
  {journal} {Phys. Rev. D}\ }\textbf {\bibinfo {volume} {13}},\ \bibinfo
  {pages} {2739} (\bibinfo {year} {1976})}\BibitemShut {NoStop}%
\bibitem [{\citenamefont {Coleman}(1985)}]{Coleman:1985ki}%
  \BibitemOpen
  \bibfield  {author} {\bibinfo {author} {\bibfnamefont {S.~R.}\ \bibnamefont
  {Coleman}},\ }\bibfield  {title} {\bibinfo {title} {{Q-balls}},\ }\href
  {https://doi.org/10.1016/0550-3213(86)90520-1} {\bibfield  {journal}
  {\bibinfo  {journal} {Nucl. Phys. B}\ }\textbf {\bibinfo {volume} {262}},\
  \bibinfo {pages} {263} (\bibinfo {year} {1985})},\ \bibinfo {note}
  {[Addendum: Nucl.Phys.B 269, 744 (1986)]}\BibitemShut {NoStop}%
\bibitem [{\citenamefont {Volkov}\ and\ \citenamefont
  {Wohnert}(2002)}]{Volkov:2002aj}%
  \BibitemOpen
  \bibfield  {author} {\bibinfo {author} {\bibfnamefont {M.~S.}\ \bibnamefont
  {Volkov}}\ and\ \bibinfo {author} {\bibfnamefont {E.}~\bibnamefont
  {Wohnert}},\ }\bibfield  {title} {\bibinfo {title} {{Spinning Q balls}},\
  }\href {https://doi.org/10.1103/PhysRevD.66.085003} {\bibfield  {journal}
  {\bibinfo  {journal} {Phys. Rev. D}\ }\textbf {\bibinfo {volume} {66}},\
  \bibinfo {pages} {085003} (\bibinfo {year} {2002})},\ \Eprint
  {https://arxiv.org/abs/hep-th/0205157} {arXiv:hep-th/0205157} \BibitemShut
  {NoStop}%
\bibitem [{\citenamefont {Kleihaus}\ \emph {et~al.}(2005)\citenamefont
  {Kleihaus}, \citenamefont {Kunz},\ and\ \citenamefont
  {List}}]{Kleihaus:2005me}%
  \BibitemOpen
  \bibfield  {author} {\bibinfo {author} {\bibfnamefont {B.}~\bibnamefont
  {Kleihaus}}, \bibinfo {author} {\bibfnamefont {J.}~\bibnamefont {Kunz}},\
  and\ \bibinfo {author} {\bibfnamefont {M.}~\bibnamefont {List}},\ }\bibfield
  {title} {\bibinfo {title} {{Rotating boson stars and Q-balls}},\ }\href
  {https://doi.org/10.1103/PhysRevD.72.064002} {\bibfield  {journal} {\bibinfo
  {journal} {Phys. Rev. D}\ }\textbf {\bibinfo {volume} {72}},\ \bibinfo
  {pages} {064002} (\bibinfo {year} {2005})},\ \Eprint
  {https://arxiv.org/abs/gr-qc/0505143} {arXiv:gr-qc/0505143} \BibitemShut
  {NoStop}%
\bibitem [{\citenamefont {Radu}\ and\ \citenamefont
  {Volkov}(2008)}]{Radu:2008pp}%
  \BibitemOpen
  \bibfield  {author} {\bibinfo {author} {\bibfnamefont {E.}~\bibnamefont
  {Radu}}\ and\ \bibinfo {author} {\bibfnamefont {M.~S.}\ \bibnamefont
  {Volkov}},\ }\bibfield  {title} {\bibinfo {title} {{Existence of stationary,
  non-radiating ring solitons in field theory: knots and vortons}},\ }\href
  {https://doi.org/10.1016/j.physrep.2008.07.002} {\bibfield  {journal}
  {\bibinfo  {journal} {Phys. Rept.}\ }\textbf {\bibinfo {volume} {468}},\
  \bibinfo {pages} {101} (\bibinfo {year} {2008})},\ \Eprint
  {https://arxiv.org/abs/0804.1357} {arXiv:0804.1357 [hep-th]} \BibitemShut
  {NoStop}%
\bibitem [{\citenamefont {Kleihaus}\ \emph {et~al.}(2012)\citenamefont
  {Kleihaus}, \citenamefont {Kunz},\ and\ \citenamefont
  {Schneider}}]{Kleihaus:2011sx}%
  \BibitemOpen
  \bibfield  {author} {\bibinfo {author} {\bibfnamefont {B.}~\bibnamefont
  {Kleihaus}}, \bibinfo {author} {\bibfnamefont {J.}~\bibnamefont {Kunz}},\
  and\ \bibinfo {author} {\bibfnamefont {S.}~\bibnamefont {Schneider}},\
  }\bibfield  {title} {\bibinfo {title} {{Stable Phases of Boson Stars}},\
  }\href {https://doi.org/10.1103/PhysRevD.85.024045} {\bibfield  {journal}
  {\bibinfo  {journal} {Phys. Rev. D}\ }\textbf {\bibinfo {volume} {85}},\
  \bibinfo {pages} {024045} (\bibinfo {year} {2012})},\ \Eprint
  {https://arxiv.org/abs/1109.5858} {arXiv:1109.5858 [gr-qc]} \BibitemShut
  {NoStop}%
\bibitem [{\citenamefont {Herdeiro}\ \emph {et~al.}(2014)\citenamefont
  {Herdeiro}, \citenamefont {Radu},\ and\ \citenamefont
  {Runarsson}}]{Herdeiro:2014pka}%
  \BibitemOpen
  \bibfield  {author} {\bibinfo {author} {\bibfnamefont {C.}~\bibnamefont
  {Herdeiro}}, \bibinfo {author} {\bibfnamefont {E.}~\bibnamefont {Radu}},\
  and\ \bibinfo {author} {\bibfnamefont {H.}~\bibnamefont {Runarsson}},\
  }\bibfield  {title} {\bibinfo {title} {{Non-linear $Q$-clouds around Kerr
  black holes}},\ }\href {https://doi.org/10.1016/j.physletb.2014.11.005}
  {\bibfield  {journal} {\bibinfo  {journal} {Phys. Lett. B}\ }\textbf
  {\bibinfo {volume} {739}},\ \bibinfo {pages} {302} (\bibinfo {year}
  {2014})},\ \Eprint {https://arxiv.org/abs/1409.2877} {arXiv:1409.2877
  [gr-qc]} \BibitemShut {NoStop}%
\bibitem [{\citenamefont {Almumin}\ \emph {et~al.}(2023)\citenamefont
  {Almumin}, \citenamefont {Heeck}, \citenamefont {Rajaraman},\ and\
  \citenamefont {Verhaaren}}]{Almumin:2023wwi}%
  \BibitemOpen
  \bibfield  {author} {\bibinfo {author} {\bibfnamefont {Y.}~\bibnamefont
  {Almumin}}, \bibinfo {author} {\bibfnamefont {J.}~\bibnamefont {Heeck}},
  \bibinfo {author} {\bibfnamefont {A.}~\bibnamefont {Rajaraman}},\ and\
  \bibinfo {author} {\bibfnamefont {C.~B.}\ \bibnamefont {Verhaaren}},\
  }\bibfield  {title} {\bibinfo {title} {{Rotating Q-balls}},\ }\href@noop {}
  {\  (\bibinfo {year} {2023})},\ \Eprint {https://arxiv.org/abs/2302.11589}
  {arXiv:2302.11589 [hep-th]} \BibitemShut {NoStop}%
\bibitem [{\citenamefont {Copeland}\ \emph {et~al.}(2014)\citenamefont
  {Copeland}, \citenamefont {Saffin},\ and\ \citenamefont
  {Zhou}}]{Copeland:2014qra}%
  \BibitemOpen
  \bibfield  {author} {\bibinfo {author} {\bibfnamefont {E.~J.}\ \bibnamefont
  {Copeland}}, \bibinfo {author} {\bibfnamefont {P.~M.}\ \bibnamefont
  {Saffin}},\ and\ \bibinfo {author} {\bibfnamefont {S.-Y.}\ \bibnamefont
  {Zhou}},\ }\bibfield  {title} {\bibinfo {title} {{Charge-Swapping Q-balls}},\
  }\href {https://doi.org/10.1103/PhysRevLett.113.231603} {\bibfield  {journal}
  {\bibinfo  {journal} {Phys. Rev. Lett.}\ }\textbf {\bibinfo {volume} {113}},\
  \bibinfo {pages} {231603} (\bibinfo {year} {2014})},\ \Eprint
  {https://arxiv.org/abs/1409.3232} {arXiv:1409.3232 [hep-th]} \BibitemShut
  {NoStop}%
\bibitem [{\citenamefont {Xie}\ \emph {et~al.}(2021)\citenamefont {Xie},
  \citenamefont {Saffin},\ and\ \citenamefont {Zhou}}]{Xie:2021glp}%
  \BibitemOpen
  \bibfield  {author} {\bibinfo {author} {\bibfnamefont {Q.-X.}\ \bibnamefont
  {Xie}}, \bibinfo {author} {\bibfnamefont {P.~M.}\ \bibnamefont {Saffin}},\
  and\ \bibinfo {author} {\bibfnamefont {S.-Y.}\ \bibnamefont {Zhou}},\
  }\bibfield  {title} {\bibinfo {title} {{Charge-Swapping Q-balls and Their
  Lifetimes}},\ }\href {https://doi.org/10.1007/JHEP07(2021)062} {\bibfield
  {journal} {\bibinfo  {journal} {JHEP}\ }\textbf {\bibinfo {volume} {07}},\
  \bibinfo {pages} {062}},\ \Eprint {https://arxiv.org/abs/2101.06988}
  {arXiv:2101.06988 [hep-th]} \BibitemShut {NoStop}%
\bibitem [{\citenamefont {Hou}\ \emph {et~al.}(2022)\citenamefont {Hou},
  \citenamefont {Saffin}, \citenamefont {Xie},\ and\ \citenamefont
  {Zhou}}]{Hou:2022jcd}%
  \BibitemOpen
  \bibfield  {author} {\bibinfo {author} {\bibfnamefont {S.-Y.}\ \bibnamefont
  {Hou}}, \bibinfo {author} {\bibfnamefont {P.~M.}\ \bibnamefont {Saffin}},
  \bibinfo {author} {\bibfnamefont {Q.-X.}\ \bibnamefont {Xie}},\ and\ \bibinfo
  {author} {\bibfnamefont {S.-Y.}\ \bibnamefont {Zhou}},\ }\bibfield  {title}
  {\bibinfo {title} {{Charge-swapping Q-balls in a logarithmic potential and
  Affleck-Dine condensate fragmentation}},\ }\href
  {https://doi.org/10.1007/JHEP07(2022)060} {\bibfield  {journal} {\bibinfo
  {journal} {JHEP}\ }\textbf {\bibinfo {volume} {07}},\ \bibinfo {pages}
  {060}},\ \Eprint {https://arxiv.org/abs/2202.08392} {arXiv:2202.08392
  [hep-ph]} \BibitemShut {NoStop}%
\bibitem [{\citenamefont {Tranberg}\ and\ \citenamefont
  {Weir}(2014)}]{Tranberg:2013cka}%
  \BibitemOpen
  \bibfield  {author} {\bibinfo {author} {\bibfnamefont {A.}~\bibnamefont
  {Tranberg}}\ and\ \bibinfo {author} {\bibfnamefont {D.~J.}\ \bibnamefont
  {Weir}},\ }\bibfield  {title} {\bibinfo {title} {{On the quantum stability of
  Q-balls}},\ }\href {https://doi.org/10.1007/JHEP04(2014)184} {\bibfield
  {journal} {\bibinfo  {journal} {JHEP}\ }\textbf {\bibinfo {volume} {04}},\
  \bibinfo {pages} {184}},\ \Eprint {https://arxiv.org/abs/1310.7487}
  {arXiv:1310.7487 [hep-ph]} \BibitemShut {NoStop}%
\bibitem [{\citenamefont {Xie}\ \emph {et~al.}(2023)\citenamefont {Xie},
  \citenamefont {Saffin}, \citenamefont {Tranberg},\ and\ \citenamefont
  {Zhou}}]{Xie:2023psz}%
  \BibitemOpen
  \bibfield  {author} {\bibinfo {author} {\bibfnamefont {Q.-X.}\ \bibnamefont
  {Xie}}, \bibinfo {author} {\bibfnamefont {P.~M.}\ \bibnamefont {Saffin}},
  \bibinfo {author} {\bibfnamefont {A.}~\bibnamefont {Tranberg}},\ and\
  \bibinfo {author} {\bibfnamefont {S.-Y.}\ \bibnamefont {Zhou}},\ }\bibfield
  {title} {\bibinfo {title} {{Quantum corrected Q-ball dynamics}},\ }\href@noop
  {} {\  (\bibinfo {year} {2023})},\ \Eprint {https://arxiv.org/abs/2312.01139}
  {arXiv:2312.01139 [hep-th]} \BibitemShut {NoStop}%
\bibitem [{\citenamefont {Kovtun}\ and\ \citenamefont
  {Zantedeschi}(2022)}]{Kovtun:2020udn}%
  \BibitemOpen
  \bibfield  {author} {\bibinfo {author} {\bibfnamefont {A.}~\bibnamefont
  {Kovtun}}\ and\ \bibinfo {author} {\bibfnamefont {M.}~\bibnamefont
  {Zantedeschi}},\ }\bibfield  {title} {\bibinfo {title} {{Breaking BEC:
  Quantum evolution of unstable condensates}},\ }\href
  {https://doi.org/10.1103/PhysRevD.105.085019} {\bibfield  {journal} {\bibinfo
   {journal} {Phys. Rev. D}\ }\textbf {\bibinfo {volume} {105}},\ \bibinfo
  {pages} {085019} (\bibinfo {year} {2022})},\ \Eprint
  {https://arxiv.org/abs/2008.02187} {arXiv:2008.02187 [hep-th]} \BibitemShut
  {NoStop}%
\bibitem [{\citenamefont {Kusenko}\ and\ \citenamefont
  {Shaposhnikov}(1998)}]{Kusenko:1997si}%
  \BibitemOpen
  \bibfield  {author} {\bibinfo {author} {\bibfnamefont {A.}~\bibnamefont
  {Kusenko}}\ and\ \bibinfo {author} {\bibfnamefont {M.~E.}\ \bibnamefont
  {Shaposhnikov}},\ }\bibfield  {title} {\bibinfo {title} {{Supersymmetric Q
  balls as dark matter}},\ }\href
  {https://doi.org/10.1016/S0370-2693(97)01375-0} {\bibfield  {journal}
  {\bibinfo  {journal} {Phys. Lett. B}\ }\textbf {\bibinfo {volume} {418}},\
  \bibinfo {pages} {46} (\bibinfo {year} {1998})},\ \Eprint
  {https://arxiv.org/abs/hep-ph/9709492} {arXiv:hep-ph/9709492} \BibitemShut
  {NoStop}%
\bibitem [{\citenamefont {Enqvist}\ and\ \citenamefont
  {McDonald}(1998)}]{Enqvist:1997si}%
  \BibitemOpen
  \bibfield  {author} {\bibinfo {author} {\bibfnamefont {K.}~\bibnamefont
  {Enqvist}}\ and\ \bibinfo {author} {\bibfnamefont {J.}~\bibnamefont
  {McDonald}},\ }\bibfield  {title} {\bibinfo {title} {{Q balls and
  baryogenesis in the MSSM}},\ }\href
  {https://doi.org/10.1016/S0370-2693(98)00271-8} {\bibfield  {journal}
  {\bibinfo  {journal} {Phys. Lett. B}\ }\textbf {\bibinfo {volume} {425}},\
  \bibinfo {pages} {309} (\bibinfo {year} {1998})},\ \Eprint
  {https://arxiv.org/abs/hep-ph/9711514} {arXiv:hep-ph/9711514} \BibitemShut
  {NoStop}%
\bibitem [{\citenamefont {Fujii}\ and\ \citenamefont
  {Hamaguchi}(2002)}]{Fujii:2002kr}%
  \BibitemOpen
  \bibfield  {author} {\bibinfo {author} {\bibfnamefont {M.}~\bibnamefont
  {Fujii}}\ and\ \bibinfo {author} {\bibfnamefont {K.}~\bibnamefont
  {Hamaguchi}},\ }\bibfield  {title} {\bibinfo {title} {{Nonthermal dark matter
  via Affleck-Dine baryogenesis and its detection possibility}},\ }\href
  {https://doi.org/10.1103/PhysRevD.66.083501} {\bibfield  {journal} {\bibinfo
  {journal} {Phys. Rev. D}\ }\textbf {\bibinfo {volume} {66}},\ \bibinfo
  {pages} {083501} (\bibinfo {year} {2002})},\ \Eprint
  {https://arxiv.org/abs/hep-ph/0205044} {arXiv:hep-ph/0205044} \BibitemShut
  {NoStop}%
\bibitem [{\citenamefont {Enqvist}\ and\ \citenamefont
  {Mazumdar}(2003)}]{Enqvist:2003gh}%
  \BibitemOpen
  \bibfield  {author} {\bibinfo {author} {\bibfnamefont {K.}~\bibnamefont
  {Enqvist}}\ and\ \bibinfo {author} {\bibfnamefont {A.}~\bibnamefont
  {Mazumdar}},\ }\bibfield  {title} {\bibinfo {title} {{Cosmological
  consequences of MSSM flat directions}},\ }\href
  {https://doi.org/10.1016/S0370-1573(03)00119-4} {\bibfield  {journal}
  {\bibinfo  {journal} {Phys. Rept.}\ }\textbf {\bibinfo {volume} {380}},\
  \bibinfo {pages} {99} (\bibinfo {year} {2003})},\ \Eprint
  {https://arxiv.org/abs/hep-ph/0209244} {arXiv:hep-ph/0209244} \BibitemShut
  {NoStop}%
\bibitem [{\citenamefont {Roszkowski}\ and\ \citenamefont
  {Seto}(2007)}]{Roszkowski:2006kw}%
  \BibitemOpen
  \bibfield  {author} {\bibinfo {author} {\bibfnamefont {L.}~\bibnamefont
  {Roszkowski}}\ and\ \bibinfo {author} {\bibfnamefont {O.}~\bibnamefont
  {Seto}},\ }\bibfield  {title} {\bibinfo {title} {{Axino dark matter from
  Q-balls in Affleck-Dine baryogenesis and the Omega(b) - Omega(DM) coincidence
  problem}},\ }\href {https://doi.org/10.1103/PhysRevLett.98.161304} {\bibfield
   {journal} {\bibinfo  {journal} {Phys. Rev. Lett.}\ }\textbf {\bibinfo
  {volume} {98}},\ \bibinfo {pages} {161304} (\bibinfo {year} {2007})},\
  \Eprint {https://arxiv.org/abs/hep-ph/0608013} {arXiv:hep-ph/0608013}
  \BibitemShut {NoStop}%
\bibitem [{\citenamefont {Shoemaker}\ and\ \citenamefont
  {Kusenko}(2009)}]{Shoemaker:2009kg}%
  \BibitemOpen
  \bibfield  {author} {\bibinfo {author} {\bibfnamefont {I.~M.}\ \bibnamefont
  {Shoemaker}}\ and\ \bibinfo {author} {\bibfnamefont {A.}~\bibnamefont
  {Kusenko}},\ }\bibfield  {title} {\bibinfo {title} {{Gravitino dark matter
  from Q-ball decays}},\ }\href {https://doi.org/10.1103/PhysRevD.80.075021}
  {\bibfield  {journal} {\bibinfo  {journal} {Phys. Rev. D}\ }\textbf {\bibinfo
  {volume} {80}},\ \bibinfo {pages} {075021} (\bibinfo {year} {2009})},\
  \Eprint {https://arxiv.org/abs/0909.3334} {arXiv:0909.3334 [hep-ph]}
  \BibitemShut {NoStop}%
\bibitem [{\citenamefont {Zhou}(2015)}]{Zhou:2015yfa}%
  \BibitemOpen
  \bibfield  {author} {\bibinfo {author} {\bibfnamefont {S.-Y.}\ \bibnamefont
  {Zhou}},\ }\bibfield  {title} {\bibinfo {title} {{Gravitational waves from
  Affleck-Dine condensate fragmentation}},\ }\href
  {https://doi.org/10.1088/1475-7516/2015/06/033} {\bibfield  {journal}
  {\bibinfo  {journal} {JCAP}\ }\textbf {\bibinfo {volume} {06}},\ \bibinfo
  {pages} {033}},\ \Eprint {https://arxiv.org/abs/1501.01217} {arXiv:1501.01217
  [astro-ph.CO]} \BibitemShut {NoStop}%
\bibitem [{\citenamefont {Kawasaki}\ and\ \citenamefont
  {Nakatsuka}(2020)}]{Kawasaki:2019ywz}%
  \BibitemOpen
  \bibfield  {author} {\bibinfo {author} {\bibfnamefont {M.}~\bibnamefont
  {Kawasaki}}\ and\ \bibinfo {author} {\bibfnamefont {H.}~\bibnamefont
  {Nakatsuka}},\ }\bibfield  {title} {\bibinfo {title} {{Q-ball decay through
  A-term in the gauge-mediated SUSY breaking scenario}},\ }\href
  {https://doi.org/10.1088/1475-7516/2020/04/017} {\bibfield  {journal}
  {\bibinfo  {journal} {JCAP}\ }\textbf {\bibinfo {volume} {04}},\ \bibinfo
  {pages} {017}},\ \Eprint {https://arxiv.org/abs/1912.06993} {arXiv:1912.06993
  [hep-ph]} \BibitemShut {NoStop}%
\bibitem [{\citenamefont {Gouttenoire}\ \emph {et~al.}(2021)\citenamefont
  {Gouttenoire}, \citenamefont {Servant},\ and\ \citenamefont
  {Simakachorn}}]{Gouttenoire:2021jhk}%
  \BibitemOpen
  \bibfield  {author} {\bibinfo {author} {\bibfnamefont {Y.}~\bibnamefont
  {Gouttenoire}}, \bibinfo {author} {\bibfnamefont {G.}~\bibnamefont
  {Servant}},\ and\ \bibinfo {author} {\bibfnamefont {P.}~\bibnamefont
  {Simakachorn}},\ }\bibfield  {title} {\bibinfo {title} {{Kination cosmology
  from scalar fields and gravitational-wave signatures}},\ }\href@noop {} {\
  (\bibinfo {year} {2021})},\ \Eprint {https://arxiv.org/abs/2111.01150}
  {arXiv:2111.01150 [hep-ph]} \BibitemShut {NoStop}%
\bibitem [{\citenamefont {Kasai}\ \emph {et~al.}(2022)\citenamefont {Kasai},
  \citenamefont {Kawasaki},\ and\ \citenamefont {Murai}}]{Kasai:2022vhq}%
  \BibitemOpen
  \bibfield  {author} {\bibinfo {author} {\bibfnamefont {K.}~\bibnamefont
  {Kasai}}, \bibinfo {author} {\bibfnamefont {M.}~\bibnamefont {Kawasaki}},\
  and\ \bibinfo {author} {\bibfnamefont {K.}~\bibnamefont {Murai}},\ }\bibfield
   {title} {\bibinfo {title} {{Revisiting the Affleck-Dine mechanism for
  primordial black hole formation}},\ }\href
  {https://doi.org/10.1088/1475-7516/2022/10/048} {\bibfield  {journal}
  {\bibinfo  {journal} {JCAP}\ }\textbf {\bibinfo {volume} {10}},\ \bibinfo
  {pages} {048}},\ \Eprint {https://arxiv.org/abs/2205.10148} {arXiv:2205.10148
  [astro-ph.CO]} \BibitemShut {NoStop}%
\bibitem [{\citenamefont {El~Bourakadi}\ \emph {et~al.}(2023)\citenamefont
  {El~Bourakadi}, \citenamefont {Ferricha-Alami}, \citenamefont {Sakhi},
  \citenamefont {Bennai},\ and\ \citenamefont {Chakir}}]{ElBourakadi:2023pue}%
  \BibitemOpen
  \bibfield  {author} {\bibinfo {author} {\bibfnamefont {K.}~\bibnamefont
  {El~Bourakadi}}, \bibinfo {author} {\bibfnamefont {M.}~\bibnamefont
  {Ferricha-Alami}}, \bibinfo {author} {\bibfnamefont {Z.}~\bibnamefont
  {Sakhi}}, \bibinfo {author} {\bibfnamefont {M.}~\bibnamefont {Bennai}},\ and\
  \bibinfo {author} {\bibfnamefont {H.}~\bibnamefont {Chakir}},\ }\bibfield
  {title} {\bibinfo {title} {{Dark matter via Baryogenesis: Affleck-Dine
  Mechanism in the Minimal Supersymmetric Standard Model}},\ }\href@noop {} {\
  (\bibinfo {year} {2023})},\ \Eprint {https://arxiv.org/abs/2307.15541}
  {arXiv:2307.15541 [hep-ph]} \BibitemShut {NoStop}%
\bibitem [{\citenamefont {Kaup}(1968)}]{Kaup:1968zz}%
  \BibitemOpen
  \bibfield  {author} {\bibinfo {author} {\bibfnamefont {D.~J.}\ \bibnamefont
  {Kaup}},\ }\bibfield  {title} {\bibinfo {title} {{Klein-Gordon Geon}},\
  }\href {https://doi.org/10.1103/PhysRev.172.1331} {\bibfield  {journal}
  {\bibinfo  {journal} {Phys. Rev.}\ }\textbf {\bibinfo {volume} {172}},\
  \bibinfo {pages} {1331} (\bibinfo {year} {1968})}\BibitemShut {NoStop}%
\bibitem [{\citenamefont {Colpi}\ \emph {et~al.}(1986)\citenamefont {Colpi},
  \citenamefont {Shapiro},\ and\ \citenamefont {Wasserman}}]{Colpi:1986ye}%
  \BibitemOpen
  \bibfield  {author} {\bibinfo {author} {\bibfnamefont {M.}~\bibnamefont
  {Colpi}}, \bibinfo {author} {\bibfnamefont {S.~L.}\ \bibnamefont {Shapiro}},\
  and\ \bibinfo {author} {\bibfnamefont {I.}~\bibnamefont {Wasserman}},\
  }\bibfield  {title} {\bibinfo {title} {{Boson Stars: Gravitational Equilibria
  of Selfinteracting Scalar Fields}},\ }\href
  {https://doi.org/10.1103/PhysRevLett.57.2485} {\bibfield  {journal} {\bibinfo
   {journal} {Phys. Rev. Lett.}\ }\textbf {\bibinfo {volume} {57}},\ \bibinfo
  {pages} {2485} (\bibinfo {year} {1986})}\BibitemShut {NoStop}%
\bibitem [{\citenamefont {Schunck}\ and\ \citenamefont
  {Mielke}(2003)}]{Schunck:2003kk}%
  \BibitemOpen
  \bibfield  {author} {\bibinfo {author} {\bibfnamefont {F.~E.}\ \bibnamefont
  {Schunck}}\ and\ \bibinfo {author} {\bibfnamefont {E.~W.}\ \bibnamefont
  {Mielke}},\ }\bibfield  {title} {\bibinfo {title} {{General relativistic
  boson stars}},\ }\href {https://doi.org/10.1088/0264-9381/20/20/201}
  {\bibfield  {journal} {\bibinfo  {journal} {Class. Quant. Grav.}\ }\textbf
  {\bibinfo {volume} {20}},\ \bibinfo {pages} {R301} (\bibinfo {year}
  {2003})},\ \Eprint {https://arxiv.org/abs/0801.0307} {arXiv:0801.0307
  [astro-ph]} \BibitemShut {NoStop}%
\bibitem [{\citenamefont {Chavanis}\ and\ \citenamefont
  {Delfini}(2011)}]{Chavanis:2011zm}%
  \BibitemOpen
  \bibfield  {author} {\bibinfo {author} {\bibfnamefont {P.~H.}\ \bibnamefont
  {Chavanis}}\ and\ \bibinfo {author} {\bibfnamefont {L.}~\bibnamefont
  {Delfini}},\ }\bibfield  {title} {\bibinfo {title} {{Mass-radius relation of
  Newtonian self-gravitating Bose-Einstein condensates with short-range
  interactions: II. Numerical results}},\ }\href
  {https://doi.org/10.1103/PhysRevD.84.043532} {\bibfield  {journal} {\bibinfo
  {journal} {Phys. Rev. D}\ }\textbf {\bibinfo {volume} {84}},\ \bibinfo
  {pages} {043532} (\bibinfo {year} {2011})},\ \Eprint
  {https://arxiv.org/abs/1103.2054} {arXiv:1103.2054 [astro-ph.CO]}
  \BibitemShut {NoStop}%
\bibitem [{\citenamefont {Liebling}\ and\ \citenamefont
  {Palenzuela}(2023)}]{Liebling:2012fv}%
  \BibitemOpen
  \bibfield  {author} {\bibinfo {author} {\bibfnamefont {S.~L.}\ \bibnamefont
  {Liebling}}\ and\ \bibinfo {author} {\bibfnamefont {C.}~\bibnamefont
  {Palenzuela}},\ }\bibfield  {title} {\bibinfo {title} {{Dynamical boson
  stars}},\ }\href {https://doi.org/10.1007/s41114-023-00043-4} {\bibfield
  {journal} {\bibinfo  {journal} {Living Rev. Rel.}\ }\textbf {\bibinfo
  {volume} {26}},\ \bibinfo {pages} {1} (\bibinfo {year} {2023})},\ \Eprint
  {https://arxiv.org/abs/1202.5809} {arXiv:1202.5809 [gr-qc]} \BibitemShut
  {NoStop}%
\bibitem [{\citenamefont {Herdeiro}\ and\ \citenamefont
  {Radu}(2014)}]{Herdeiro:2014goa}%
  \BibitemOpen
  \bibfield  {author} {\bibinfo {author} {\bibfnamefont {C.~A.~R.}\
  \bibnamefont {Herdeiro}}\ and\ \bibinfo {author} {\bibfnamefont
  {E.}~\bibnamefont {Radu}},\ }\bibfield  {title} {\bibinfo {title} {{Kerr
  black holes with scalar hair}},\ }\href
  {https://doi.org/10.1103/PhysRevLett.112.221101} {\bibfield  {journal}
  {\bibinfo  {journal} {Phys. Rev. Lett.}\ }\textbf {\bibinfo {volume} {112}},\
  \bibinfo {pages} {221101} (\bibinfo {year} {2014})},\ \Eprint
  {https://arxiv.org/abs/1403.2757} {arXiv:1403.2757 [gr-qc]} \BibitemShut
  {NoStop}%
\bibitem [{\citenamefont {Cardoso}\ and\ \citenamefont
  {Pani}(2019)}]{Cardoso:2019rvt}%
  \BibitemOpen
  \bibfield  {author} {\bibinfo {author} {\bibfnamefont {V.}~\bibnamefont
  {Cardoso}}\ and\ \bibinfo {author} {\bibfnamefont {P.}~\bibnamefont {Pani}},\
  }\bibfield  {title} {\bibinfo {title} {{Testing the nature of dark compact
  objects: a status report}},\ }\href
  {https://doi.org/10.1007/s41114-019-0020-4} {\bibfield  {journal} {\bibinfo
  {journal} {Living Rev. Rel.}\ }\textbf {\bibinfo {volume} {22}},\ \bibinfo
  {pages} {4} (\bibinfo {year} {2019})},\ \Eprint
  {https://arxiv.org/abs/1904.05363} {arXiv:1904.05363 [gr-qc]} \BibitemShut
  {NoStop}%
\bibitem [{\citenamefont {Visinelli}(2021)}]{Visinelli:2021uve}%
  \BibitemOpen
  \bibfield  {author} {\bibinfo {author} {\bibfnamefont {L.}~\bibnamefont
  {Visinelli}},\ }\bibfield  {title} {\bibinfo {title} {{Boson stars and
  oscillatons: A review}},\ }\href {https://doi.org/10.1142/S0218271821300068}
  {\bibfield  {journal} {\bibinfo  {journal} {Int. J. Mod. Phys. D}\ }\textbf
  {\bibinfo {volume} {30}},\ \bibinfo {pages} {2130006} (\bibinfo {year}
  {2021})},\ \Eprint {https://arxiv.org/abs/2109.05481} {arXiv:2109.05481
  [gr-qc]} \BibitemShut {NoStop}%
\bibitem [{\citenamefont {Enqvist}\ and\ \citenamefont
  {Laine}(2003)}]{Enqvist:2003zb}%
  \BibitemOpen
  \bibfield  {author} {\bibinfo {author} {\bibfnamefont {K.}~\bibnamefont
  {Enqvist}}\ and\ \bibinfo {author} {\bibfnamefont {M.}~\bibnamefont
  {Laine}},\ }\bibfield  {title} {\bibinfo {title} {{Q-ball dynamics from
  atomic Bose-Einstein condensates}},\ }\href
  {https://doi.org/10.1088/1475-7516/2003/08/003} {\bibfield  {journal}
  {\bibinfo  {journal} {JCAP}\ }\textbf {\bibinfo {volume} {08}},\ \bibinfo
  {pages} {003}},\ \Eprint {https://arxiv.org/abs/cond-mat/0304355}
  {arXiv:cond-mat/0304355} \BibitemShut {NoStop}%
\bibitem [{\citenamefont {Bunkov}\ and\ \citenamefont
  {Volovik}(2007)}]{Bunkov:2007fe}%
  \BibitemOpen
  \bibfield  {author} {\bibinfo {author} {\bibfnamefont {Y.}~\bibnamefont
  {Bunkov}}\ and\ \bibinfo {author} {\bibfnamefont {G.}~\bibnamefont
  {Volovik}},\ }\bibfield  {title} {\bibinfo {title} {{Magnons condensation
  into Q-ball in He-3 - B}},\ }\href
  {https://doi.org/10.1103/PhysRevLett.98.265302} {\bibfield  {journal}
  {\bibinfo  {journal} {Phys. Rev. Lett.}\ }\textbf {\bibinfo {volume} {98}},\
  \bibinfo {pages} {265302} (\bibinfo {year} {2007})},\ \Eprint
  {https://arxiv.org/abs/cond-mat/0703183} {arXiv:cond-mat/0703183}
  \BibitemShut {NoStop}%
\bibitem [{\citenamefont {Dicke}(1954)}]{Dicke:1954zz}%
  \BibitemOpen
  \bibfield  {author} {\bibinfo {author} {\bibfnamefont {R.~H.}\ \bibnamefont
  {Dicke}},\ }\bibfield  {title} {\bibinfo {title} {{Coherence in Spontaneous
  Radiation Processes}},\ }\href {https://doi.org/10.1103/PhysRev.93.99}
  {\bibfield  {journal} {\bibinfo  {journal} {Phys. Rev.}\ }\textbf {\bibinfo
  {volume} {93}},\ \bibinfo {pages} {99} (\bibinfo {year} {1954})}\BibitemShut
  {NoStop}%
\bibitem [{\citenamefont {Zel'dovich}({\natexlab{a}})}]{Zeld1}%
  \BibitemOpen
  \bibfield  {author} {\bibinfo {author} {\bibfnamefont {Y.~B.}\ \bibnamefont
  {Zel'dovich}},\ }\bibfield  {title} {\bibinfo {title} {{GENERATION OF WAVES
  BY A ROTATING BODY}},\ }\href@noop {} {\bibfield  {journal} {\bibinfo
  {journal} {Zh. Eksp. Teor. Fiz. Pis'ma {\bf 14}, 270 (1971) [JETP Letters
  {\bf 14}, 180 (1971)]}\ } ({\natexlab{a}})}\BibitemShut {NoStop}%
\bibitem [{\citenamefont {Zel'dovich}({\natexlab{b}})}]{Zeld2}%
  \BibitemOpen
  \bibfield  {author} {\bibinfo {author} {\bibfnamefont {Y.~B.}\ \bibnamefont
  {Zel'dovich}},\ }\bibfield  {title} {\bibinfo {title} {{Amplification of
  Cylindrical Electromagnetic Waves Reflected from a Rotating Body}},\
  }\href@noop {} {\bibfield  {journal} {\bibinfo  {journal} {Zh. Eksp. Teor.
  Fiz. {\bf 62}, 2076 (1971) [JETP {\bf 35}, 1085 (1971)]}\ }
  ({\natexlab{b}})}\BibitemShut {NoStop}%
\bibitem [{\citenamefont {Cardoso}\ and\ \citenamefont
  {Dias}(2004)}]{Cardoso:2004hs}%
  \BibitemOpen
  \bibfield  {author} {\bibinfo {author} {\bibfnamefont {V.}~\bibnamefont
  {Cardoso}}\ and\ \bibinfo {author} {\bibfnamefont {O.~J.~C.}\ \bibnamefont
  {Dias}},\ }\bibfield  {title} {\bibinfo {title} {{Small Kerr-anti-de Sitter
  black holes are unstable}},\ }\href
  {https://doi.org/10.1103/PhysRevD.70.084011} {\bibfield  {journal} {\bibinfo
  {journal} {Phys. Rev. D}\ }\textbf {\bibinfo {volume} {70}},\ \bibinfo
  {pages} {084011} (\bibinfo {year} {2004})},\ \Eprint
  {https://arxiv.org/abs/hep-th/0405006} {arXiv:hep-th/0405006} \BibitemShut
  {NoStop}%
\bibitem [{\citenamefont {Dolan}(2007)}]{Dolan:2007mj}%
  \BibitemOpen
  \bibfield  {author} {\bibinfo {author} {\bibfnamefont {S.~R.}\ \bibnamefont
  {Dolan}},\ }\bibfield  {title} {\bibinfo {title} {{Instability of the massive
  Klein-Gordon field on the Kerr spacetime}},\ }\href
  {https://doi.org/10.1103/PhysRevD.76.084001} {\bibfield  {journal} {\bibinfo
  {journal} {Phys. Rev. D}\ }\textbf {\bibinfo {volume} {76}},\ \bibinfo
  {pages} {084001} (\bibinfo {year} {2007})},\ \Eprint
  {https://arxiv.org/abs/0705.2880} {arXiv:0705.2880 [gr-qc]} \BibitemShut
  {NoStop}%
\bibitem [{\citenamefont {Arvanitaki}\ \emph {et~al.}(2010)\citenamefont
  {Arvanitaki}, \citenamefont {Dimopoulos}, \citenamefont {Dubovsky},
  \citenamefont {Kaloper},\ and\ \citenamefont
  {March-Russell}}]{Arvanitaki:2009fg}%
  \BibitemOpen
  \bibfield  {author} {\bibinfo {author} {\bibfnamefont {A.}~\bibnamefont
  {Arvanitaki}}, \bibinfo {author} {\bibfnamefont {S.}~\bibnamefont
  {Dimopoulos}}, \bibinfo {author} {\bibfnamefont {S.}~\bibnamefont
  {Dubovsky}}, \bibinfo {author} {\bibfnamefont {N.}~\bibnamefont {Kaloper}},\
  and\ \bibinfo {author} {\bibfnamefont {J.}~\bibnamefont {March-Russell}},\
  }\bibfield  {title} {\bibinfo {title} {{String Axiverse}},\ }\href
  {https://doi.org/10.1103/PhysRevD.81.123530} {\bibfield  {journal} {\bibinfo
  {journal} {Phys. Rev. D}\ }\textbf {\bibinfo {volume} {81}},\ \bibinfo
  {pages} {123530} (\bibinfo {year} {2010})},\ \Eprint
  {https://arxiv.org/abs/0905.4720} {arXiv:0905.4720 [hep-th]} \BibitemShut
  {NoStop}%
\bibitem [{\citenamefont {Bredberg}\ \emph {et~al.}(2010)\citenamefont
  {Bredberg}, \citenamefont {Hartman}, \citenamefont {Song},\ and\
  \citenamefont {Strominger}}]{Bredberg:2009pv}%
  \BibitemOpen
  \bibfield  {author} {\bibinfo {author} {\bibfnamefont {I.}~\bibnamefont
  {Bredberg}}, \bibinfo {author} {\bibfnamefont {T.}~\bibnamefont {Hartman}},
  \bibinfo {author} {\bibfnamefont {W.}~\bibnamefont {Song}},\ and\ \bibinfo
  {author} {\bibfnamefont {A.}~\bibnamefont {Strominger}},\ }\bibfield  {title}
  {\bibinfo {title} {{Black Hole Superradiance From Kerr/CFT}},\ }\href
  {https://doi.org/10.1007/JHEP04(2010)019} {\bibfield  {journal} {\bibinfo
  {journal} {JHEP}\ }\textbf {\bibinfo {volume} {04}},\ \bibinfo {pages}
  {019}},\ \Eprint {https://arxiv.org/abs/0907.3477} {arXiv:0907.3477 [hep-th]}
  \BibitemShut {NoStop}%
\bibitem [{\citenamefont {Arvanitaki}\ and\ \citenamefont
  {Dubovsky}(2011)}]{Arvanitaki:2010sy}%
  \BibitemOpen
  \bibfield  {author} {\bibinfo {author} {\bibfnamefont {A.}~\bibnamefont
  {Arvanitaki}}\ and\ \bibinfo {author} {\bibfnamefont {S.}~\bibnamefont
  {Dubovsky}},\ }\bibfield  {title} {\bibinfo {title} {{Exploring the String
  Axiverse with Precision Black Hole Physics}},\ }\href
  {https://doi.org/10.1103/PhysRevD.83.044026} {\bibfield  {journal} {\bibinfo
  {journal} {Phys. Rev. D}\ }\textbf {\bibinfo {volume} {83}},\ \bibinfo
  {pages} {044026} (\bibinfo {year} {2011})},\ \Eprint
  {https://arxiv.org/abs/1004.3558} {arXiv:1004.3558 [hep-th]} \BibitemShut
  {NoStop}%
\bibitem [{\citenamefont {Pani}\ \emph {et~al.}(2012)\citenamefont {Pani},
  \citenamefont {Cardoso}, \citenamefont {Gualtieri}, \citenamefont {Berti},\
  and\ \citenamefont {Ishibashi}}]{Pani:2012vp}%
  \BibitemOpen
  \bibfield  {author} {\bibinfo {author} {\bibfnamefont {P.}~\bibnamefont
  {Pani}}, \bibinfo {author} {\bibfnamefont {V.}~\bibnamefont {Cardoso}},
  \bibinfo {author} {\bibfnamefont {L.}~\bibnamefont {Gualtieri}}, \bibinfo
  {author} {\bibfnamefont {E.}~\bibnamefont {Berti}},\ and\ \bibinfo {author}
  {\bibfnamefont {A.}~\bibnamefont {Ishibashi}},\ }\bibfield  {title} {\bibinfo
  {title} {{Black hole bombs and photon mass bounds}},\ }\href
  {https://doi.org/10.1103/PhysRevLett.109.131102} {\bibfield  {journal}
  {\bibinfo  {journal} {Phys. Rev. Lett.}\ }\textbf {\bibinfo {volume} {109}},\
  \bibinfo {pages} {131102} (\bibinfo {year} {2012})},\ \Eprint
  {https://arxiv.org/abs/1209.0465} {arXiv:1209.0465 [gr-qc]} \BibitemShut
  {NoStop}%
\bibitem [{\citenamefont {Berti}\ \emph {et~al.}(2015)\citenamefont {Berti}
  \emph {et~al.}}]{Berti:2015itd}%
  \BibitemOpen
  \bibfield  {author} {\bibinfo {author} {\bibfnamefont {E.}~\bibnamefont
  {Berti}} \emph {et~al.},\ }\bibfield  {title} {\bibinfo {title} {{Testing
  General Relativity with Present and Future Astrophysical Observations}},\
  }\href {https://doi.org/10.1088/0264-9381/32/24/243001} {\bibfield  {journal}
  {\bibinfo  {journal} {Class. Quant. Grav.}\ }\textbf {\bibinfo {volume}
  {32}},\ \bibinfo {pages} {243001} (\bibinfo {year} {2015})},\ \Eprint
  {https://arxiv.org/abs/1501.07274} {arXiv:1501.07274 [gr-qc]} \BibitemShut
  {NoStop}%
\bibitem [{\citenamefont {Marsh}(2016)}]{Marsh:2015xka}%
  \BibitemOpen
  \bibfield  {author} {\bibinfo {author} {\bibfnamefont {D.~J.~E.}\
  \bibnamefont {Marsh}},\ }\bibfield  {title} {\bibinfo {title} {{Axion
  Cosmology}},\ }\href {https://doi.org/10.1016/j.physrep.2016.06.005}
  {\bibfield  {journal} {\bibinfo  {journal} {Phys. Rept.}\ }\textbf {\bibinfo
  {volume} {643}},\ \bibinfo {pages} {1} (\bibinfo {year} {2016})},\ \Eprint
  {https://arxiv.org/abs/1510.07633} {arXiv:1510.07633 [astro-ph.CO]}
  \BibitemShut {NoStop}%
\bibitem [{\citenamefont {Cardoso}\ \emph {et~al.}(2015)\citenamefont
  {Cardoso}, \citenamefont {Brito},\ and\ \citenamefont
  {Rosa}}]{Cardoso:2015zqa}%
  \BibitemOpen
  \bibfield  {author} {\bibinfo {author} {\bibfnamefont {V.}~\bibnamefont
  {Cardoso}}, \bibinfo {author} {\bibfnamefont {R.}~\bibnamefont {Brito}},\
  and\ \bibinfo {author} {\bibfnamefont {J.~L.}\ \bibnamefont {Rosa}},\
  }\bibfield  {title} {\bibinfo {title} {{Superradiance in stars}},\ }\href
  {https://doi.org/10.1103/PhysRevD.91.124026} {\bibfield  {journal} {\bibinfo
  {journal} {Phys. Rev. D}\ }\textbf {\bibinfo {volume} {91}},\ \bibinfo
  {pages} {124026} (\bibinfo {year} {2015})},\ \Eprint
  {https://arxiv.org/abs/1505.05509} {arXiv:1505.05509 [gr-qc]} \BibitemShut
  {NoStop}%
\bibitem [{\citenamefont {East}\ and\ \citenamefont
  {Pretorius}(2017)}]{East:2017ovw}%
  \BibitemOpen
  \bibfield  {author} {\bibinfo {author} {\bibfnamefont {W.~E.}\ \bibnamefont
  {East}}\ and\ \bibinfo {author} {\bibfnamefont {F.}~\bibnamefont
  {Pretorius}},\ }\bibfield  {title} {\bibinfo {title} {{Superradiant
  Instability and Backreaction of Massive Vector Fields around Kerr Black
  Holes}},\ }\href {https://doi.org/10.1103/PhysRevLett.119.041101} {\bibfield
  {journal} {\bibinfo  {journal} {Phys. Rev. Lett.}\ }\textbf {\bibinfo
  {volume} {119}},\ \bibinfo {pages} {041101} (\bibinfo {year} {2017})},\
  \Eprint {https://arxiv.org/abs/1704.04791} {arXiv:1704.04791 [gr-qc]}
  \BibitemShut {NoStop}%
\bibitem [{\citenamefont {Baryakhtar}\ \emph {et~al.}(2017)\citenamefont
  {Baryakhtar}, \citenamefont {Lasenby},\ and\ \citenamefont
  {Teo}}]{Baryakhtar:2017ngi}%
  \BibitemOpen
  \bibfield  {author} {\bibinfo {author} {\bibfnamefont {M.}~\bibnamefont
  {Baryakhtar}}, \bibinfo {author} {\bibfnamefont {R.}~\bibnamefont
  {Lasenby}},\ and\ \bibinfo {author} {\bibfnamefont {M.}~\bibnamefont {Teo}},\
  }\bibfield  {title} {\bibinfo {title} {{Black Hole Superradiance Signatures
  of Ultralight Vectors}},\ }\href {https://doi.org/10.1103/PhysRevD.96.035019}
  {\bibfield  {journal} {\bibinfo  {journal} {Phys. Rev. D}\ }\textbf {\bibinfo
  {volume} {96}},\ \bibinfo {pages} {035019} (\bibinfo {year} {2017})},\
  \Eprint {https://arxiv.org/abs/1704.05081} {arXiv:1704.05081 [hep-ph]}
  \BibitemShut {NoStop}%
\bibitem [{\citenamefont {Baumann}\ \emph {et~al.}(2019)\citenamefont
  {Baumann}, \citenamefont {Chia},\ and\ \citenamefont
  {Porto}}]{Baumann:2018vus}%
  \BibitemOpen
  \bibfield  {author} {\bibinfo {author} {\bibfnamefont {D.}~\bibnamefont
  {Baumann}}, \bibinfo {author} {\bibfnamefont {H.~S.}\ \bibnamefont {Chia}},\
  and\ \bibinfo {author} {\bibfnamefont {R.~A.}\ \bibnamefont {Porto}},\
  }\bibfield  {title} {\bibinfo {title} {{Probing Ultralight Bosons with Binary
  Black Holes}},\ }\href {https://doi.org/10.1103/PhysRevD.99.044001}
  {\bibfield  {journal} {\bibinfo  {journal} {Phys. Rev. D}\ }\textbf {\bibinfo
  {volume} {99}},\ \bibinfo {pages} {044001} (\bibinfo {year} {2019})},\
  \Eprint {https://arxiv.org/abs/1804.03208} {arXiv:1804.03208 [gr-qc]}
  \BibitemShut {NoStop}%
\bibitem [{\citenamefont {Berti}\ \emph {et~al.}(2019)\citenamefont {Berti},
  \citenamefont {Brito}, \citenamefont {Macedo}, \citenamefont {Raposo},\ and\
  \citenamefont {Rosa}}]{Berti:2019wnn}%
  \BibitemOpen
  \bibfield  {author} {\bibinfo {author} {\bibfnamefont {E.}~\bibnamefont
  {Berti}}, \bibinfo {author} {\bibfnamefont {R.}~\bibnamefont {Brito}},
  \bibinfo {author} {\bibfnamefont {C.~F.~B.}\ \bibnamefont {Macedo}}, \bibinfo
  {author} {\bibfnamefont {G.}~\bibnamefont {Raposo}},\ and\ \bibinfo {author}
  {\bibfnamefont {J.~L.}\ \bibnamefont {Rosa}},\ }\bibfield  {title} {\bibinfo
  {title} {{Ultralight boson cloud depletion in binary systems}},\ }\href
  {https://doi.org/10.1103/PhysRevD.99.104039} {\bibfield  {journal} {\bibinfo
  {journal} {Phys. Rev. D}\ }\textbf {\bibinfo {volume} {99}},\ \bibinfo
  {pages} {104039} (\bibinfo {year} {2019})},\ \Eprint
  {https://arxiv.org/abs/1904.03131} {arXiv:1904.03131 [gr-qc]} \BibitemShut
  {NoStop}%
\bibitem [{\citenamefont {Zhu}\ \emph {et~al.}(2020)\citenamefont {Zhu},
  \citenamefont {Baryakhtar}, \citenamefont {Papa}, \citenamefont {Tsuna},
  \citenamefont {Kawanaka},\ and\ \citenamefont {Eggenstein}}]{Zhu:2020tht}%
  \BibitemOpen
  \bibfield  {author} {\bibinfo {author} {\bibfnamefont {S.~J.}\ \bibnamefont
  {Zhu}}, \bibinfo {author} {\bibfnamefont {M.}~\bibnamefont {Baryakhtar}},
  \bibinfo {author} {\bibfnamefont {M.~A.}\ \bibnamefont {Papa}}, \bibinfo
  {author} {\bibfnamefont {D.}~\bibnamefont {Tsuna}}, \bibinfo {author}
  {\bibfnamefont {N.}~\bibnamefont {Kawanaka}},\ and\ \bibinfo {author}
  {\bibfnamefont {H.-B.}\ \bibnamefont {Eggenstein}},\ }\bibfield  {title}
  {\bibinfo {title} {{Characterizing the continuous gravitational-wave signal
  from boson clouds around Galactic isolated black holes}},\ }\href
  {https://doi.org/10.1103/PhysRevD.102.063020} {\bibfield  {journal} {\bibinfo
   {journal} {Phys. Rev. D}\ }\textbf {\bibinfo {volume} {102}},\ \bibinfo
  {pages} {063020} (\bibinfo {year} {2020})},\ \Eprint
  {https://arxiv.org/abs/2003.03359} {arXiv:2003.03359 [gr-qc]} \BibitemShut
  {NoStop}%
\bibitem [{\citenamefont {Zhang}\ \emph {et~al.}(2020)\citenamefont {Zhang},
  \citenamefont {Zhang}, \citenamefont {Li},\ and\ \citenamefont
  {Guo}}]{Zhang:2020sjh}%
  \BibitemOpen
  \bibfield  {author} {\bibinfo {author} {\bibfnamefont {C.-Y.}\ \bibnamefont
  {Zhang}}, \bibinfo {author} {\bibfnamefont {S.-J.}\ \bibnamefont {Zhang}},
  \bibinfo {author} {\bibfnamefont {P.-C.}\ \bibnamefont {Li}},\ and\ \bibinfo
  {author} {\bibfnamefont {M.}~\bibnamefont {Guo}},\ }\bibfield  {title}
  {\bibinfo {title} {{Superradiance and stability of the regularized 4D charged
  Einstein-Gauss-Bonnet black hole}},\ }\href
  {https://doi.org/10.1007/JHEP08(2020)105} {\bibfield  {journal} {\bibinfo
  {journal} {JHEP}\ }\textbf {\bibinfo {volume} {08}},\ \bibinfo {pages}
  {105}},\ \Eprint {https://arxiv.org/abs/2004.03141} {arXiv:2004.03141
  [gr-qc]} \BibitemShut {NoStop}%
\bibitem [{\citenamefont {Stott}(2020)}]{Stott:2020gjj}%
  \BibitemOpen
  \bibfield  {author} {\bibinfo {author} {\bibfnamefont {M.~J.}\ \bibnamefont
  {Stott}},\ }\bibfield  {title} {\bibinfo {title} {{Ultralight Bosonic Field
  Mass Bounds from Astrophysical Black Hole Spin}},\ }\href@noop {} {\
  (\bibinfo {year} {2020})},\ \Eprint {https://arxiv.org/abs/2009.07206}
  {arXiv:2009.07206 [hep-ph]} \BibitemShut {NoStop}%
\bibitem [{\citenamefont {Baryakhtar}\ \emph {et~al.}(2021)\citenamefont
  {Baryakhtar}, \citenamefont {Galanis}, \citenamefont {Lasenby},\ and\
  \citenamefont {Simon}}]{Baryakhtar:2020gao}%
  \BibitemOpen
  \bibfield  {author} {\bibinfo {author} {\bibfnamefont {M.}~\bibnamefont
  {Baryakhtar}}, \bibinfo {author} {\bibfnamefont {M.}~\bibnamefont {Galanis}},
  \bibinfo {author} {\bibfnamefont {R.}~\bibnamefont {Lasenby}},\ and\ \bibinfo
  {author} {\bibfnamefont {O.}~\bibnamefont {Simon}},\ }\bibfield  {title}
  {\bibinfo {title} {{Black hole superradiance of self-interacting scalar
  fields}},\ }\href {https://doi.org/10.1103/PhysRevD.103.095019} {\bibfield
  {journal} {\bibinfo  {journal} {Phys. Rev. D}\ }\textbf {\bibinfo {volume}
  {103}},\ \bibinfo {pages} {095019} (\bibinfo {year} {2021})},\ \Eprint
  {https://arxiv.org/abs/2011.11646} {arXiv:2011.11646 [hep-ph]} \BibitemShut
  {NoStop}%
\bibitem [{\citenamefont {Mehta}\ \emph {et~al.}(2021)\citenamefont {Mehta},
  \citenamefont {Demirtas}, \citenamefont {Long}, \citenamefont {Marsh},
  \citenamefont {McAllister},\ and\ \citenamefont {Stott}}]{Mehta:2021pwf}%
  \BibitemOpen
  \bibfield  {author} {\bibinfo {author} {\bibfnamefont {V.~M.}\ \bibnamefont
  {Mehta}}, \bibinfo {author} {\bibfnamefont {M.}~\bibnamefont {Demirtas}},
  \bibinfo {author} {\bibfnamefont {C.}~\bibnamefont {Long}}, \bibinfo {author}
  {\bibfnamefont {D.~J.~E.}\ \bibnamefont {Marsh}}, \bibinfo {author}
  {\bibfnamefont {L.}~\bibnamefont {McAllister}},\ and\ \bibinfo {author}
  {\bibfnamefont {M.~J.}\ \bibnamefont {Stott}},\ }\bibfield  {title} {\bibinfo
  {title} {{Superradiance in string theory}},\ }\href
  {https://doi.org/10.1088/1475-7516/2021/07/033} {\bibfield  {journal}
  {\bibinfo  {journal} {JCAP}\ }\textbf {\bibinfo {volume} {07}},\ \bibinfo
  {pages} {033}},\ \Eprint {https://arxiv.org/abs/2103.06812} {arXiv:2103.06812
  [hep-th]} \BibitemShut {NoStop}%
\bibitem [{\citenamefont {Roy}\ \emph {et~al.}(2022)\citenamefont {Roy},
  \citenamefont {Vagnozzi},\ and\ \citenamefont {Visinelli}}]{Roy:2021uye}%
  \BibitemOpen
  \bibfield  {author} {\bibinfo {author} {\bibfnamefont {R.}~\bibnamefont
  {Roy}}, \bibinfo {author} {\bibfnamefont {S.}~\bibnamefont {Vagnozzi}},\ and\
  \bibinfo {author} {\bibfnamefont {L.}~\bibnamefont {Visinelli}},\ }\bibfield
  {title} {\bibinfo {title} {{Superradiance evolution of black hole shadows
  revisited}},\ }\href {https://doi.org/10.1103/PhysRevD.105.083002} {\bibfield
   {journal} {\bibinfo  {journal} {Phys. Rev. D}\ }\textbf {\bibinfo {volume}
  {105}},\ \bibinfo {pages} {083002} (\bibinfo {year} {2022})},\ \Eprint
  {https://arxiv.org/abs/2112.06932} {arXiv:2112.06932 [astro-ph.HE]}
  \BibitemShut {NoStop}%
\bibitem [{\citenamefont {Chen}\ \emph {et~al.}(2022)\citenamefont {Chen},
  \citenamefont {Roy}, \citenamefont {Vagnozzi},\ and\ \citenamefont
  {Visinelli}}]{Chen:2022nbb}%
  \BibitemOpen
  \bibfield  {author} {\bibinfo {author} {\bibfnamefont {Y.}~\bibnamefont
  {Chen}}, \bibinfo {author} {\bibfnamefont {R.}~\bibnamefont {Roy}}, \bibinfo
  {author} {\bibfnamefont {S.}~\bibnamefont {Vagnozzi}},\ and\ \bibinfo
  {author} {\bibfnamefont {L.}~\bibnamefont {Visinelli}},\ }\bibfield  {title}
  {\bibinfo {title} {{Superradiant evolution of the shadow and photon ring of
  Sgr A\ensuremath{\star}}},\ }\href
  {https://doi.org/10.1103/PhysRevD.106.043021} {\bibfield  {journal} {\bibinfo
   {journal} {Phys. Rev. D}\ }\textbf {\bibinfo {volume} {106}},\ \bibinfo
  {pages} {043021} (\bibinfo {year} {2022})},\ \Eprint
  {https://arxiv.org/abs/2205.06238} {arXiv:2205.06238 [astro-ph.HE]}
  \BibitemShut {NoStop}%
\bibitem [{\citenamefont {Siemonsen}\ \emph {et~al.}(2023)\citenamefont
  {Siemonsen}, \citenamefont {May},\ and\ \citenamefont
  {East}}]{Siemonsen:2022yyf}%
  \BibitemOpen
  \bibfield  {author} {\bibinfo {author} {\bibfnamefont {N.}~\bibnamefont
  {Siemonsen}}, \bibinfo {author} {\bibfnamefont {T.}~\bibnamefont {May}},\
  and\ \bibinfo {author} {\bibfnamefont {W.~E.}\ \bibnamefont {East}},\
  }\bibfield  {title} {\bibinfo {title} {{Modeling the black hole superradiance
  gravitational waveform}},\ }\href
  {https://doi.org/10.1103/PhysRevD.107.104003} {\bibfield  {journal} {\bibinfo
   {journal} {Phys. Rev. D}\ }\textbf {\bibinfo {volume} {107}},\ \bibinfo
  {pages} {104003} (\bibinfo {year} {2023})},\ \Eprint
  {https://arxiv.org/abs/2211.03845} {arXiv:2211.03845 [gr-qc]} \BibitemShut
  {NoStop}%
\bibitem [{\citenamefont {Bekenstein}\ and\ \citenamefont
  {Schiffer}(1998)}]{Bekenstein:1998nt}%
  \BibitemOpen
  \bibfield  {author} {\bibinfo {author} {\bibfnamefont {J.~D.}\ \bibnamefont
  {Bekenstein}}\ and\ \bibinfo {author} {\bibfnamefont {M.}~\bibnamefont
  {Schiffer}},\ }\bibfield  {title} {\bibinfo {title} {{The Many faces of
  superradiance}},\ }\href {https://doi.org/10.1103/PhysRevD.58.064014}
  {\bibfield  {journal} {\bibinfo  {journal} {Phys. Rev. D}\ }\textbf {\bibinfo
  {volume} {58}},\ \bibinfo {pages} {064014} (\bibinfo {year} {1998})},\
  \Eprint {https://arxiv.org/abs/gr-qc/9803033} {arXiv:gr-qc/9803033}
  \BibitemShut {NoStop}%
\bibitem [{\citenamefont {Brito}\ \emph {et~al.}(2015)\citenamefont {Brito},
  \citenamefont {Cardoso},\ and\ \citenamefont {Pani}}]{Brito:2015oca}%
  \BibitemOpen
  \bibfield  {author} {\bibinfo {author} {\bibfnamefont {R.}~\bibnamefont
  {Brito}}, \bibinfo {author} {\bibfnamefont {V.}~\bibnamefont {Cardoso}},\
  and\ \bibinfo {author} {\bibfnamefont {P.}~\bibnamefont {Pani}},\ }\bibfield
  {title} {\bibinfo {title} {{Superradiance}: {New Frontiers in Black Hole
  Physics}},\ }\href {https://doi.org/10.1007/978-3-319-19000-6} {\bibfield
  {journal} {\bibinfo  {journal} {Lect. Notes Phys.}\ }\textbf {\bibinfo
  {volume} {906}},\ \bibinfo {pages} {pp.1} (\bibinfo {year} {2015})},\ \Eprint
  {https://arxiv.org/abs/1501.06570} {arXiv:1501.06570 [gr-qc]} \BibitemShut
  {NoStop}%
\bibitem [{\citenamefont {Saffin}\ \emph {et~al.}(2023)\citenamefont {Saffin},
  \citenamefont {Xie},\ and\ \citenamefont {Zhou}}]{Saffin:2022tub}%
  \BibitemOpen
  \bibfield  {author} {\bibinfo {author} {\bibfnamefont {P.~M.}\ \bibnamefont
  {Saffin}}, \bibinfo {author} {\bibfnamefont {Q.-X.}\ \bibnamefont {Xie}},\
  and\ \bibinfo {author} {\bibfnamefont {S.-Y.}\ \bibnamefont {Zhou}},\
  }\bibfield  {title} {\bibinfo {title} {{$Q$-ball Superradiance}},\ }\href
  {https://doi.org/10.1103/PhysRevLett.131.111601} {\bibfield  {journal}
  {\bibinfo  {journal} {Phys. Rev. Lett.}\ }\textbf {\bibinfo {volume} {131}},\
  \bibinfo {pages} {111601} (\bibinfo {year} {2023})},\ \Eprint
  {https://arxiv.org/abs/2212.03269} {arXiv:2212.03269 [hep-th]} \BibitemShut
  {NoStop}%
\bibitem [{\citenamefont {Cardoso}\ \emph {et~al.}(2023)\citenamefont
  {Cardoso}, \citenamefont {Vicente},\ and\ \citenamefont
  {Zhong}}]{Cardoso:2023dtm}%
  \BibitemOpen
  \bibfield  {author} {\bibinfo {author} {\bibfnamefont {V.}~\bibnamefont
  {Cardoso}}, \bibinfo {author} {\bibfnamefont {R.}~\bibnamefont {Vicente}},\
  and\ \bibinfo {author} {\bibfnamefont {Z.}~\bibnamefont {Zhong}},\ }\bibfield
   {title} {\bibinfo {title} {{Energy Extraction from Q-balls and Other
  Fundamental Solitons}},\ }\href
  {https://doi.org/10.1103/PhysRevLett.131.111602} {\bibfield  {journal}
  {\bibinfo  {journal} {Phys. Rev. Lett.}\ }\textbf {\bibinfo {volume} {131}},\
  \bibinfo {pages} {111602} (\bibinfo {year} {2023})},\ \Eprint
  {https://arxiv.org/abs/2307.13734} {arXiv:2307.13734 [hep-th]} \BibitemShut
  {NoStop}%
\bibitem [{\citenamefont {Gao}\ \emph {et~al.}(2023)\citenamefont {Gao},
  \citenamefont {Saffin}, \citenamefont {Wang}, \citenamefont {Xie},\ and\
  \citenamefont {Zhou}}]{Gao:2023gof}%
  \BibitemOpen
  \bibfield  {author} {\bibinfo {author} {\bibfnamefont {H.-Y.}\ \bibnamefont
  {Gao}}, \bibinfo {author} {\bibfnamefont {P.~M.}\ \bibnamefont {Saffin}},
  \bibinfo {author} {\bibfnamefont {Y.-J.}\ \bibnamefont {Wang}}, \bibinfo
  {author} {\bibfnamefont {Q.-X.}\ \bibnamefont {Xie}},\ and\ \bibinfo {author}
  {\bibfnamefont {S.-Y.}\ \bibnamefont {Zhou}},\ }\bibfield  {title} {\bibinfo
  {title} {{Boson Star Superradiance}},\ }\href@noop {} {\  (\bibinfo {year}
  {2023})},\ \Eprint {https://arxiv.org/abs/2306.01868} {arXiv:2306.01868
  [gr-qc]} \BibitemShut {NoStop}%
\bibitem [{\citenamefont {Volkov}\ and\ \citenamefont
  {Wöhnert}(2002)}]{Volkov_2002}%
  \BibitemOpen
  \bibfield  {author} {\bibinfo {author} {\bibfnamefont {M.~S.}\ \bibnamefont
  {Volkov}}\ and\ \bibinfo {author} {\bibfnamefont {E.}~\bibnamefont
  {Wöhnert}},\ }\bibfield  {title} {\bibinfo {title} {{Spinning $Q$-balls}},\
  }\bibfield  {journal} {\bibinfo  {journal} {Physical Review D}\ }\textbf
  {\bibinfo {volume} {66}},\ \href {https://doi.org/10.1103/physrevd.66.085003}
  {10.1103/physrevd.66.085003} (\bibinfo {year} {2002}),\ \Eprint
  {https://arxiv.org/abs/hep-th/0205157} {arXiv:hep-th/0205157 [hep-th]}
  \BibitemShut {NoStop}%
\bibitem [{\citenamefont {Press}\ \emph {et~al.}(1992)\citenamefont {Press},
  \citenamefont {Teukolsky}, \citenamefont {Vetterling},\ and\ \citenamefont
  {Flannery}}]{NRC}%
  \BibitemOpen
  \bibfield  {author} {\bibinfo {author} {\bibfnamefont {W.~H.}\ \bibnamefont
  {Press}}, \bibinfo {author} {\bibfnamefont {S.~A.}\ \bibnamefont
  {Teukolsky}}, \bibinfo {author} {\bibfnamefont {W.~T.}\ \bibnamefont
  {Vetterling}},\ and\ \bibinfo {author} {\bibfnamefont {B.~P.}\ \bibnamefont
  {Flannery}},\ }\href@noop {} {\emph {\bibinfo {title} {Numerical Recipes in
  C: The Art of Scientific Computing Second Edition}}}\ (\bibinfo  {publisher}
  {Cambridge University Press},\ \bibinfo {year} {1992})\BibitemShut {NoStop}%
\bibitem [{\citenamefont {Dvali}\ \emph {et~al.}(2023)\citenamefont {Dvali},
  \citenamefont {Kaikov}, \citenamefont {Kuhnel}, \citenamefont
  {Valbuena-Bermudez},\ and\ \citenamefont {Zantedeschi}}]{Dvali:2023qlk}%
  \BibitemOpen
  \bibfield  {author} {\bibinfo {author} {\bibfnamefont {G.}~\bibnamefont
  {Dvali}}, \bibinfo {author} {\bibfnamefont {O.}~\bibnamefont {Kaikov}},
  \bibinfo {author} {\bibfnamefont {F.}~\bibnamefont {Kuhnel}}, \bibinfo
  {author} {\bibfnamefont {J.~S.}\ \bibnamefont {Valbuena-Bermudez}},\ and\
  \bibinfo {author} {\bibfnamefont {M.}~\bibnamefont {Zantedeschi}},\
  }\bibfield  {title} {\bibinfo {title} {{Vortex Effects in Merging Black Holes
  and Saturons}},\ }\href@noop {} {\  (\bibinfo {year} {2023})},\ \Eprint
  {https://arxiv.org/abs/2310.02288} {arXiv:2310.02288 [hep-ph]} \BibitemShut
  {NoStop}%
\end{thebibliography}%

\end{document}